\documentclass[traditabstract]{aa}

\newcommand{\lsim}{\lower 2pt \hbox{$\, \buildrel {\scriptstyle
<}\over {\scriptstyle \sim}\,$}}  \newcommand{\gsim}{\lower 2pt
\hbox{$\, \buildrel {\scriptstyle >}\over {\scriptstyle \sim}\,$}}
\usepackage{appendix}
\usepackage{amsmath}
\usepackage{graphicx}
\usepackage{txfonts}
\usepackage{natbib}
\usepackage{longtable}
\usepackage{rotating}
\usepackage{lscape}
\usepackage{float}
\usepackage{subfig}
\usepackage{multirow}
\usepackage{array}
\usepackage{fixltx2e}
\usepackage{hyperref}
\usepackage{breakurl}

\newcommand{\Ha}{H$\alpha$}
\newcommand{\Pa}{Pa$\alpha$}
\newcommand{\Brd}{Br$\delta$}
\newcommand{\Brg}{Br$\gamma$}

\newcommand{\Pal}{Pa$\alpha\,\lambda$1.876${\mu}$m}

\newcommand{\Brgl}{Br$\gamma\,\rm\lambda$2.166${\mu}$m}

\newcommand{\Hcerounol}{\mbox{H$_{2}$ 1-0S(1)${\lambda}$2.122${\mu}$m}}

\newcommand{\FeII}{[FeII]$\rm\lambda$1.644\,$\mu$m}
\newcommand{\HeI}{HeI$\rm\lambda$2.059\,$\mu$m}

\newcommand{\LTIR}{L$_{\rm IR}$}
\newcommand{\Lsolar}{L$_{\odot}$}

\newcommand{\dg}{\ensuremath{^\circ}}

\newcommand\nodata{ ~$\cdots$~ }

\newcounter{subfig}
\newcounter{fake_fig}

\newcounter{subtab}
\newcounter{fake_table}

\newcolumntype{x}[1]{%
>{\raggedleft\hspace{0pt}}p{#1}}%

\newcolumntype{z}[1]{%
>{\raggedright\hspace{0pt}}p{#1}}%

\begin{document}

\title{VLT-SINFONI integral field spectroscopy of low-z luminous and ultraluminous infrared galaxies}

\subtitle{I. Atlas of the 2D gas structure}

\author{J. Piqueras L\'opez\inst{1}
\and L. Colina\inst{1}
\and S. Arribas\inst{1}
\and A. Alonso-Herrero\inst{1,2}
\and A. G. Bedregal\inst{3,4}}
\offprints{Javier Piqueras L\'opez\\ {\tt piqueraslj@cab.inta-csic.es} \smallskip}

\institute{
Centro de Astrobiolog\'ia (INTA-CSIC), Ctra de Torrej\'on a Ajalvir, km 4, 28850, Torrej\'on de Ardoz, Madrid, Spain
\and Instituto de F\'isica de Cantabria, CSIC-UC, Avenida de los Castros S/N, 39005 Santander, Spain
\and Departamento de Astrof\'isica, facultad de F\'isicas, Universidad Complutense de Madrid, 28040 Madrid, Spain
\and Minnesota Institute for Astrophysics, University of Minnesota, 116 Church Street SE, Minneapolis, MN55455, USA
}

\date{Received 9 April 2012 /
Accepted 9 August 2012}

\abstract{We present an atlas of a sample of local ($z<0.1$) LIRGs (10) and ULIRGs (7) covering the luminosity range log(\LTIR/\Lsolar)$=11.1-12.4$. The atlas is based on near-infrared H (1.45 - 1.85 ${\mu}$m) and K-band (1.95 - 2.45 ${\mu}$m) VLT-SINFONI integral field spectroscopy (IFS). The atlas presents the ionised, partially ionised, and warm molecular gas two-dimensional flux distributions and kinematics over a FoV of $\sim3\times3$\,kpc (LIRGs) and $\sim12\times12$\,kpc (ULIRGs) and with average linear resolutions of $\sim$0.2\,kpc and $\sim$0.9\,kpc, respectively. The different phases of the gas show a wide morphological variety with the nucleus as the brightest \Brg\ source for $\sim$33\% of the LIRGs and $\sim$71\% of the ULIRGs, whereas all the LIRGs and ULIRGs have their maximum H$_2$ emission in their nuclear regions. In LIRGs, the ionised gas distribution is dominated by the emission from the star-forming rings or giant HII regions in the spiral arms. The \Brg\ and [FeII] line at 1.644\,$\mu$m trace the same structures, although the emission peaks at different locations in some of the objects, and the [FeII] seems to be more extended and diffuse. The ULIRG subsample is at larger distances and contains mainly pre-coalescence interacting systems. Although the peaks of the molecular gas emission and the continuum coincide in $\sim71$\% of the ULIRGs,  regions with intense \Pa\ (\Brg) emission tracing luminous star-forming regions located at distances of 2--4\,kpc away from the nucleus are also detected, usually associated with secondary nuclei or tidal tails. LIRGs have mean observed (i.e. uncorrected for internal extinction) SFR surface densities of about 0.4 to 0.9 M$_{\odot}$ yr$^{-1}$ kpc$^{-2}$ over large areas (4--9 kpc$^2$) with peaks of about 2$-$2.5 M$_{\odot}$ yr$^{-1}$ kpc$^{-2}$ in the smaller regions (0.16\,kpc$^2$) associated with the nucleus of the galaxy or the brightest \Brg\ region. ULIRGs do have similar average SFR surface densities for the integrated emitting regions of $\sim$0.4\,M$_{\odot}$ yr$^{-1}$ kpc$^{-2}$  in somewhat larger areas (100$-$200 kpc$^2$) and for the \Pa\ peak ($\sim$ 2\,M$_{\odot}$ yr$^{-1}$ kpc$^{-2}$ in 4 kpc$^2$).  The observed gas kinematics in LIRGs is primarily due to rotational motions around the centre of the galaxy, although local deviations associated with radial flows and/or regions of higher velocity dispersions are present. The ionised and molecular gas share the same kinematics (velocity field and velocity dispersion) to first order, showing slight differences in the velocity amplitudes (peak-to-peak) in some cases, whereas the average velocity dispersions are compatible within uncertainties. As expected, the kinematics of the ULIRG subsample is more complex, owing to the interacting nature of the objects of the sample.}


\maketitle 
%

\section{Introduction}

The \emph{Infrared Astronomical Satellite} (\emph{IRAS}) discovered  a population of galaxies with their bolometric luminosities dominated by its mid- and far-infrared emission (\citealt{Soifer:1984p8204}, \citealt{Sanders:1996p845}). Although the number density of these luminous (LIRGs; $10^{11}$\Lsolar$<$\LTIR$< 10^{12}$\Lsolar) and ultraluminous (ULIRGs; $10^{12}$\Lsolar$<$\LTIR$< 10^{13}$\Lsolar) infrared galaxies is low locally (\citealt{Sanders:1996p845}), their number increases steadily up to redshift of $\sim$ 2.5 and dominates at redshifts  $\sim$ 1.5 and above (\citealt{PerezGonzalez:2005p4031}, \citealt{Lonsdale:2006p4228}, \citealt{Sargent:2012ApJ747}). The energy output of the local U/LIRGs is now established as mainly due to massive starbursts with a small AGN contribution for LIRGs, whereas the contribution from the AGN increases with \LTIR\ and dominates bolometrically at the very high \LTIR\ end of ULIRGs (e.g. \citealt{Nardini:2010p405}, \citealt{Alonso-Herrero:2012p744} and references therein). Morphological studies show that most/all local ULIRGs show clear signs of on going interactions or recent mergers between two or more gas-rich spirals (e.g. \citealt{Murphy:1996p8249}, \citealt{Borne:2000ApJ529}, \citealt{Veilleux:2002p760}, \citealt{Dasyra:2006ApJ651}). LIRGs are, on the other hand, mostly normal spirals where some are involved in interactions (\citealt{Arribas:2004p127}, \citealt{Haan:2011p141}). 

In recent years, optical integral field spectroscopy (IFS) of representative samples of local LIRGs (\citealt{Arribas:2008p4403}, \citealt{AlonsoHerrero:2009p3373}) and ULIRGs \citep{GarciaMarin:2009p4348} have been performed with the goal of investigating the nature of the ionisation sources \citep{Monreal-Ibero:2010p517}, the structure of the star-forming regions (\citealt{2011A&A...527A..60R}, \citealt{Arribas:2012p1203}), the 2D internal dust/extinction distribution \citep{GarciaMarin:2009p8459}, and the gas kinematics (\citealt{Colina:2005p3767}, \citealt{AlonsoHerrero:2009p3373}). In parallel, a considerable effort has been made to investigate the nature of star-forming galaxies at redshifts between 1 and 3 (e.g. \citealt{ForsterSchreiber:2006ApJ645}, \citeyear{ForsterSchreiber:2009p706}, \citeyear{ForsterSchreiber:2011p731}, \citealt{Law:2009p697}, \citealt{Wright:2009p699}, \citealt{Wisnioski:2011p417}, \citealt{Epinat:2012ga}, \citealt{2012arXiv1202.3107V}). The advent of IFS has allowed spatially and spectrally resolved studies of optically/UV selected galaxies at early stages in their evolution. Such studies map the morphologies and kinematics of the gas and stars and have demonstrated that massive star-forming galaxies either appear to be large massive rotating disks (\citealt{ForsterSchreiber:2011p731}, \citealt{Wisnioski:2011p417}, \citealt{Epinat:2012ga}) or are found in highly disturbed mergers (\citealt{ForsterSchreiber:2006ApJ645}, \citealt{Epinat:2012ga}). 

The local population of LIRGs and ULIRGs therefore represents the closest examples of the two modes of formation of massive star-forming galaxies at high redshifts, during the peak of star formation in the history of the Universe. Their distances offer the possibility of investigating their physical processes, taking advantage of the high spatial resolution and S/N achieved. The detailed study of these mechanisms on physical scales of a few hundred parsecs can then be applied in more distant galaxies, where such a level of detail is extremely challenging, or not even possible. This is the first paper in a series presenting new H- and K-band SINFONI (\emph{Spectrograph for INtegral Field Observations in the Near Infrared}, \citealt{Eisenhauer:2003p8484}), seeing-limited observations of a sample of local LIRGs and ULIRGs ($z<0.1$), for which previous optical IFS is already available (see references above). The aim of this paper is to describe the general 2D properties of the whole sample and to lay the foundations for further detailed studies. These studies, to be addressed in forthcoming publications, will focus on the structure and excitation mechanisms of the ionised, partially-ionised and warm molecular gas, the distribution of the different stellar populations, and the stellar and multi-phase gas kinematics. 

The paper is organised as follows. Section \ref{section:sample} gives details about the sample.  Section \ref{section:observations} contains the description of the observations and techniques that have been used to reduce and calibrate the data, and the procedures applied to obtain the maps of the emission lines. Section \ref{section:overview} includes a general overview of the data and the physical processes of the line emitting gas and stellar populations. In Section \ref{section:gas}, we discuss the 2D properties as inferred from the SINFONI spectral maps, focussing on the general aspects of the morphology and kinematics of the gas emission. Finally, Section \ref{section:summary} includes a brief summary of the paper, and Appendix \ref{notes} comprises the notes on individual sources of the galaxies of our sample.

\section{The sample}
\label{section:sample}

The sample is part of a larger survey \citep{Arribas:2008p4403} of local LIRGs and ULIRGs observed with different optical IFS facilities including INTEGRAL+WYFFOS \citep{Arribas:1998p8476} at the 4.2\,m William Herschel Telescope, VLT-VIMOS (\emph{VIsible MultiObject Spectrograph}, \citealt{LeFevre:2003p8480}), and PMAS (\emph{Potsdam MultiAperture Spectrophotometer}, \citealt{Roth:2005p4504}). It covers the whole range of LIRG and ULIRG infrared luminosities and the different morphologies observed in this class of objects, by sampling galaxies in both hemispheres. 

The present SINFONI sample comprises a set of ten LIRGs and seven ULIRGs covering a range in luminosity of log(\LTIR/\Lsolar)$=11.10-12.43$ (see Table~\ref{table:sample}). The objects were selected to cover a representative sample of the different morphological types of LIRGs and ULIRGs, although this is not complete in either flux or distance. All the LIRGs of the sample were selected from the volume-limited sample of \cite{AlonsoHerrero:2006p4703}, whereas all the ULIRGs with the exception of \object{IRAS 06206-6315} and \object{IRAS 21130-4446} come from the \emph{IRAS Bright Galaxy Survey} (\citealt{Soifer:1989AJ98}, \citealt{Sanders:1995AJ110}). Our sample contains objects with intense star formation, AGN activity, isolated galaxies, strongly interacting systems, and mergers. The mean redshift of the LIRGs and ULIRGs subsamples is $z_{\rm LIRGs}=0.014$ and $z_{\rm ULIRGs}=0.072$, and the mean luminosities are log(\LTIR/\Lsolar)$=11.33$ and log(\LTIR/\Lsolar)$=12.29$, respectively.

\begin{table*}
\caption{\label{table:sample}The SINFONI sample}
{\tiny
\centering
\resizebox{1.0\textwidth}{!}{
\begin{tabular}{cccccccccc}
\hline
\hline
     ID1      &   ID2   &$\alpha$& $\delta$& z &   D       &     Scale    & log \LTIR & Classification        &  References\\
Common  &   IRAS  &(J2000) & (J2000) &    & (Mpc)    &(pc/arcsec) & (L$_\odot$)  &   & \\
     (1)       &   (2)     &   (3)      &   (4)      &  (5)   &    (6)      &      (7)     &        (8)         &    (9)                 & (10) \\
\hline
\object{IRAS 06206-6315} & \object{IRAS 06206-6315} & 06h21m01.21s & -63\dg17'23\farcs5 & 0.092441& 425 & 1726 & 12.31 & Sy2 & 1,2,3\\
\object{NGC 2369} & \object{IRAS 07160-6215} & 07h16m37.73s & -62\dg20'37\farcs4 & 0.010807& 48.6 & 230 & 11.17 & Composite & 4\\
\object{NGC 3110} & \object{IRAS 10015-0614} & 10h04m02.11s & -06\dg28'29\farcs2 & 0.016858& 78.4 & 367 & 11.34 &  Composite & 4\\
\object{NGC 3256} & \object{IRAS 10257-4338} & 10h27m51.27s & -43\dg54'13\farcs8 & 0.009354 & 44.6 & 212 & 11.74 & HII, Starburst & 1,5\\
\object{ESO 320-G030} & \object{IRAS 11506-3851} & 11h53m11.72s & -39\dg07'48\farcs9 & 0.010781& 51.1 & 242 & 11.35 &  HII & 4\\
\object{IRAS 12112+0305} & \object{IRAS 12112+0305} & 12h13m46.00s & +02\dg48'38\farcs0 & 0.073317& 337 & 1416 & 12.38 &  LINER & 1,2\\
\object{IRASF 12115-4656} & \object{IRAS 12115-4657} & 12h14m12.84s & -47\dg13'43\farcs2 & 0.018489 & 84.4 & 394 & 11.10 & HII & 1\\
\object{NGC 5135} & \object{IRAS 13229-2934} & 13h25m44.06s & -29\dg50'01\farcs2 & 0.013693& 63.5 & 299 & 11.33 & HII, Sy2 & 1,6\\
\object{IRAS 14348-1447} & \object{IRAS 14348-1447} & 14h37m38.40s & -15\dg00'20\farcs0 & 0.083000& 382 & 1575 & 12.41 & LINER & 1,2\\
\object{IRASF 17138-1017} & \object{IRAS 17138-1017} & 17h16m35.79s & -10\dg20'39\farcs4 & 0.017335& 75.3 & 353 & 11.42 & HII & 1\\
\object{IRAS 17208-0014} & \object{IRAS 17208-0014} & 17h23m21.95s & -00\dg17'00\farcs9 & 0.042810& 189 & 844 & 12.43 & LINER & 1,6\\
\object{IC 4687} & \object{IRAS 18093-5744} & 18h13m39.63s & -57\dg43'31\farcs3 & 0.017345 & 75.1 & 352 & 11.44 & HII & 1, 6 \\
\object{IRAS 21130-4446} & \object{IRAS 21130-4446} & 21h16m18.52s & -44\dg33'38\farcs0 & 0.092554& 421 & 1712 & 12.22 & HII & 7\\
\object{NGC 7130} & \object{IRAS 21453-3511} & 21h48m19.50s & -34\dg57'04\farcs7 & 0.016151 & 66.3 & 312 & 11.34 & HII, Sy2, LINER & 1,6\\
\object{IC 5179} & \object{IRAS 22132-3705} & 22h16m09.10s & -36\dg50'37\farcs4 & 0.011415& 45.6 & 216 & 11.12 &  HII & 4\\
\object{IRAS 22491-1808} & \object{IRAS 22491-1808} & 22h51m49.26s & -17\dg52'23\farcs5 & 0.077760& 347 & 1453 & 12.23 & HII & 1,2\\
\object{IRAS 23128-5919} & \object{IRAS 23128-5919} & 23h15m46.78s & -59\dg03'15\farcs6 & 0.044601& 195 & 869 & 12.04 & Sy2, LINER &1,3,6\\
\hline
\hline
\end{tabular}}
\tablefoot{Cols. (3) and (4): right ascension and declination from the NASA Extragalactic Database (NED). Col. (5): redshift from NED. Cols. (6) and (7): Luminosity distance and scale from Ned Wright's Cosmology Calculator \citep{Wright:2006p4236} given h$_{0}$ = 0.70, $\Omega_{\rm M}$ = 0.7, $\Omega_{\rm M}$ = 0.3. Col. (8): \LTIR (8--1000$\mu$m) calculated from the IRAS flux densities $f_{12}$, $f_{25}$, $f_{60}$ and $f_{100}$ \citep{Sanders:2003p1433}, using the expression in \cite{Sanders:1996p845} Col. (9): Spectroscopic classification based on the nuclear optical spectra from the literature. Galaxies classified as composite are likely to be a combination of AGN activity and star formation.} \tablebib{ 1:\cite{Sanders:2003p1433}, 2: \cite{Lutz:1999ApJ517L.13L}, 3:\cite{Duc:1997p7880}, 4:\cite{PereiraSantaella:2011p535}, 5:\cite{Wamsteker:1985p7881}, 6:\cite{VeronCetty:2006p4241},  7:\cite{Farrah:2003iy}}
}
\end{table*}

\section{Observations, data reduction, and analysis}
\label{section:observations}
\subsection{SINFONI observations}

The observations were obtained in service mode using the near-infrared spectrometer SINFONI of the VLT, during the periods 77B, 78B, and 81B (from April 2006 to July 2008). All the galaxies in the sample were observed in the K band (1.95--2.45\,$\mu$m) with a plate scale of 0\farcs125$\times$0\farcs250\,pixel$^{-1}$ yielding an FoV of 8"x8" in a 2D 64x64 spaxel frame\footnote{A detailed description of the correspondence between the pixels in the focal plane of the instrument and the reconstructed cube can be found in the user manual of the instrument. \url{http://www.eso.org/sci/facilities/paranal/instruments/sinfoni/doc/}}. The subsample of LIRGs was also observed in the H band (1.45--1.85\,$\mu$m) with the same scale, so the [FeII] line at 1.64\,$\mu$m rest frame could be observed. The spectral resolution for this configuration is R$\sim$3000 for H-band and R$\sim$4000 for K-band, and the full-width-at-half-maximum (FWHM) as measured from the OH sky lines is $6.6\pm0.3$\,\AA\ for the H band and$6.0\pm0.6$\,\AA\ for the K band with a dispersion of 1.95\,\AA/pix and 2.45\,\AA/pix, respectively.

Given the limited field of view (FoV) of 8"x8" in the 250\,mas configuration provided by SINFONI , we are sampling the central regions of the objects. However, owing to the jittering process and the different pointings used in some objects, the final FoV of the observations extends beyond that value, typically from $\lsim$9"$\times$9" up to $\lsim$12"$\times$12" or more.  That is translated to an average coverage of the central regions of $\sim3\times3$\,kpc for the LIRGs and of $\sim12\times12$\,kpc for the ULIRGs subsample. Due to this constraint, some of the more extended galaxies or those with multiple nuclei were observed in different pointings, each located in regions of interest. Our seeing-limited observations have an average resolution of $\sim$0.63\,arcsec (FWHM) that corresponds to $\sim$0.2\,kpc and $\sim$0.9\,kpc.

Owing to the strong and quick variation of the IR sky emission, the observations were split into short exposures of 150\,s each, following a jittering O-S-S-O pattern for sky and on-source frames. The detailed information about the observed bands and integration time for each object is shown in Table \ref{table:observations}. Besides the objects of the sample, a set of spectrophotometric standard stars and their respective sky frames were observed to correct for the instrument response and to flux-calibrate the data. As shown in Table \ref{table:observations},  \object{NGC 3256} was observed in different pointings for the different bands because of an error during the implementation of the Phase 2 template.

\begin{table}
\caption{Observed bands and integration times}
\centering
{
\begin{tabular}{ccc}
\hline
\hline
Object / Pointing & Observed Bands & t$_{\rm exp}$(s) per band\\
(1) & (2) & (3)\\
\hline
IRAS 06206-6315 & K & 2550 \\
NGC 2369 & H, K & 2250, 2250 \\
NGC 3110 & H, K & 2250, 1200 \\
NGC 3256-N & H & 2250 \\
NGC 3256-S & K & 1950 \\
NGC 3256-W & H, K & 750, 950 \\
ESO 320-G030 & H, K & 2700, 1650 \\
IRAS 12112+0305-N & K & 2550 \\
IRAS 12112+0305-S & K & 2550 \\
IRASF 12115-4656-E & K & 2400 \\
IRASF 12115-4656-W & K & 2400 \\
NGC 5135 & H, K & 2400, 1500 \\
IRAS 14348-1447-N & K & 2550 \\
IRAS 14348-1447-S & K & 1050 \\
IRASF 17138-1017 & H, K & 5550, 2850 \\
IRAS 17208-0014 & K & 3450 \\
IC 4687 & H, K & 3000, 2400 \\
IRAS 21130-4446 & K & 3000 \\
NGC 7130-N & H, K & 2400, 2550 \\
NGC 7130-S & H, K & 2400, 2400\\
IC 5179-E & H, K & 2400, 2400 \\
IC 5179-W & H, K & 2400, 2400 \\
IRAS 22491-1808 & K & 4350 \\
IRAS 23128-5919-N & K & 2700 \\
IRAS 23128-5919-S & K & 2700 \\
\hline
\hline
\end{tabular}}
\tablefoot{Col. (3): Total integration time on-target for each band in seconds.}
\label{table:observations}
\end{table}

\subsection{Data reduction}

The calibration process was performed using the standard ESO pipeline ESOREX (version 2.0.5). The usual corrections of dark subtraction, flat fielding, detector linearity, geometrical distortion, and wavelength calibration were applied to each object and sky frame, prior to the sky subtraction from each object frame. The method used to remove the background sky emission is outlined in \cite{Davies:2007p2525}. We used our own IDL routines to perform the flux calibration on every single cube and to reconstruct a final data cube for each pointing, while taking the relative shifts in the jittering pattern into account. For those objects with different pointings, the final data cubes were combined to build a final mosaic.

The flux calibration was performed in two steps. Firstly, to obtain the atmospheric transmission curves, we extracted the spectra of the standard stars with an aperture of 5$\sigma$ of the best 2D Gaussian fit of a collapsed image. The spectra were then normalised by a black body profile at the T$_{\rm eff}$ listed in the Tycho-2 Spectral Type Catalog \citep{Wright:2003p4322}, taking the more relevant absorption spectral features of the stars into account. As discussed in \cite{Bedregal:2009p2426}, in most cases the only spectral features in absorption are the Brackett series so we modelled them using a Lorentzian profile. The result is a ``sensitivity function" that accounts for the atmospheric transmission.

Secondly, the spectra of the star was converted from counts to physical units. We made use of the response curves of 2MASS filters \citep{Cohen:2003p3372} to obtain the magnitude in counts of the standard stars and the H and K magnitudes from the 2MASS catalogue \citep{Skrutskie:2006p5780} to translate these values to physical units. Every individual cube was then divided by the ``sensitivity function" and multiplied by the conversion factor to obtain a full-calibrated data cube. The typical relative uncertainty for the conversion factor is $\sim$5\% for both bands.

\subsection{Line fitting}

The maps of the brightest emission lines were constructed by fitting a Gaussian profile on a spaxel-by-spaxel basis. We made use of the IDL routine MPFIT \citep{Markwardt:2009p7399} and developed our own routines to perform the fitting of the cubes in an automated fashion. For each object and every line, we obtained the integrated flux, equivalent width, radial velocity, and velocity dispersion maps. To account for the instrumental broadening, we made use of OH sky lines for each band at 1.690\,$\mu$m and 2.190\,$\mu$m.

\subsection{Voronoi binning}

\begin{table*}[t]
\caption{Minimum signal-to-noise (S/N) thresholds per bin used for the Voronoi binning}
\centering
{
\begin{tabular}{cccccccc}
\hline
\hline
Object & \Pa & \Brg & H$_{2}$ 1-0S(1) & HeI & [FeII] & H-band FoV & K-band FoV\\
 &  1.876\,$\mu$m&  2.166\,$\mu$m&  2.122\,$\mu$m&  2.059\,$\mu$m&  1.644\,$\mu$m& (kpc$^2$) & (kpc$^2$)\\
\hline
IRAS 06206-6315 & 15 & \nodata & 12 & \nodata & \nodata & \nodata & 186.1\\
NGC 2369 & \nodata & 13 & 15 & 10 & 15 & 2.8 & 3.0\\
NGC 3110 & \nodata & 15 & 15 & 10 & 10& 8.3 & 7.4\\
NGC 3256 & \nodata & 25 & 20 & 15 & 30 & 4.9 & 4.6\\
ESO 320-G030 & \nodata & 25 & 10 & 15 & 30& 3.2& 3.1\\
IRAS 12112+0305 & 20 & \nodata & 12 & 10 & \nodata & \nodata & 101.3\\
IRASF 12115-4656 & \nodata & 15 & 13 & 10 & \nodata& \nodata & 14.4\\
NGC 5135 & \nodata & 20 & 15 & 15 & 25 & 5.4 & 4.7\\
IRAS 14348-1447 & 20 & \nodata & 13 & 10 & \nodata & \nodata &169.7\\
IRASF 17138-1017 & \nodata & 17 & 20 & 15 & 20 & 6.3 & 6.1\\
IRAS 17208-0014 & 25 & \nodata & 13 & 10 & \nodata & \nodata & 32.3\\
IC 4687 & \nodata & 25 & 20 & 20 & 25 & 7.7 & 6.9\\
IRAS 21130-4446 & 25 & \nodata & 8 & 10 & \nodata & \nodata & 38.3\\
NGC 7130 & \nodata & 15 & 13 & 12 & 12& 10.0 & 9.5\\
IC 5179 & \nodata & 20 & 15 & 13 & 15 & 5.2 & 5.2 \\
IRAS 22491-1808 & 20 & \nodata & 9 & 10 & \nodata & \nodata & 82.5\\
IRAS 23128-5919 & 20 & \nodata & 15 & 15 & \nodata & \nodata & 34.3\\
\hline
\hline
\end{tabular}}
\tablefoot{The last two columns give the total area of the FoV used to derive the integrated fluxes for the emission lines and the stacked spectra for the H- and K-bands, respectively.}
\label{table:S/N}
\end{table*}

Before extracting the kinematics, the data were binned using the Voronoi method by \cite{Cappellari:2003p4908} to achieve a minimum S/N over the entire FoV. This technique employs bins of approximately circular shape to divide the FoV, which is described in terms of a set of points called \emph{generators}. Every spaxel of the field is accreted to the bin described by the closest generator, until the S/N threshold is reached. This set of generators is refined to satisfy different topological and morphological criteria and to ensure that the scatter of the S/N of each bin is reduced to a minimum. This method ensures that the spatial resolution of the regions with high S/N is preserved, since these bins are reduced to a single spaxel.

The maps from different lines are binned independently since the spatial distribution of the emission is different and the S/N is line dependent (see Fig.~\ref{figure:LIRG} and \ref{figure:ULIRG}). Every S/N threshold has been chosen to achieve roughly the same number of bins in each map of every object and are listed in Table \ref{table:S/N}.

\subsection{Spectral maps and aperture normalised spectra}

As mentioned above, the maps of the emission lines were constructed by fitting a single Gaussian profile to the spectra. Figure \ref{figure:LIRG} shows, for the subsample of 10 LIRGs, the \Brg\ and H$_{2}$ 1-0S(1) emission and equivalent width maps, together with the velocity dispersion and radial velocity ones. The figures also include emission maps of the HeI at 2.059\,$\mu$m and, for those objects observed in the H band, [FeII] line emission maps at 1.644\,$\mu$m. We have also constructed a K band map from the SINFONI data by integrating the flux along the response curve of the 2MASS K-band filter, to compare with archival HST images when available. Figure \ref{figure:ULIRG} shows the maps of the subsample of 7 ULIRGs but with the \Pa\ emission line instead of \Brg. All the line emission maps are shown in arbitrary units on a logarithmic scale to maximise the contrast between the bright and diffuse regions and are oriented following the standard criterium that situates the north up and the east to the left\footnote{The only exception to this criterium is \object{IC 5179} (Fig. \ref{figure:IC5179}) where we have adopted the original orientation of the data to maximise the size of the maps. The axes' orientation is plotted for reference.}. 

The radial velocity maps are scaled to the velocity measured at the brightest spaxel in the K band image. This spaxel is marked with a cross in all the maps and usually coincides with the nucleus of the galaxy or with one of them for the interacting systems. The measured systemic radial velocities are similar to the NED published values within less than $\sim$1\%. Although the main nucleus of \object{NGC 3256} was observed in the H band,  the reference spaxel corresponds to its southern nucleus, which is highly extinguished (\citealt{Kotilainen:1996p305}, \citealt{AlonsoHerrero:2006p4703}, \citealt{DiazSantos:2008p685}), since the main one was not observed in the K band (see Fig.~\ref{figure:NGC3256}). The values of the reference radial velocities are shown in Table \ref{table:cz}.

\begin{table}[h]
\caption{Systemic radial velocities}
\centering
{
\begin{tabular}{c r@{ $\pm$ }l}
\hline
\hline
Object & \multicolumn{2}{c}{cz}\\
 & \multicolumn{2}{c}{(km\,s$^{-1}$)}\\
\hline
IRAS 06206-6315 &27625 & 15\\
NGC 2369 & 3465 & 5\\
NGC 3110 & 5151 & 4 \\
NGC 3256 & 2780 & 2\\
ESO 320-G030 & 3132 & 24\\
IRAS 12112+0305 & 21973 & 3\\
IRASF 12115-4656 & 5519 & 52\\
NGC 5135 & 4115 & 6\\
IRAS 14348-1447 & 24799 & 4\\
IRASF 17138-1017 & 5184 & 1\\
IRAS 17208-0014 & 12805 & 6\\
IC 4687 & 5226 & 3\\
IRAS 21130-4446 & 28016 & 17\\
NGC 7130 & 4895 & 2\\
IC 5179 & 3384 & 1\\
IRAS 22491-1808 & 23287 & 5\\
IRAS 23128-5919 &13437 & 2\\
\hline
\hline
\end{tabular}}
\tablefoot{Velocities derived for the sample of galaxies using the \Brg\ line at the spaxel with the brightest K-band flux. The spaxels are marked with a cross in Figs. \ref{figure:LIRG} and \ref{figure:ULIRG}. The differences between these and NED published values are typically less than $\sim$1\%.}
\label{table:cz}
\end{table}

Besides the spectral maps, Figs. \ref{figure:LIRG} and \ref{figure:ULIRG} show, for illustrative purposes, the integrated spectra in the K band of two regions of the FoV. The apertures used to extract the spectra are drawn on the maps and are labelled with the letters ``A" and ``B". Aperture ``A" is centred on the brightest spaxel of the K band image, which usually corresponds to the nucleus of the galaxy. On the other hand, aperture ``B" is centred in regions of interest that differ from object to object. In the LIRG subsample, it covers the brightest region in the \Brg\ equivalent width map. The same criterion is used for the ULIRG subsample, except in those objects with two distinct nuclei, where aperture ``B" covers the secondary nucleus.

The spectra are normalised to the continuum, measured between 2.080\,$\mu$m and 2.115\,$\mu$m and between 2.172\,$\mu$m and 2.204\,$\mu$m.  We also stacked the spectra of one of our sky cubes into a single spectrum and plotted it to illustrate the typical sky emission. This is useful for identifying the residuals from sky lines that are the result of the sky subtraction during the data reduction. Besides the OH sky lines, some of the K-band spectra show the residuals from the atmospheric absorption of water vapour. These features are easily traced along the wavelength ranges [1.991--2.035]\,$\mu$m and [2.045--2.080]\,$\mu$m, and are marked in grey in Figs. \ref{figure:LIRG} and \ref{figure:ULIRG}.

\subsection{Generation of the stacked spectra for the LIRG and ULIRG subsamples}
\label{section:stacked}

\setcounter{figure}{2}
\begin{figure*}[t!]
\begin{center}
\resizebox{\hsize}{!}{\includegraphics[angle=0]{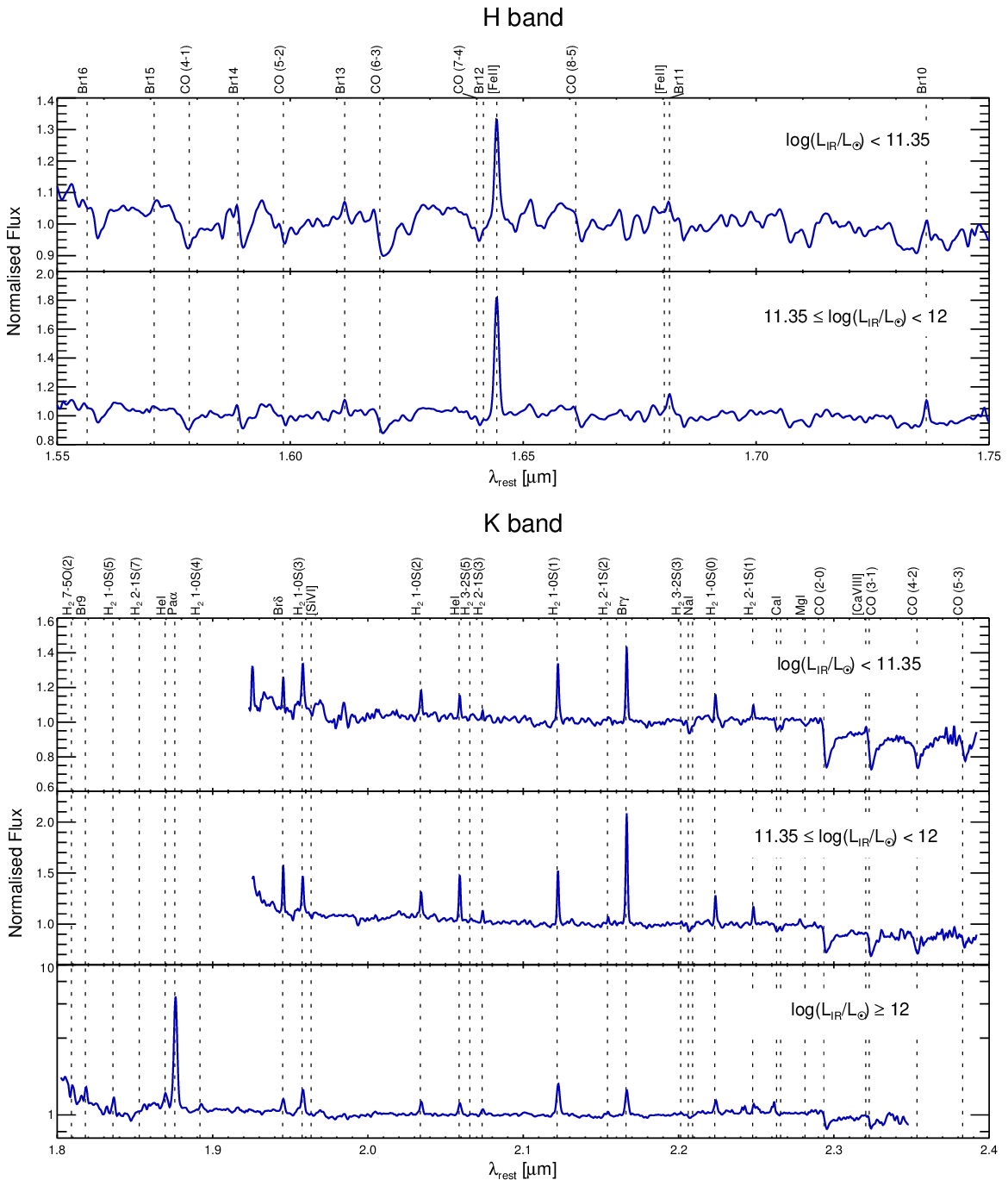}}
\end{center}
\caption{H- and K-band stacked spectra of the SINFONI sample, divided into three subsets with log(\LTIR/\Lsolar)\,$<$\,11.35, 11.35\,$\leq$\,log(\LTIR/\Lsolar)\,$<$\,12,  and  log(\LTIR/\Lsolar)\,$\geq$\,12. The spectra are normalised to a linear fit of the continuum measured within the intervals [1.600, 1.610]\,$\mu$m and [1.690, 1.700]\,$\mu$m for the H band and [2.080, 2.115]\,$\mu$m and [2.172, 2.204]\,$\mu$m for the K band. From top to bottom, H-band and K-band spectra of the different subsets by increasing \LTIR. The spectra are available in electronic form at the CDS via anonymous ftp to \url{ftp://cdsarc.u-strasbg.fr} (130.79.128.5) or via \url{http://cdsweb.u-strasbg.fr/cgi-bin/qcat?J/A+A/}.}
\label{figure:stacked_spectra}
\end{figure*}

Figure \ref{figure:stacked_spectra} shows the stacked spectra of three different subsets of the sample defined according to the \LTIR\ range. As discussed in \cite{RosalesOrtega:2012A&A539}, there are different techniques for optimising the S/N within IFS data. We have adopted a flux-based criterium to exclude those spaxels with low surface brightness that may contribute to increase the noise of the resulting spectra. For each object, we considered the continuum images for each band and ordered the spaxels by decreasing flux. We then selected a set of spectra from those spaxels that contain at least the $\gsim90\%$ of the total continuum flux. By assuming this criterium, we assure that typically between the $\sim85-95\%$ of the flux in the lines is also taken into account.  

Before the stacking, every individual spectrum is \emph{de-rotated}, i.e. shifted to the same rest frame. This procedure decorrelates the noise due to imperfect sky subtraction, since the residuals are no longer aligned in the spectral axis, and prevents the smearing of the lines due to the stacking along wide apertures. To derotate the spectra, we focussed on the [FeII] line for the H band and on \Brg\ (\Pa\ for the ULIRG subset) and the H$_{2}$ 1-0S(1) line for the K band, since the relative shifts in the spectral  axis could be different for each phase of the gas. For the ULIRG subset, we have only considered the \Pa\ line, since the H$_{2}$ 1-0S(1) line is not bright enough in all the spaxels where the spectra are extracted. For the K band spectra of the LIRG subsample, we measured the difference between the relative shifts obtained for the \Brg\ and H$_{2}$ lines, to assure that no artificial broadening is introduced  if only one phase is considered as reference for the whole spectra. Given that only less than $\sim$10\% of the spaxels have more than one spectral pixel of difference between the relative shifts measured with both emission lines, we considered that the effect in the width of the lines is negligible so we have adopted the \Brg\ line as reference for the whole LIRG subset.

After the derotation procedure, every spectrum of each object is normalised to a linear fit of the continuum, measured within the intervals [1.600, 1.610]\,$\mu$m and [1.690, 1.700]\,$\mu$m for the H-band and [2.080, 2.115]\,$\mu$m and [2.172, 2.204]\,$\mu$m for the K-band, and stacked in one single spectrum per object. Finally, the spectra of each galaxy in each luminosity bin are rebinned, stacked, and convolved to a resolution of 10\AA\ (FWHM) to achieve a homogeneous resolution. The spectra of the different luminosity bins are available as online material. 

\subsection{Gas emission and line fluxes}

We extracted the spectra of different regions of interest for all the galaxies of the sample, which comprise the nucleus (identified as the K-band continuum peak), the integrated spectrum over the FoV, and the peak of emission of \Brg\ (\Pa\ for the ULIRGs subsample), H$_{2}$ 1-0S(1), and [FeII]. For every region, we integrated the spectra within apertures of 400$\times$400\,pc for the LIRGs and 2$\times$2\,kpc  for ULIRGs, and measured the flux of the \Brg, H$_{2}$ 1-0S(1), and [FeII] lines for all the LIRGs of the sample and the \Pa\ and H$_{2}$ 1-0S(1) line flux for the ULIRGs subsample. Although the study of the ionised gas is focussed on the \Pa\ line in the ULIRG subset, we have also made measurements of the \Brg\ line to directly compare with the results obtained for the LIRGs.

To obtain the line fluxes over the FoV, we only took the brightest spaxels in the H and K-band images (for the [FeII] and \Brg\, \Pa\ and H$_{2}$ 1-0S(1) respectively) into account, to include $\gsim90\%$ of the total flux in each image. This ensures that only those spaxels with the highest S/N are included in the spectra, and removes all those with a low surface brightness that contributes significantly to increasing the noise and the sky residuals in the spectra. 

The line fitting is performed following the same procedure as in the spectral maps, by fitting a single Gaussian model to the line profile. To estimate the errors of the line fluxes, we implemented a Monte Carlo method. We measured the noise of the spectra as the \emph{rms} of the residuals after subtracting the Gaussian profile. Taking this value of the noise into account, we constructed a total of $N=1000$ simulated spectra whose lines are again fitted. The error of the measurements is obtained as the standard deviation of the fluxes of each line. The advantage of this kind of method is that the errors calculated not only consider the photon noise but also the uncertainties due to an improper line fitting or continuum level estimation.

The values of the line fluxes for the different regions in the sample of galaxies are shown in Table \ref{table:fluxes}. Besides the gas emission, we also measured the equivalent width of the CO (2--0) band at 2.293\,$\mu$m (W$_{\rm CO}$) using the definition of \citealt{ForsterSchreiber:2000AJ120}. This stellar feature is detected in all the galaxies of the LIRG subsample and in two ULIRGs (\object{IRAS 17208-0014} and \object{IRAS 23128-5919}), since it lays out of our spectral coverage for the rest of the ULIRGs.

\subsection{Stellar absorption features}

Although the study of the stellar populations and of their kinematics, derived from the CO absorption lines, will be addressed in a forthcoming paper (Azzollini et al. 2012, in preparation), we have included measurements of the equivalent width of the first CO absorption band (see Table~\ref{table:fluxes}) obtained using the penalised PiXel-Fitting (pPXF) software \citep{Cappellari:2004p4916} to fit a library of stellar templates to our data. We made use of the Near-IR Library of Spectral templates of the Gemini Observatory \citep{Winge:2009p4732}, which covers the wavelength range of 2.15\,$\mu$m -- 2.43$\,\mu$m with a spectral resolution of 1\,\AA\,pixel$^{-1}$. The library contains a total of 23 late-type stars, from F7III to M3III, and was previously convolved to our SINFONI resolution.

\section{Overview of the data}
\label{section:overview}
The wide spectral coverage of the SINFONI data allows us to study in detail a large number of spectral features that trace different phases of the interstellar medium and the stellar population \citep{Bedregal:2009p2426}. In this work we focus on the gas emission in LIRGs and ULIRGs by studying the brightest lines in the H and K bands, i.e. [FeII] at 1.644\,$\mu$m, \Pa\ at 1.876\,$\mu$m, HeI at 2.059\,$\mu$m, H$_{2}$ 1-0S(1) at 2.122\,$\mu$m and \Brg\ at 2.166\,$\mu$m. The maps of these spectral features together with the K-band spectra of the nucleus (identified as the K-band peak), and of the brightest \Brg\ (or \Pa\ for ULIRGs) region are shown in Figs. \ref{figure:LIRG} and \ref{figure:ULIRG} for the sample of LIRGs and ULIRGs, respectively. 

In the present section, we briefly describe the different physical mechanisms and processes that create the emission lines and stellar features observed in our data. The detailed study of these mechanisms are beyond the scope of the present work, but some of them will be addressed in the forthcoming papers of these series.

\subsection{Hydrogen lines and 2D extinction maps}

The overall structure of the ionised gas, mostly associated with recent star formation, is traced by the hydrogen recombination lines \Pa\ for ULIRGs and \Brg\ for LIRGs. Although \Brg\ is also observed in the ULIRGs subsample, for this group we focus the study of the ionised gas on the \Pa\ emission, since its brightness allows better measurements. The \Brd\ line at 1.945\,$\mu$m is also observed in all the galaxies of the sample; however, for the LIRG subsample, it lies in a spectral region where the atmospheric transmission is not optimal and, for the ULIRG subset, it is too weak to be mapped.

It is well known that the bulk of luminosity produced in local (U)LIRGs is due to the large amount of dust that hides a large fraction of their star formation and nuclear activity (see \citealt{AlonsoHerrero:2006p4703}, \citealt{GarciaMarin:2009p8459} and references therein). This dust is responsible for the absorption of UV photons that are then re-emitted at FIR and submillimetre wavelengths. A detailed 2D quantitative study of the internal extinction could be performed by using the \Brd/\Brg\ ratios in LIRGs and \Brg/\Pa\ in ULIRGs. This study of the objects in the sample will be presented in the next paper of this series (Paper II, Piqueras L\'opez et al. 2012a, in preparation).

The detailed characterisation of the extinction is essential to accurate measurement of the SFR in these dusty environments. This treatment of the extinction allows us to obtain maps of the SFR surface density that are corrected for extinction on a spaxel-by-spaxel basis. The analysis of the SFR in the objects of the sample, based on the \Brg\ and \Pa\ maps presented in this work, will be addressed in Piqueras-L\'opez et al 2012b (in preparation).
 
 \subsection{Emission lines and star formation}
 
 The hydrogen recombination lines have been widely used as a primary indicator of recent star-formation activity, where UV photons from massive OB stars keep the gas in an ionised state. The measurements of the \Brg\ equivalent width (EW), in combination with the stellar population synthesis models, such as STARBURST99 \citep{Leitherer:1999p6938} or Claudia Maraston's models (\citealt{Maraston:1998p2886}, \citeyear{Maraston:2005p2885}), could be used to constrain the age of the youngest stellar population. The HeI emission is also usually associated to star-forming regions, and used as a tracer of the youngest OB stars, given its high ionisation potential of 24.6\,eV. This emission may depend on different factors, such us density, temperature, dust content, and He/H relative abundance and ionisation fractions.

The [FeII] emission is usually associated with regions where the gas is partially ionised by X-rays or shocks \citep{Mouri:2000p6654}. Shocks from supernovae cause efficient grain destruction that releases the iron atoms contained in the dust. The atoms are then singly ionised by the interstellar radiation field and excited in the extended post-shock region by free electron collision on timescales of $\sim$10$^4$\,yr. The [FeII] lines at 1.257\,$\mu$m and 1.644\,$\mu$m are widely used to estimate the supernova rate in starbursts (\citealt{Colina:1993p6004}, \citealt{AlonsoHerrero:2003p3431}, \citealt{Labrie:2006p166}, \citealt{Rosenberg:2012A&A540} and references therein), whereas the EW$_{\rm [FeII]}$ could also be used, in combination with the EW$_{\rm Br\gamma}$, to constrain the age of the stellar populations in the synthesis models.

\subsection{H$_2$ lines and excitation mechanisms}

The H$_{2}$ 1-0S(1) line is used to trace the warm molecular gas, since it is the brightest H$_2$ emission line in the K band and it is well detected in all the objects with sufficient S/N. Furthermore, the presence of several roto-vibrational transitions of the molecular hydrogen within the K band allows studying the excitation mechanisms of the H$_2$: fluorescence due to the excitation by UV photons from AGB stars in PDRs (\emph{photon-dominated regions}), thermal processes like collisional excitation by SN fast shocks, or X-rays (\citealt{vanderWerf:2000vb}, \citealt{Davies:2003p3644}, \citeyear{Davies:2005p3646}). The determination of the H$_2$ excitation mechanisms in general requires measurements of several lines, usually weak lines, since the different processes mentioned may rise to similar intense and thermalised 1-0 emissions.

Based on the relative fluxes of the transitions to the brightest H$_{2}$ 1-0S(1) line, we could obtain population diagrams of the emitting regions. In these diagrams, the population of each level in an ideal thermalized PDR could be determined as a function of the excitation temperature. The presence of non-thermal processes like UV fluorescence is translated to an overpopulation of the upper levels and a deviation from the ideal thermalised model. However, the way these levels are overpopulated due to non-thermal processes is complex, and might depend on several parameters like density or the intensity of the illuminating UV field (\citealt{Davies:2003p3644}, \citeyear{Davies:2005p3646}, \citealt{Ferland:2008p8001} and references therein). The detailed study of the excitation mechanisms of the molecular hydrogen will be addressed in a future paper of this series (Piqueras L\'opez et al. 2012c, in preparation). 

\subsection{Coronal lines as AGN tracers}

The [SiVI] at 1.963\,$\mu$m and [CaVIII] at 2.321\,$\mu$m coronal lines are the main AGN tracers within the K-band. However, the [CaVIII] line is too faint (typically $\times$4 fainter than [SiVI], \citealt{RodriguezArdila2011ApJ743}) and to close to CO (3-1) to be measured easily. Given the high ionisation potential of 167\,eV for [SiVI] and 128\,eV for [CaVIII], the outskirts of the broad-line region and extended narrow-line regions have been proposed as possible locations for the formation of these lines in AGNs. Although which mechanism is responsible for the emission remains unclear, there are two main processes proposed: photoionisation due to the central source and shocks due to high-velocity clouds and the NLR gas (see \citealt{RodriguezArdila2011ApJ743} and references therein).

\subsection{Line ratios}

The interpretation of the H$_2$ 1-0S(1)/\Brg\ ratio is sometimes not straightforward since H$_2$ could be excited by both thermal and radiative processes, in contrast to the [FeII] emission that is predominantly powered by thermal mechanisms. This ratio is in principle not biased by extinction, and starburst galaxies and HII regions empirically exhibit lower H$_2$/\Brg\ ratios, whereas Seyfert galaxies and LINERs show higher values (0.6$\lsim$H$_2$ 1-0S(1)/\Brg$\lsim2.0$, \citealt{Dale:2004ApJ601}, \citealt{Rodriguez-Ardila:2005p364}, \citealt{Riffel:2010MNRAS404}, \citealt{ValenciaS:2012up}).

In combination with the \Brg\ emission that traces the photoionised regions, the [FeII]/\Brg\ ratio allows us to distinguish regions where the gas is ionised by star formation activity where the [FeII] is expected to be weak \citep{Mouri:2000p6654}, from zones where the gas is partially ionised by shocks \citep{AlonsoHerrero:1997p5041}. Since [FeII] is not expected in HII regions where iron would be in higher ionisation states, we could trace different ionisation mechanisms and efficiencies by using the [FeII]/\Brg\ ratio, and probe the excitation mechanisms that produce the [FeII] line in those regions where the emission has stellar origin. In addition, the [FeII]/\Brg\ ratio depends on the grain depletion, and a high depletion of iron would reduce the number of atoms available in the interstellar medium, hence reduce the line ratio \citep{AlonsoHerrero:1997p5041}.

 Although the HeI line could be used as a primary indicator of stellar effective temperature, interpreting the emission and the HeI/\Brg\ ratio without a detailed photoionisation model is still controversial (\citealt{Doherty:1995MNRAS277}, \citealt{Lumsden:2001MNRAS320}, \citeyear{Lumsden:2003MNRAS340}). In addition, the HeI transition is also influenced by collisional excitation, and a full photoionisation treatment is not enough to predict the line emission \citep{Shields:1993ApJ419}.

\subsection{Absorption lines and stellar populations}

Besides the emission lines, there are different absorption features that lie within the K band, such as the NaI doublet at 2.206\,$\mu$m and 2.209\,$\mu$m, the CaI doublet at 2.263\,$\mu$m, and 2.266\,$\mu$m and the CO absorption bands CO (2--0) at 2.293\,$\mu$m, CO (3--1) at 2.323\,$\mu$m or CO (4-2) at 2.354\,$\mu$m. The absorption features, such as the CO bands and the NaI doublet, are typical of K and later stellar types, and they also trace red giant and supergiant populations. Given the limited S/N of the NaI doublet, it is not possible to map the absorption with the present data, but it could be suitable for integrated analysis. On the other hand, the CO (2--0) band can be used to spatially sample the stellar component of all the LIRGs of the sample and in two of the ULIRGs. Both EW$_{\rm CO}$ and EW$_{\rm NaI}$ could be used in combination with the stellar population synthesis models to constrain the age of the stellar populations (see \citealt{Bedregal:2009p2426}).

\section{Results and discussion}
\label{section:gas}

Most of the LIRGs of the sample are spiral galaxies with some different levels of interaction, ranging from isolated galaxies as \object{ESO 320-G030} to close interacting systems as \object{IC 4687}+\object{IC 4686} or mergers like \object{NGC 3256} \citep{Lipari:2000p120}, and to objects that show long tidal tails several kiloparsecs away from its nucleus (e. g. \object{NGC 7130}, Fig. \ref{figure:NGC7130}). The emission from ionised and molecular hydrogen has different morphologies in many galaxies of the LIRGs subsample. The ULIRG subsample contains mainly interacting systems in an ongoing merging process with two well-differentiated nuclei. The \Pa\ emission extends over several kiloparsecs with bright condensations not observed in the continuum maps. The molecular hydrogen emission, on the other hand, is rather compact ($\lsim2-3$\,kpc) and is associated with the nuclei of the systems. An individual description of the more relevant features of the gas emission morphology for each galaxy can be found in Appendix \ref{notes}. We now discuss each gas phase and the stellar component separately.

\subsection{Ionised gas}

The dynamical structures as spiral arms are clearly delineated on the ionised gas maps traced by the \Brg\ (LIRGs) and \Pa\ (ULIRGs) lines. The observed ionised gas emission in LIRGs is dominated by high surface brightness clumps associated with extranuclear star-forming regions located in circumnuclear rings or spiral arms at radial distances of several hundred parses. The nuclei are also detected as bright \Brg\ sources in several galaxies but represent the maximum emission peak in only a small fraction ($\sim$33\%) of LIRGs. These results are in excellent agreement with those derived from the \Pa\ emission in \cite{AlonsoHerrero:2006p4703} using HST NICMOS images for all the objects of our LIRG sample, with the exception of \object{IRASF 12115-4656}, which was not included in their sample.  On the other hand, the main nucleus is the brightest \Pa\ emission peak in the majority ($\sim$71\%) of ULIRGs, also showing emission peaks along the tidal tails, secondary nucleus, or in extranuclear regions at distances of 2--4\,kpc from their centre. However, since ULIRGs in our sample are about four to five times more distant than our LIRGs, the typical angular resolution of our SINFONI maps for ULIRGs covers sizes of about 1.5 kpc, and therefore the \Pa\ peak emission detected in the nuclei could still be due to circumnuclear star-forming regions, as in LIRGs.

The surface density and total luminosity distributions of the \Brg\ (and \Pa\ for the ULIRG subsample) are presented in Figs. \ref{figure:histograms} and \ref{figure:histograms_area}. On average, the observed (i.e. uncorrected for internal extinction) luminosities of the \Brg\ brightest emitting region are $\sim1.2\times10^5$\,\Lsolar\ for LIRGs and $\sim2.3\times10^6$\,\Lsolar\ for ULIRGs, accounting for about $\sim10$\% and $\sim43$\% of the integrated \Brg\ luminosity, respectively. There is a factor $\sim$20 difference in luminosity between LIRGs and ULIRGs. Although the ULIRGs are intrinsically more luminous, this difference is also due to a distance effect since the angular aperture used to obtain the luminosities covers for ULIRGs an area 25 times larger than for LIRGs.  For the ULIRG subset, we have also measured peak \Pa\ luminosities of the order of $4.5\times10^7$\,\Lsolar, in agreement with the expected value derived from \Brg\ luminosities assuming case B recombination ratios and no extinction.

The observed \Brg\ surface luminosity densities of the  \Brg\ (and \Pa) brightest emitting regions are $\sim0.7$\,\Lsolar\,pc$^{-2}$ and $\sim0.6$\,\Lsolar\,pc$^{-2}$, on average, for LIRGs and ULIRGs, respectively, whereas the \Pa\ surface density for the ULIRG subset is $\sim6$\,\Lsolar\,pc$^{-2}$. Figure~\ref{figure:histograms_area} shows that the distributions of the \Brg\ surface luminosity density for the different luminosity bins are very similar for the \Brg\ (and \Pa) peak and the nucleus of the objects, and range between $\sim0.1$\,\Lsolar\,pc$^{-2}$ and $\sim3$\,\Lsolar\,pc$^{-2}$. 

As shown in Fig. \ref{figure:histograms}, the luminosity of \Brg\ ranges from $\sim1.7\times10^4$\,\Lsolar\ to $\sim5.1\times10^6$\,\Lsolar\ in the nuclear regions of LIRGs, and the \Pa\ emission reaches up to $\sim5.0\times10^7$\,\Lsolar\ in ULIRGs. Assuming the standard star formation rate to H$\alpha$ luminosity ratio given by the expression \citep{Kennicutt:1998p36},

\begin{align*}
 \rm SFR (M_{\odot} yr^{-1}) &= 7.9 \times 10^{-42} \times \rm L(H\alpha) (erg\,s^{-1}), \\
\end{align*}

\noindent an estimate of the SFR surface densities, uncorrected for internal reddening, can be directly obtained from the previous expression if the \Ha\ to \Pa\ and \Brg\  recombination factors are taken into account:

\begin{align*}
   \rm SFR (M_{\odot} yr^{-1}) &=  6.8 \times 10^{-41} \times \rm L(P\alpha)  (erg\,s^{-1}) \\
   &=  8.2 \times 10^{-40} \times \rm L(Br\gamma) (erg\,s^{-1}). 
\end{align*}

For LIRGs, the mean SFR surface densities integrated over areas of several kpc$^2$, range between 0.4 and 0.9 M$_{\odot}$ yr$^{-1}$ kpc$^{-2}$ with peaks of about 2$-$2.5 M$_{\odot}$ yr$^{-1}$ kpc$^{-2}$ in smaller regions (0.16 kpc$^2$) associated with the nucleus or the brightest \Brg\ region. For ULIRGs, the corresponding values are similar, $\sim$0.4 for the integrated emission and $\sim$ 2 M$_{\odot}$ yr$^{-1}$ kpc$^{-2}$ for peak emission. However, since ULIRGs are at distances further away than LIRGs, the sizes of the overall ionised regions and brightest emission peaks covered by the SINFONI data are greater than those in LIRGs, and they correspond to 100$-$200 kpc$^2$ and 4 kpc$^2$, respectively.

We estimated the extinction effects by comparing the observed \Brg/\Brd\ and \Brg/\Pa\ ratios (for LIRGs and ULIRGs, respectively) with the theoretical ones derived from a case B recombination. We measured A$_{V}$ values that range from $\sim$2--3\,mag up to $\sim$10--12\,mag in the nuclei of the objects (Piqueras L\'opez et al. 2012a, in preparation). This is translated to extinction values from $\sim0.3$\,mag to $\sim1.0$\,mag at \Brg\ wavelengths, and from $\sim0.4$\,mag to $\sim1.6$\,mag at \Pa, and indicates that the internal extinction in these objects still plays a role at these wavelengths. These values are similar to those obtained by \cite{AlonsoHerrero:2006p4703} from the nuclear emission in LIRGs. From the detailed 2D study of the internal extinction that will be presented in Paper II, we estimated the median visual extinction for each luminosity subsamples of LIRGs and ULIRGs. These values are A$_{\rm V,LIRGs} = 6.5$\,mag and A$_{\rm V,ULIRGs} = 7.1$\,mag, that correspond to A$_{\rm Br\gamma} = 0.6$\,mag and A$_{\rm Pa\alpha} = 1.0$\,mag respectively. 

Considering the median extinction values presented above, the \Brg\ and \Pa\ luminosities, hence the SFR surface densities, are underestimated approximately by a factor $\times1.7$ in LIRGs and $\times2.5$ in ULIRGs. However, on scales of a few kpc or less, the distribution of dust in LIRGs and ULIRGs is not uniform, and it shows a patchy structure that includes almost transparent regions and very obscured ones (see \citealt{GarciaMarin:2009p8459} and Paper II). This non-uniform distribution of the dust implies that the correction from the extinction depends on the sampling scale, so that the correction to the SFR would depend on the scales where the \Brg\ (\Pa) is sampled. For further discussion of the extinction and the implications of the sampling scale in its measurements, please see Piqueras L\'opez et al. 2012a (in preparation).
 
A detailed analysis of SFR surface densities based on the \Brg, \Pa, and H$\alpha$ emission lines will be presented elsewhere (Piqueras L\'opez et al. 2012b, in preparation). 

\subsection{Warm molecular gas}

The H$_2$ emission is associated with the nuclear regions of the objects, either to the main nucleus or to the secondary in the interacting systems. In some cases, its maximum does not coincide with the \Brg\ peak, although in all the LIRGs and $\sim71$\% of the ULIRGs it coincides with the main nucleus, identified as the brightest region in the K-band image. The typical H$_2$ 1-0S(1) luminosity of the nuclei ranges from $\sim1.3\times10^5$\,\Lsolar\ for the LIRG subsample up to $\sim4.6\times10^6$\,\Lsolar\ for the ULIRGs, and accounts for $\sim13$\% and $\sim41$\% of the total luminosity measured in the entire FoV. The H$_2$ luminosity observed in the nucleus and in the \Brg\ (\Pa) peak in both LIRGs and ULIRGs is very similar to the \Brg\ luminosity. The range of observed luminosity in the nuclei spans from $\sim4.2\times10^4$\,\Lsolar\ to $\sim6.8\times10^6$\,\Lsolar, and is also very similar to the distribution observed for the \Brg\ emission. Since the H$_2$ (1$-$0)/\Brg\ ratio is close to one (range of 0.4 to 1.4, see Fig.~\ref{figure:ratios}), the surface brightness values derived for the H$_2$ emission are similar to those obtained for \Brg\ (see Figure 8).

\subsection{Partially ionised gas}

The [FeII] maps reveal that the emission roughly traces the same structures as the \Brg\ line, although the emission peaks are not spatially coincident in some of the objects, and the [FeII] seems to be more extended and diffuse. In $\sim55$\% of the LIRGs, the peak of the [FeII] emission is measured in the nucleus, with typical luminosities of $\sim1.2\times10^5$\,\Lsolar\ on scales of $\sim0.16$\,kpc$^2$. The nuclear emission accounts on average for $\sim16$\% of the total observed luminosity. The differences in the morphology between the ionised and partially ionised gas could be understood in terms of the local distribution of the different stellar populations: although both lines trace young star-forming regions, the \Brg\ emission is enhanced by the youngest population of OB stars of $\lsim6$\,Myr, whereas the [FeII] is mainly associated with the supernova explosions of more evolved stellar populations of $\sim7.5$\,Myr (see STARBURST99 models, \citealt{Leitherer:1999p6938}).

\subsection{Coronal line emission}

\begin{figure}[t!]
\begin{center}
\resizebox{\hsize}{!}{\includegraphics[angle=90]{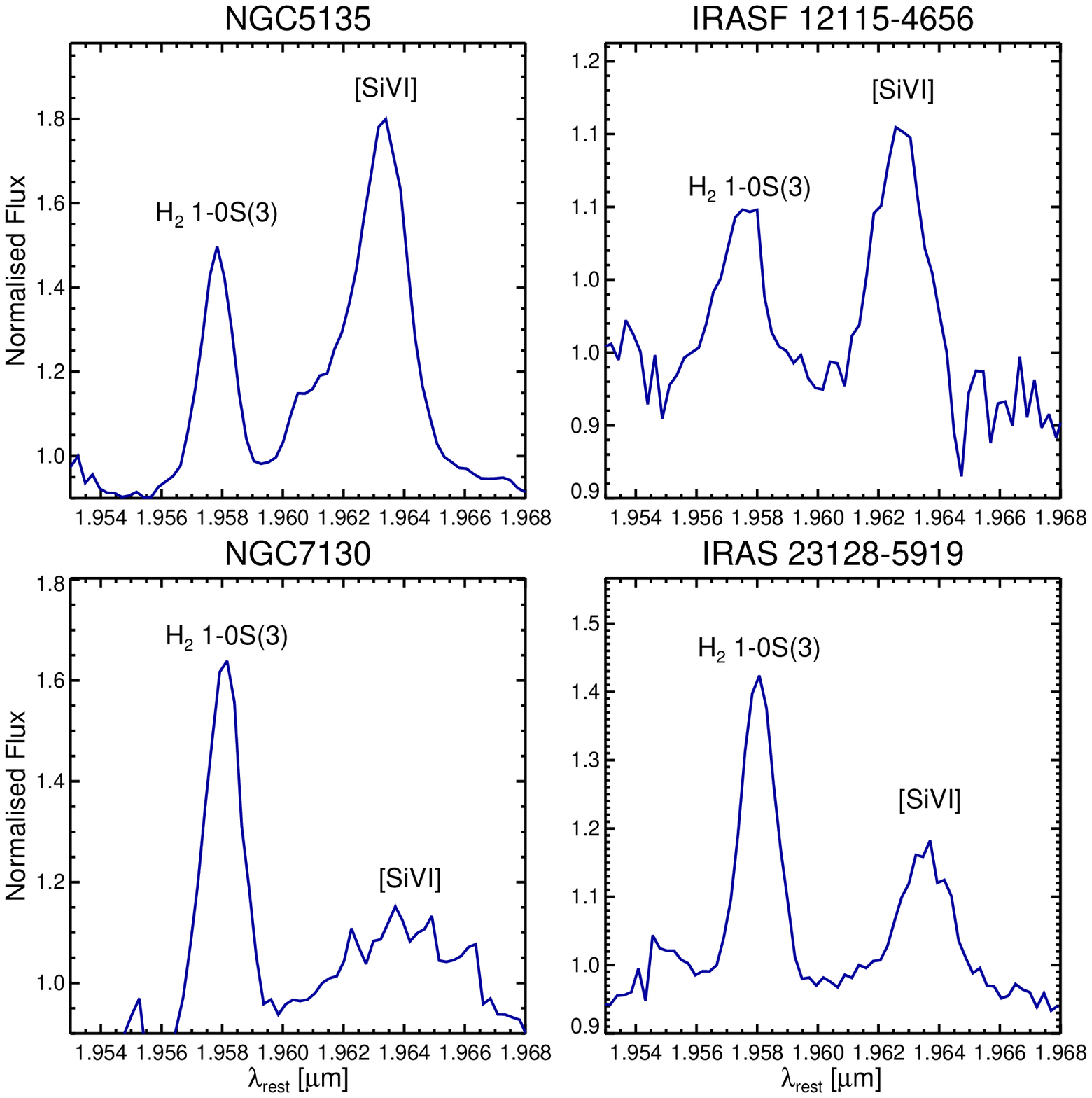}}
\caption{H$_2$ 1--0S(3) and [SiVI] normalised flux profiles of the brightest spaxel in [SiVI] for the four objects where the coronal line is detected.}
\label{figure:SiVI}
\end{center}
\end{figure}

\begin{figure}[t]
\begin{center}
\resizebox{\hsize}{!}{\includegraphics[angle=90]{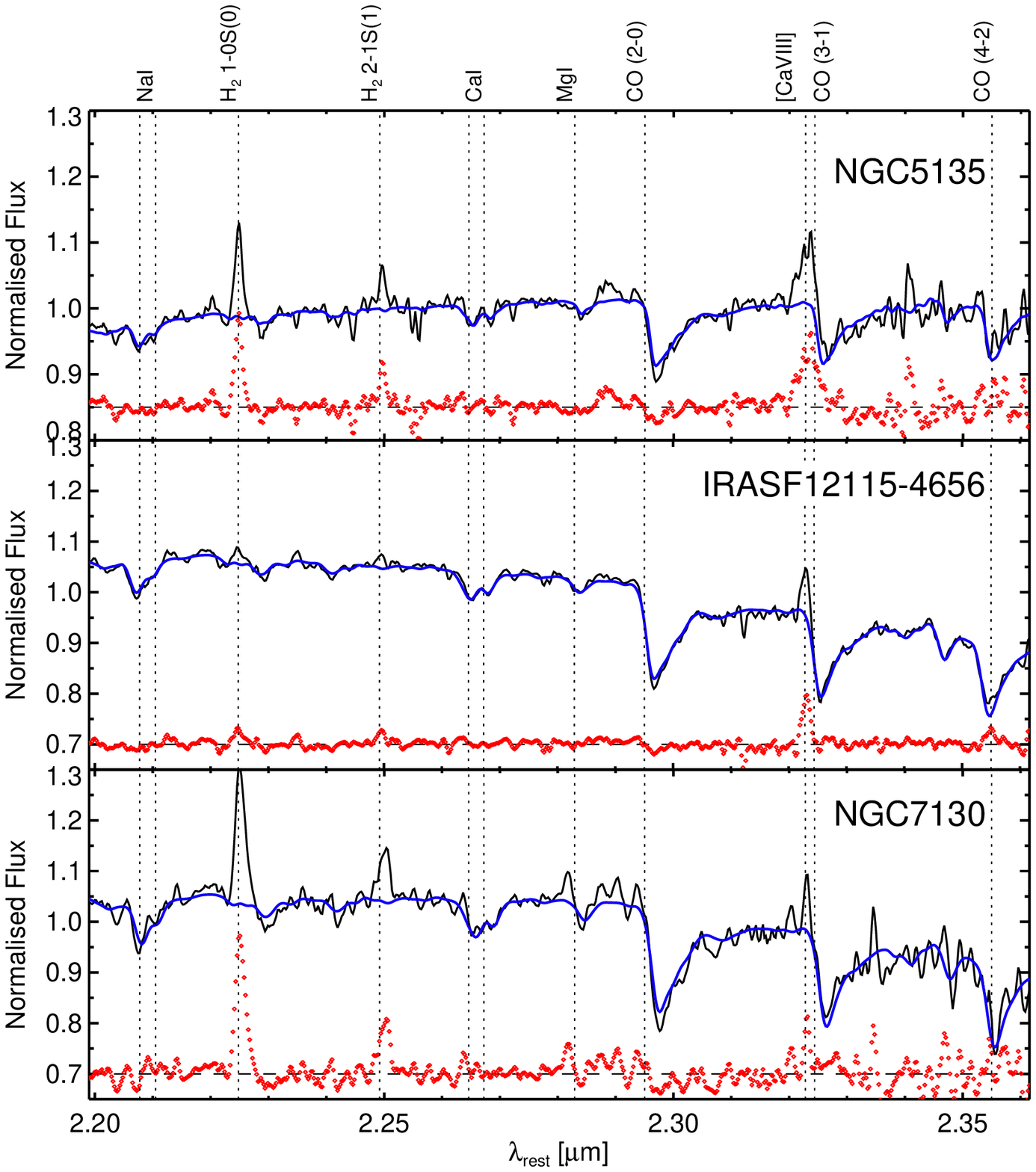}}
\caption{Zoom around the region containing the stellar absorption features and the coronal line [CaVIII] at 2.321\,$\mu$m for three of the objects where coronal emission is detected. Spectra correspond to the brightest spaxel in [SiVI] (see Fig.~\ref{figure:SiVI}). The pPXF fitting of the stellar absorptions is plotted in blue, and the residuals from the fitting are shown as a red dotted line.}
\label{figure:CaVIII}
\end{center}
\end{figure}

The [SiVI] coronal line at 1.963\,$\mu$m (see Fig.~\ref{figure:SiVI}) is detected in two LIRGs (\object{NGC 5135} and \object{IRASF 12115-4656}) and in one ULIRG (\object{IRAS 23128-5919}), with a tentative detection in another LIRG (\object{NGC 7130}). The [SiVI] line has a high ionisation potential (167\,eV) and it is associated with Seyfert activity where the gas is ionised outside the broad line region of the AGN \citep{Bedregal:2009p2426}. The [SiVI] emission is usually rather compact, concentrated around the unresolved nucleus and extending up to a few tens or a few hundred parsecs in some Seyferts (\citealt{Prieto:2005p364}, \citealt{Rodriguez-Ardila:2006ApJ653}). While in galaxies like \object{IRASF 12115-4656} and \object{IRAS 23128-5919} the emission is unresolved (i.e. sizes less than $\sim150$\,pc and $\sim550$\,pc, respectively), a relevant exception is \object{NGC 5135}, which presents a cone of emission centred on the AGN and extending $\sim$600\,pc ($\sim2$\,arcsec) from the nucleus, as discussed in \cite{Bedregal:2009p2426}. The line profiles for the four galaxies are given in Fig.~\ref{figure:SiVI}.

As shown in Fig~\ref{figure:CaVIII}, the [CaVIII] coronal line at 2.321\,$\mu$m is also detected in three of these objects, \object{NGC 5135}, \object{IRASF 12115-4656} and tentatively in \object{NGC 7130}. Although the [CaVIII] line lays also within the rest-frame spectral coverage of \object{IRAS 23128-5919}, the S/N in this region of the spectra is very low, so was not included in the figure.

\subsection{Characteristics of the near-IR stacked spectra of LIRGs and ULIRGs}

To obtain representative spectra of LIRGs and ULIRGs, we divided the sample into three subsamples according to their total infrared luminosity (see Section~\ref{section:stacked}), i.e. low luminosity bin, log(\LTIR/\Lsolar)\,$<$\,11.35; intermediate, 11.35\,$\leq$\,log(\LTIR/\Lsolar)\,$<$\,12; and high, the ULIRGs subsample, log(\LTIR/\Lsolar)\,$\geq$\,12. The average luminosities for each bin are log(\LTIR/\Lsolar)$=11.23$, log(\LTIR/\Lsolar)$=$11.48, and log(\LTIR/\Lsolar)$=$12.29. Each luminosity bin contains a similar number of objects in each subsample, six, four, and seven, respectively. The stacked spectra for each luminosity bin are presented in Fig. \ref{figure:stacked_spectra}

\addtocounter{table}{2}
\begin{table*}[t]
\caption{\Brg\ (and \Pa) average luminosities and surface brightness for the U/LIRGs according to their \LTIR}
{\tiny
\centering
{
\begin{tabular}{c c c c c c c c c c c}
\hline
\hline
 \multicolumn{11}{c}{ \Brg\ and \Pa\ Emission}\\
 \hline
  \multicolumn{1}{c}{\multirow{2}{*}{Region}}& \multicolumn{2}{c}{Low}& &  \multicolumn{2}{c}{Intermediate} & & \multicolumn{4}{c}{High}\\
  \cline{2-3} \cline{5-6} \cline{8-11}
 & L$_{\rm Br\gamma}$($\times 10^5$L$_{\odot}$) & S$_{\rm Br\gamma}$(L$_{\odot}$\,pc$^{-2}$) & & 
  L$_{\rm Br\gamma}$($\times 10^5$L$_{\odot}$) & S$_{\rm Br\gamma}$(L$_{\odot}$\,pc$^{-2}$) & & 
  L$_{\rm Br\gamma}$($\times 10^5$L$_{\odot}$) & S$_{\rm Br\gamma}$(L$_{\odot}$\,pc$^{-2}$) & L$_{\rm Pa\alpha}$($\times 10^5$L$_{\odot}$) & S$_{\rm Pa\alpha}$(L$_{\odot}$\,pc$^{-2}$)\\
\hline
\hline
HII max  & 0.958 (0.785) & 0.599 (0.490) & & 1.292 (0.744) & 0.807 (0.465) & & 22.92 (15.12) & 0.573 (0.378) & 236.5 (148.3) & 5.912 (3.708)\\
H$_2$  max & 0.977 (0.722)& 0.611  (0.451) & & 0.585  (0.452)& 0.366 (0.282)& & \multicolumn{2}{c}{\nodata} & 241.1 (143.8) & 6.026 (3.596)\\
$[$FeII] max & 0.873 (0.801)&  0.546 (0.501)& & 0.709 (0.700)& 0.443 (0.438)& & \multicolumn{2}{c}{\nodata} & \multicolumn{2}{c}{\nodata}\\
Integrated & 7.829 (0.405)& 0.120 (0.048)& & 15.450 (7.931)& 0.286 (0.067)& & 97.76 (116.8)&0.119 (0.072)& 521.5 (207.9)& 0.972 (0.792)\\
Nuclear & 0.991  (0.746)& 0.620  (0.467)& & 0.909  (0.512)& 0.568  (0.320)& & 23.14  (15.44)& 0.578  (0.386)& 233.8  (154.8)& 5.846 (3.869)\\
\hline
\hline

\end{tabular}}
\tablefoot{The low-luminosity bin is defined as log(\LTIR/\Lsolar)\,$<$\,11.35, the intermediate is defined as 11.35\,$\leq$\,log(\LTIR/\Lsolar)\,$<$\,12, and the high-luminosity bin corresponds to the ULIRGs subsample, log(\LTIR/\Lsolar)\,$\geq$\,12. The standard deviation of the values within each bin is shown in brackets. The number of objects is six, four, and seven for the low, intermediate, and high luminosity bins, respectively.}
\label{table:stacked_spectra_table}
}
\end{table*}

\begin{figure}
\begin{center}
\resizebox{\hsize}{!}{\includegraphics[angle=90]{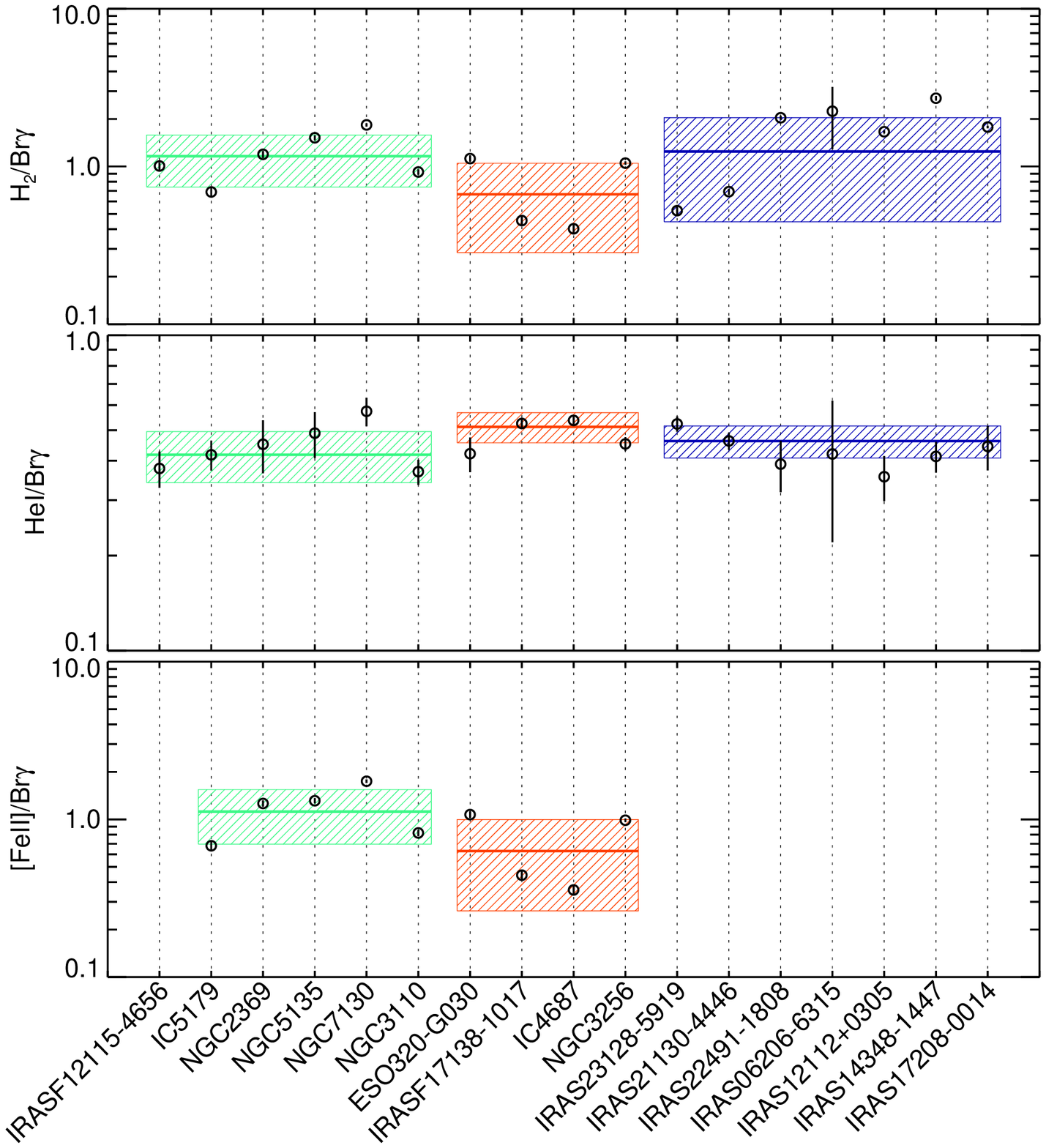}}
\caption{H$_2$/\Brg\ (top), HeI/\Brg\ (centre), and [FeII]/\Brg\ (bottom) line ratios of the galaxies of the sample, ordered by increasing \LTIR. The values are measured in the integrated spectra. The weighted mean of each luminosity bin (low, log(\LTIR/\Lsolar)\,$<$\,11.35; intermediate, 11.35\,$\leq$\,log(\LTIR/\Lsolar)\,$<$\,12 and high, log(\LTIR/\Lsolar)\,$\geq$\,12) is plotted as a thick line, whereas the box represents the standard deviation of the values. Since all the ULIRGs and one LIRG were not observed in the H-band, [FeII]/\Brg\ data are presented for only nine LIRGs.}
\label{figure:ratios}
\end{center}
\end{figure}

The H-band spectrum of LIRGs is dominated by the stellar continuum, with pronounced absorption features from water vapour and CO. The main emission feature is the [FeII] line at 1.644\,$\mu$m, while the faint high-order hydrogen Brackett lines (Br10 to Br14) are also present. The K-band spectra show various stellar absorption features like the faint NaI, CaI, MgI, lines and the strong CO bands. The emission line spectra contain the hydrogen (\Brg\ and \Brd) and He recombination lines, as well as a series of the H$_2$ lines covering different transitions. While the [SiVI] coronal line is detected in some LIRGs and ULIRGs, it is a weak line and thus not visible in the stacked spectra at any luminosity. The average luminosity and surface brightness of the \Brg\ and \Pa\ emission for each bin is shown in Table \ref{table:stacked_spectra_table}.

Considering only the brightest emission lines that trace different phases of the gas and/or excitation conditions, there appears to be some small differences ($\sim1$\,$\rm\sigma$) in their ratios with the \LTIR (see Fig \ref{figure:ratios} for the H$_{2}$ 1-0S(1)/\Brg, HeI/\Brg, and [FeII]/\Brg\ line ratios measured for the different luminosity bins). The HeI/\Brg\ ratio is slightly higher ($\times1.3$) in intermediate and high luminosity galaxies than in low luminosity objects. A plausible interpretation could be that young and massive stars in low luminosity LIRGs represent a lower fraction than in more luminous infrared galaxies. Whether this could be caused by age effects or by lower IMF upper mass limits remains to be investigated in more detail. Some differences are also identified in the H$_{2}$ 1-0S(1)/\Brg\ line ratio. While this ratio is close to unity for low luminosity LIRGs, it drops to about 0.6 for intermediate luminosity LIRGs, and increases up to about 1.3 for the most luminous galaxies. However, given the large dispersion of the values for the individual objects, and the low number of galaxies in each bin, we could not draw any significant conclusion about these differences. While low H$_{2}$ 1-0S(1)/\Brg\ values appear to be characteristic of starbursts (H$_{2}$ 1-0S(1)/\Brg$\lsim0.6$), classical Seyfert 1 and 2 galaxies also display a range of values (\citealt{Rodriguez-Ardila:2004p425}, \citeyear{Rodriguez-Ardila:2005p364}, \citealt{Riffel:2010MNRAS404}) that are compatible with those measured in our sample.

The [FeII]/\Brg\ ratio also shows values compatible with those observed in starbursts, although higher values would have been expected for Seyfert galaxies, such as \object{NGC 5135} and \object{NGC 7130} (\citealt {Rodriguez-Ardila:2004p425}, \citealt{Riffel:2010MNRAS404}, \citealt{ValenciaS:2012up}). These galaxies show the highest values of the [FeII]/\Brg\ ratio of the whole sample, close to $\sim2.0$, and are similar to values reported for other Seyfert 2 galaxies \citep{Blietz:1994ApJ421}. These differences could be related to the different apertures used to extract the integrated values of the ratios. Even more, the reported ratios are observed values (not corrected for extinction). Although the H$_{2}$ 1-0S(1)/\Brg\ ratio is almost unaffected by extinction, the [FeII]/\Brg\ ratio could be affected by obscuration, so that an accurate study of the extinction is needed to confirm or dismiss these discrepancies between both ratios.

Based on our current survey, no evidence of relevant differences in the emission line spectra of LIRGs and ULIRGs appear as a function of \LTIR. A larger sample would be required to confirm the differences in the emission line ratios presented here, since they are still compatible within the uncertainties. The full two-dimensional study of the line ratios and the ionisation and excitation mechanisms of the gas will be addressed in a future paper of the series (Piqueras L\'opez et al. 2013, in preparation), since its detailed analysis is beyond the scope of the present work.

\subsection{Stellar component}

In Table \ref{table:fluxes}, we have included the EW of the first absorption band of the CO at 2.293\,$\mu$m. In most of the objects, the values correspond to the \Brg\ (\Pa) peak and nucleus, which are the regions with enough S/N in the continuum to detect the band. The nuclear values of the LIRGs are 7.1\,\AA\ $\le \rm EW_{\rm CO}\le$  12.3\,\AA, with typical uncertainties of $\sim10$\% and an average of 10.6\,\AA\ whereas the values measured at the \Brg\ (\Pa) peak cover the range 8.3\,\AA\ $\le \rm EW_{\rm CO}\le$ 12.2\,\AA, with the same uncertainties and a mean value of 10.7\,\AA. According to the stellar population synthesis models like STARBURST99 \citep{Leitherer:1999p6938}, these values of the EW correspond to stellar populations older than $\log \rm T(yr)\sim6.8$ and up to $\log \rm T(yr)\sim8.2$ or more, depending on whether we consider a instantaneous burst or a continuum star-formation activity. The detailed study of the 2D distribution of the stellar populations using the CO stellar absorption, the H, and He emission lines will be addressed in forthcoming papers.

\subsection{Kinematics of the gas}

Besides the general morphology and luminosities of the different emission lines, their 2D kinematics (velocity field and velocity dispersion maps) are also presented (Figs.~\ref{figure:LIRG} and \ref{figure:ULIRG}). We obtained the velocity dispersion of the different regions of interest described above, i.e. nucleus, the emission peaks of the \Brg\ (\Pa), H$_2$ 1-0S(1) and [FeII] lines, and the entire FoV. The values of the velocity dispersion, corrected for the instrumental broadening, are shown in Table \ref{table:sigma}. The errors are obtained following the same Monte Carlo method implemented to estimate the flux error. Figure~\ref{figure:histograms_sigma} shows the distributions of the velocity dispersion obtained from the values of Table \ref{table:sigma}. The distributions show that there is no clear relationship between the \LTIR\ of the objects and the velocity dispersion of the different regions, although the highest values of velocity dispersion tend to come from high-luminosity objects where measurements come from scales $\times$4--5 larger, so could be affected by beam smearing.

The average velocity dispersion of the \Brg\ and H$_{2}$ 1-0S(1) lines in the LIRGs is $\sim90$\,km\,s$^{-1}$, whereas the measured average values for the ULIRG subsample are $\sim140$\,km\,s$^{-1}$ and $\sim120$\,km\,s$^{-1}$, respectively. The higher velocity dispersion measured in the ULIRG subset can be explained mainly as a distance effect: the contribution from the unresolved velocity field to the width of the line is larger since the physical scales are also larger. We estimated this effect by extracting several spectra over apertures of increasing radius and measuring the width of the \Brg\ and H$_{2}$ 1-0S(1) lines in one of the objects of the LIRG subset. We used NGC~3110 since the \Brg\ and H$_{2}$ emitting gas is extended and well sampled in almost the entire FoV, and their velocity fields show a well defined rotation pattern. The velocity dispersion of the unresolved nuclear \Brg\ emission is $\sigma_{v}\sim75$\,km\,s$^{-1}$ at a distance of 78.4\,Mpc. We then simulated the observed spectra of the object at increasing distances up to 500\,Mpc and found that, at the average distance of the ULIRG subsample of $\sim328$\,Mpc, the measured velocity dispersion of the \Brg\ line rises up to $\sigma_{v}\sim105$\,km\,s$^{-1}$,  yielding an increase of $\sim30$\,km\,s$^{-1}$. The results obtained with the H$_{2}$ 1-0S(1) line are equivalent and yield a difference of $\sim28$\,km\,s$^{-1}$, since the amplitude of the velocity fields of both phases of the gas are almost identical. Based on these estimates, the difference in velocity dispersion observed between LIRGs and ULIRGs appears to mainly be due to distance effects. However, galaxies with steeper velocity gradients, radial gas flows, turbulence, or massive regions with intrinsically high velocity dispersion would produced an additional increase in the value of the dispersion. 

The velocity fields of the gas observed in the rotating LIRGs have the typical spider pattern characteristic of a thin disk, with a well identified kinematic centre that coincides in most cases with the K-band photometric centre, and with a major kinematic axis close to the major photometric axis. These results are similar to those derived from the H$\alpha$ emission in the central regions of LIRGs \citep{AlonsoHerrero:2009p3373} and from the mid-infrared [NeII] and H$_2$ emission \citep{PereiraSantaella:2010p4662} for larger FoVs. These characteristics indicate that the velocity fields of both the ionised and the warm molecular gas are dominated by the rotation of a disk around the centre of the galaxy,  as expected given that almost all of the objects of the subsample are spiral galaxies with different levels of inclination. Local deviations and irregularities from rotation, as well as regions of higher velocity dispersion, are present in most/all the LIRGs, suggesting the presence of additional radial flows and/or regions of higher turbulence or outflows outside and close to the nucleus (e.g. \object{NGC 3256}, \object{NGC 5135}). Besides these local deviations, the ionised and molecular gas of the LIRGs show the same velocity field on almost all scales, from regions of a few hundred parsecs to scales of several kpc.

In the ULIRG subsample, since all the objects but one are mergers in a pre-coalescence phase, the kinematics of the gas show a more complex structure, with signs of strong velocity gradients associated with the different progenitors of the systems, and asymmetric line profiles that indicate there are outflows of gas associated with AGN or starburst activity. It is interesting to note that, as in LIRGs, the ionised and warm molecular gas in ULIRGs show the same overall kinematics on scales of a few to several kpc. Whether the kinematics in U/LIRGs is dominated by rotation, radial starbursts/AGN flows, tidal-induced flows, or a combination of these, the ionised and molecular gas share the same kinematics on physical scales ranging from a few hundred parsecs (LIRGs) to several kpc (ULIRGs). A detailed study of the gas kinematics of the sample is beyond the aim of this work and will be addressed in a forthcoming paper (Azzollini et al. 2012, in preparation); however, a brief individual description of the most relevant features of the gas kinematics is included in Appendix \ref{notes}.

\begin{figure*}
\begin{center}
\resizebox{\hsize}{!}{\includegraphics[angle=90]{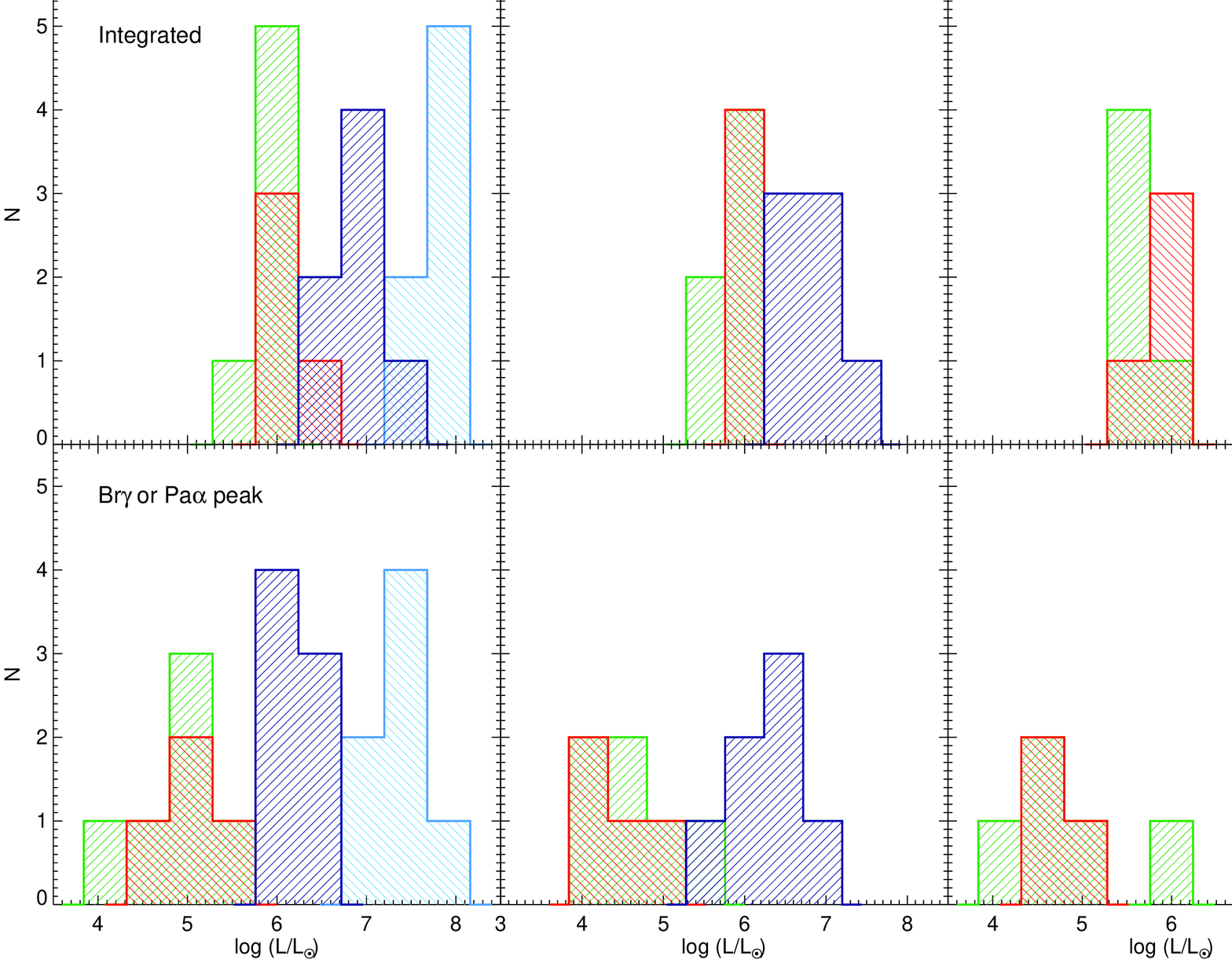}}
\caption{Luminosity distribution of the \Brg, H$_{2}$ 1-0S(1), and [FeII] emission. From top to bottom, the histograms show the distribution of the total luminosity in solar units of the lines measured in the nucleus (defined by aperture ``A'' in Figs. \ref{figure:LIRG} and \ref{figure:ULIRG}), the integrated FoV, and the peak of the \Brg\ (\Pa) emission. For the distributions of the ionised gas (first column), we have also included the \Pa\ emission in light blue for the ULIRGs. Note: \Brg\ (\Pa) peak coincides with the nucleus in $\sim33$\% of the LIRGs and in the $\sim$71\% of the ULIRGs.}
\label{figure:histograms}
\end{center}
\end{figure*}

\begin{figure*}
\begin{center}
\resizebox{\hsize}{!}{\includegraphics[angle=90]{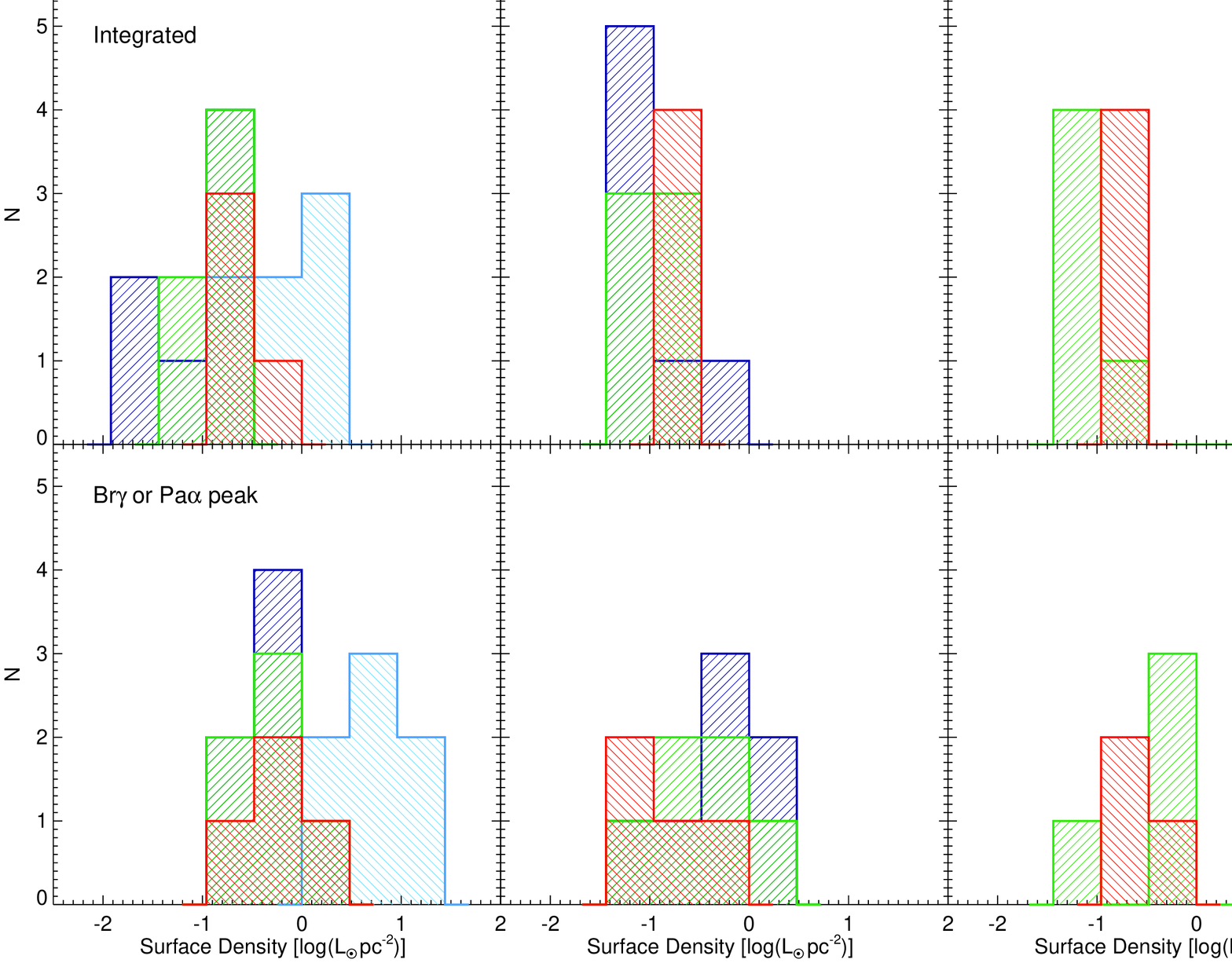}}
\caption{Surface density distribution of the \Brg, H$_{2}$ 1-0S(1), and [FeII] emission. From top to bottom, the histograms show the distribution of the surface density in solar units per unit of area (pc$^2$) of the lines measured in the nucleus (defined by aperture ``A'' in Figs. \ref{figure:LIRG} and \ref{figure:ULIRG}), the integrated FoV, and the peak of the \Brg\ (\Pa) emission. For the distributions of the ionised gas (first column), we have also included the \Pa\ emission in light blue for the ULIRGs. The \Brg\ (\Pa) peak coincides with the nucleus in $\sim33$\% of the LIRGs and in the $\sim$71\% of the ULIRGs.}
\label{figure:histograms_area}
\end{center}
\end{figure*}

\begin{figure*}
\begin{center}
\resizebox{\hsize}{!}{\includegraphics[angle=90]{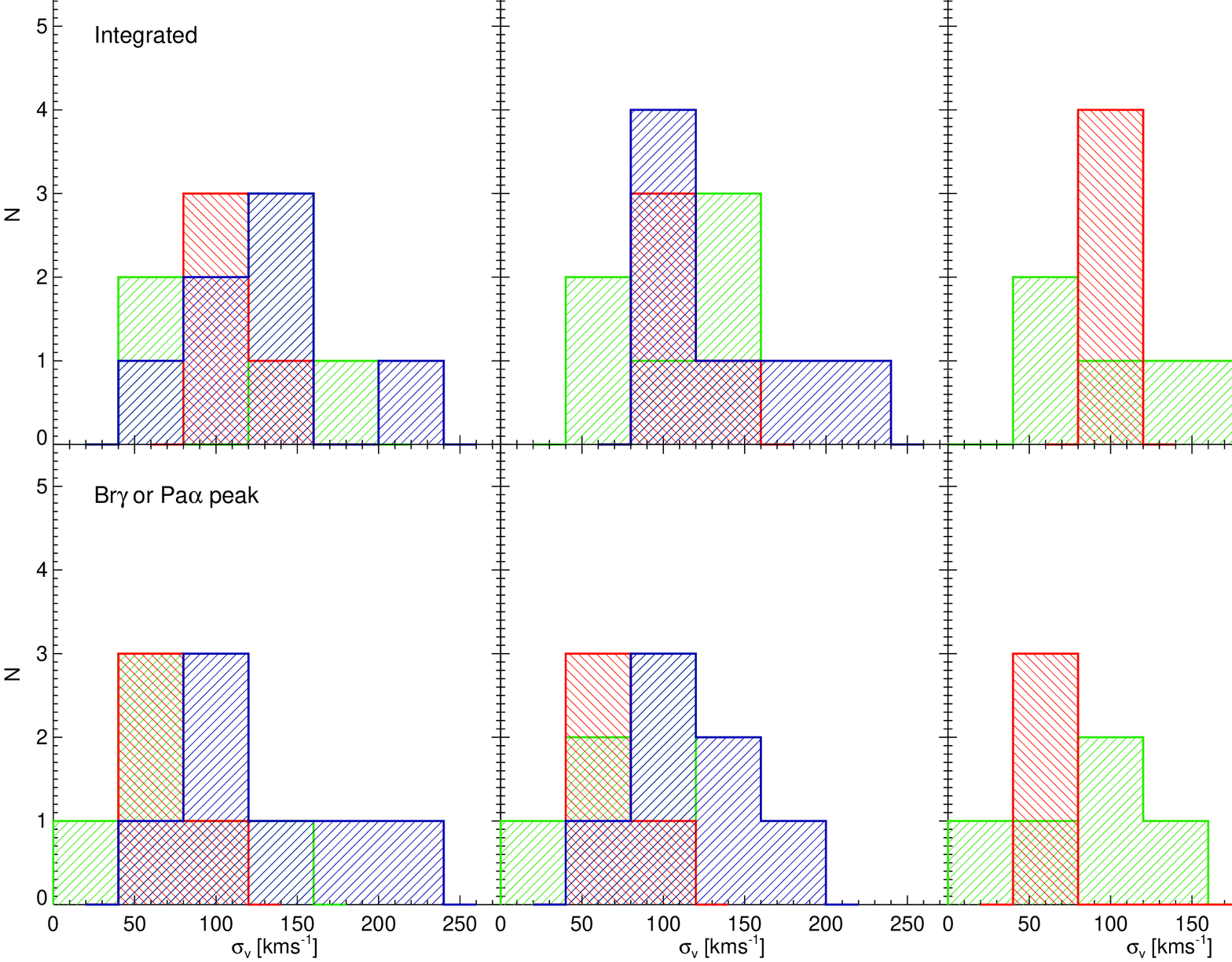}}
\caption{Distributions of the velocity dispersion of the ionised gas (\Brg\ for LIRGs \Pa\ for ULIRGs), H$_{2}$ 1-0S(1), and [FeII] emission. From top to bottom, the histograms show the distributions of the velocity dispersion measured in the nucleus (defined by aperture ``A'' in Figs. \ref{figure:LIRG} and \ref{figure:ULIRG}), the integrated FoV, and the peak of the \Brg\ (\Pa) emission. The \Brg\ (\Pa) peak coincides with the nucleus in $\sim33$\% of the LIRGs and in the $\sim$71\% of the ULIRGs.}
\label{figure:histograms_sigma}
\end{center}
\end{figure*}

\section{Summary}
\label{section:summary}

We have obtained K-band SINFONI seeing limited observations of a sample of local LIRGs and ULIRGs ($z<0.1$), together with H-band SINFONI spectroscopy for the LIRG subsample. The luminosity range covered by the observations is  log(\LTIR/\Lsolar)$=11.1-12.4$, with an average redshift of $z_{\rm LIRGs}=0.014$ and $z_{\rm ULIRGs}=0.072$ ($\sim$63\,Mpc and $\sim$328\,Mpc) for LIRGs and ULIRGs, respectively. The IFS maps cover the central $\sim3\times3$\,kpc of the LIRGs and the central $\sim12\times12$\,kpc of the ULIRGs with a scale of $0\farcs125$ per spaxel. We present the 2D distribution of the emitting line gas of the whole sample and some general results of the morphology, luminosities, and kinematics of the line-emitting gas as traced by different emission lines. The detailed studies of the excitation mechanisms, extinction, stellar populations, and stellar and gas kinematics of the entire sample will be presented in forthcoming papers.

In a third of LIRGs, the peaks of the ionised and molecular gas coincide with the stellar nucleus of the galaxy (distances of less than $0\farcs25$), and the \Brg\ line typically shows luminosities of $\sim1.2\times10^5$\,\Lsolar. However, in galaxies with star-forming rings or giant HII regions in the spiral arms, the emission of ionised gas is dominated by such structures. The warm molecular shows very similar luminosities to the \Brg\ emission and is highly concentrated in the nucleus, where it reaches its maximum in all the objects of our sample. The \Brg\ and  [FeII] emission traces the same structures, although their emission peaks are not spatially coincident in some of the objects, and the [FeII] seems to be more extended and diffuse.

The ULIRG subsample is at greater distances ($\sim$4--5 times) and mainly contains pre-coalescence interacting systems. Although the peaks of the molecular gas emission and the main nucleus of the objects coincide in $\sim71$\% of the galaxies, we also detect regions with intense \Pa\ emission up to $\sim1.1\times10^7$\,\Lsolar\ which trace luminous star-forming regions located at distances of 2--4\,kpc away from the nucleus.

LIRGs have mean observed (i.e. uncorrected for internal extinction) SFR surface densities of about 0.4 to 0.9 M$_{\odot}$ yr$^{-1}$ kpc$^{-2}$ over extended areas of 4$-$9 kpc$^2$ with peaks of about 2$-$2.5 M$_{\odot}$ yr$^{-1}$ kpc$^{-2}$ in compact regions (0.16 kpc$^2$) associated with the nucleus of the galaxy or the brightest \Brg\ region. ULIRGs do have similar values ($\sim$0.4 and $\sim$ 2 M$_{\odot}$ yr$^{-1}$ kpc$^{-2}$) over much larger areas, 100$-$200 kpc$^2$ and 4 kpc$^2$ for the integrated and peak emission, respectively.
To correct the above values from extinction, we applied a median A$_{\rm V}$ value of $\sim6.5$\,mag for LIRGs and $\sim7.1$\,mag for ULIRGs, and found that the SFR measurements should increase by a factor $\sim1.7$ in LIRGs and $\sim2.5$ in ULIRGs, when dereddened luminosities are considered.

The observed gas kinematics in LIRGs is primarily due to rotational motions around the centre of the galaxy, although local deviations associated with radial flows and/or regions of higher velocity dispersions are present.
The ionised and molecular gas share the same kinematics (velocity field and velocity dispersion), showing in some cases slight differences in the velocity amplitudes (peak-to-peak). Given the interacting nature of the objects of the subsample, the kinematics of the ULIRG show complex velocity fields with different gradients associated with the progenitors of the system and tidal tails.

\begin{acknowledgements}
We thank the anonymous referee for his/her useful comments and suggestions that improved the final content of this paper.
This work was supported by the Spanish Ministry of Science and Innovation (MICINN) under grants BES-2008-007516, ESP2007-65475-C02-01, and AYA2010-21161-C02-01.
JPL thanks Ric Davies for his valuable support and enlightening discussions about the reduction and data analysis.
JPL wants to thank Miguel Pereira-Santaella, Ruyman Azzollini, Daniel Miralles, Javier Rodr\'iguez, and Alvaro Labiano for fruitful discussions and assistance.
This paper made use of the plotting package \emph{jmaplot}, developed by Jes\'us Ma\'iz-Apell\'aniz \url{http://jmaiz.iaa.es/software/jmaplot/current/html/jmaplot_overview.html}.
Based on observations collected at the European Organisation for Astronomical Research in the Southern Hemisphere, Chile, programmes 077.B-0151A, 078.B-0066A, and 081.B-0042A.
This research made use of the NASA/IPAC Extragalactic Database (NED), which is operated by the Jet Propulsion Laboratory, California Institute of Technology, under contract with the National Aeronautics and Space Administration. 
Some of the data presented in this paper were obtained from the Multimission Archive at the Space Telescope Science Institute (MAST). STScI is operated by the Association of Universities for Research in Astronomy, Inc., under NASA contract NAS5-26555. Support for MAST for non-HST data is provided by the NASA Office of Space Science via grant NNX09AF08G and by other grants and contracts.
\end{acknowledgements}

\setcounter{table}{4}
\setcounter{fake_table}{\value{table}}
\refstepcounter{fake_table}
\label{table:fluxes}
\renewcommand{\thetable}{\arabic{table}\alph{subtab}}
\setcounter{subtab}{1}

\begin{landscape}
\begin{table}
\caption{\Brg, H$_2$ 1-0S(1), and [FeII] integrated observed fluxes and CO (2-0) equivalent widths of the LIRG subsample}
\begin{center}
\tiny
\begin{tabular}{c x{1cm}@{ $\pm$ }z{1cm} x{1cm}@{ $\pm$ }z{1cm} x{1cm}@{ $\pm$ }z{1cm} r@{ $\pm$ }l p{0.01mm} c x{1cm}@{ $\pm$ }z{1cm} x{1cm}@{ $\pm$ }z{1cm} x{1cm}@{ $\pm$ }z{1cm} r@{ $\pm$ }l}

\cline{1-9} \cline{11-19}
\noalign{\vspace{0.5mm}}
\cline{1-9} \cline{11-19}
\noalign{\smallskip}
\multicolumn{9}{c}{\object{NGC 2369}} & & \multicolumn{9}{c}{\object{NGC 3110}}\\
\cline{1-9} \cline{11-19}
\noalign{\smallskip}
\multicolumn{1}{c}{\multirow{2}{*}{Region}} & \multicolumn{2}{c}{\Brg\ Flux} & \multicolumn{2}{c}{H$_{2}$ 1-0S(1) Flux} & \multicolumn{2}{c}{[Fe II] Flux} & \multicolumn{2}{c}{W$_{\rm CO}$} & &\multicolumn{1}{c}{\multirow{2}{*}{Region}} & \multicolumn{2}{c}{\Brg\ Flux} & \multicolumn{2}{c}{H$_{2}$ 1-0S(1) Flux} & \multicolumn{2}{c}{[Fe II] Flux} & \multicolumn{2}{c}{W$_{\rm CO}$}\\
& \multicolumn{2}{m{2.5cm}}{(erg\,s$^{-1}$cm$^{-2}$)$\times$10$^{-16}$} & \multicolumn{2}{m{2.5cm}}{(erg\,s$^{-1}$cm$^{-2}$)$\times$10$^{-16}$} &\multicolumn{2}{m{2.5cm}}{(erg\,s$^{-1}$cm$^{-2}$)$\times$10$^{-16}$} & \multicolumn{2}{c}{(\AA)} & & & \multicolumn{2}{m{2.5cm}}{(erg\,s$^{-1}$cm$^{-2}$)$\times$10$^{-16}$} & \multicolumn{2}{m{2.5cm}}{(erg\,s$^{-1}$cm$^{-2}$)$\times$10$^{-16}$} &\multicolumn{2}{m{2.5cm}}{(erg\,s$^{-1}$cm$^{-2}$)$\times$10$^{-16}$} & \multicolumn{2}{c}{(\AA)} \\
\noalign{\smallskip}
\cline{1-9} \cline{11-19}
\noalign{\vspace{0.5mm}}
\cline{1-9} \cline{11-19}
\noalign{\smallskip}
Nuclear&  14.71 &   0.59 &  10.68 &   0.57 &   9.40 &   0.58 &   12.3 &    2.1& & Nuclear &   3.05 &   0.08 &   2.98 &   0.15 &   1.82 &   0.22 &   10.5 &    0.9\\
Integrated &  78.46 &  12.5 &  75.51 &   8.54 &  91.13 &  14.8 & \multicolumn{2}{c}{\nodata} & & Integrated &  45.92 &   1.48 &  52.45 &   2.77 &  22.66 &   8.20 & \multicolumn{2}{c}{\nodata}\\
\Brg\ max$^{\dag}$ &  15.27 &   0.63 &  10.92 &   0.57 &  10.21 &   0.61 &   11.9 &    1.8& & \Brg\ max &   1.66 &   0.03 &   1.08 &   0.03 &   0.68 &   0.06 &    8.2 &    0.9\\
H$_{2}$ 1-0S(1)$^{\dag}$&  14.66 &   0.59 &  10.58 &   0.56 &   9.85 &   0.59 &   12.0 &    1.7 & & H$_{2}$ 1-0S(1)$^{\dag}$ &   3.05 &   0.08 &   2.98 &   0.15 &   1.82 &   0.22 &   10.5 &    0.9\\
$[$FeII] max &  11.16 &   0.58 &   9.15 &   0.45 &   9.72 &   0.81 &   11.5 &    1.3& & [FeII] max$^{\dag}$ &   2.70 &   0.06 &   2.50 &   0.12 &   1.59 &   0.19 &   10.5 &    1.1\\
\noalign{\smallskip}
\cline{1-9} \cline{11-19}
\noalign{\vspace{0.5mm}}
\cline{1-9} \cline{11-19}
\noalign{\smallskip}

\multicolumn{9}{c}{\object{NGC 3256}$^{\ddag}$} & & \multicolumn{9}{c}{\object{ESO 320-G030}}\\
\cline{1-9} \cline{11-19}
\noalign{\smallskip}
\multicolumn{1}{c}{\multirow{2}{*}{Region}} & \multicolumn{2}{c}{\Brg\ Flux} & \multicolumn{2}{c}{H$_{2}$ 1-0S(1) Flux} & \multicolumn{2}{c}{[Fe II] Flux} & \multicolumn{2}{c}{W$_{\rm CO}$} & &\multicolumn{1}{c}{\multirow{2}{*}{Region}} & \multicolumn{2}{c}{\Brg\ Flux} & \multicolumn{2}{c}{H$_{2}$ 1-0S(1) Flux} & \multicolumn{2}{c}{[Fe II] Flux} & \multicolumn{2}{c}{W$_{\rm CO}$}\\
& \multicolumn{2}{m{2.5cm}}{(erg\,s$^{-1}$cm$^{-2}$)$\times$10$^{-16}$} & \multicolumn{2}{m{2.5cm}}{(erg\,s$^{-1}$cm$^{-2}$)$\times$10$^{-16}$} &\multicolumn{2}{m{2.5cm}}{(erg\,s$^{-1}$cm$^{-2}$)$\times$10$^{-16}$} & \multicolumn{2}{c}{(\AA)} & & & \multicolumn{2}{m{2.5cm}}{(erg\,s$^{-1}$cm$^{-2}$)$\times$10$^{-16}$} & \multicolumn{2}{m{2.5cm}}{(erg\,s$^{-1}$cm$^{-2}$)$\times$10$^{-16}$} &\multicolumn{2}{m{2.5cm}}{(erg\,s$^{-1}$cm$^{-2}$)$\times$10$^{-16}$} & \multicolumn{2}{c}{(\AA)} \\
\noalign{\smallskip}
\cline{1-9} \cline{11-19}
\noalign{\vspace{0.5mm}}
\cline{1-9} \cline{11-19}
\noalign{\smallskip}
Nuclear&  22.21 &   0.33 &  19.13 &   0.21 &   4.36 &   0.47 &   11.3 &    0.7& & Nuclear &   2.35 &   0.29 &  16.62 &   0.50 &   2.30 &   0.77 &   11.4 &    1.2\\
Integrated & 213.92 &   4.45 & 206.98 &   3.40 & 132.85 &   6.39 & \multicolumn{2}{c}{\nodata}& & Integrated &  82.80 &   7.59 &  78.57 &   6.83 &  49.03 &   7.65 & \multicolumn{2}{c}{\nodata}\\
\Brg\ max$^{\dag}$ &  22.40 &   0.36 &  18.85 &   0.24 & \multicolumn{2}{c}{\nodata} &   11.6 &    0.5& & \Brg\ max &   4.15 &   0.13 &   2.06 &   0.13 &   2.72 &   0.21 &   11.1 &    0.9\\
H$_{2}$ 1-0S(1) &   4.17 &   0.21 &  26.98 &   0.75 & \multicolumn{2}{c}{\nodata} &   10.3 &    0.8& & H$_{2}$ 1-0S(1)$^{\dag}$ &   2.07 &   0.29 &  16.14 &   0.49 &   2.12 &   0.79 &   11.6 &    1.1 \\
$[$FeII] max & \multicolumn{2}{c}{\nodata} & \multicolumn{2}{c}{\nodata} &   4.02 &   0.46 & \multicolumn{2}{c}{\nodata}& & [FeII] max$^{\dag}$ &   2.67 &   0.28 &  13.27 &   0.43 &   2.41 &   0.71 &   11.6 &    1.1\\
\noalign{\smallskip}
\cline{1-9} \cline{11-19}
\noalign{\vspace{0.5mm}}
\cline{1-9} \cline{11-19}
\noalign{\smallskip}
\multicolumn{9}{c}{\object{IRASF 12115-4656}} & & \multicolumn{9}{c}{\object{NGC 5135}}\\
\cline{1-9} \cline{11-19}
\noalign{\smallskip}
\multicolumn{1}{c}{\multirow{2}{*}{Region}} & \multicolumn{2}{c}{\Brg\ Flux} & \multicolumn{2}{c}{H$_{2}$ 1-0S(1) Flux} & \multicolumn{2}{c}{[Fe II] Flux} & \multicolumn{2}{c}{W$_{\rm CO}$} & &\multicolumn{1}{c}{\multirow{2}{*}{Region}} & \multicolumn{2}{c}{\Brg\ Flux} & \multicolumn{2}{c}{H$_{2}$ 1-0S(1) Flux} & \multicolumn{2}{c}{[Fe II] Flux} & \multicolumn{2}{c}{W$_{\rm CO}$}\\
& \multicolumn{2}{m{2.5cm}}{(erg\,s$^{-1}$cm$^{-2}$)$\times$10$^{-16}$} & \multicolumn{2}{m{2.5cm}}{(erg\,s$^{-1}$cm$^{-2}$)$\times$10$^{-16}$} &\multicolumn{2}{m{2.5cm}}{(erg\,s$^{-1}$cm$^{-2}$)$\times$10$^{-16}$} & \multicolumn{2}{c}{(\AA)} & & & \multicolumn{2}{m{2.5cm}}{(erg\,s$^{-1}$cm$^{-2}$)$\times$10$^{-16}$} & \multicolumn{2}{m{2.5cm}}{(erg\,s$^{-1}$cm$^{-2}$)$\times$10$^{-16}$} &\multicolumn{2}{m{2.5cm}}{(erg\,s$^{-1}$cm$^{-2}$)$\times$10$^{-16}$} & \multicolumn{2}{c}{(\AA)} \\
\noalign{\smallskip}
\cline{1-9} \cline{11-19}
\noalign{\vspace{0.5mm}}
\cline{1-9} \cline{11-19}
\noalign{\smallskip}
Nuclear&   0.85 &   0.09 &   1.89 &   0.18 & \multicolumn{2}{c}{\nodata} &    8.9 &    0.7& & Nuclear &   6.45 &   0.20 &   9.65 &   0.22 &   5.17 &   0.25 &    7.1 &    0.7\\
Integrated &  69.71 &   6.95 &  59.83 &   6.20 & \multicolumn{2}{c}{\nodata} & \multicolumn{2}{c}{\nodata}& & Integrated &  49.39 &   2.54 &  69.23 &   4.13 &  42.71 &   3.69 & \multicolumn{2}{c}{\nodata}\\
\Brg\ max &   0.80 &   0.03 &   0.37 &   0.03 & \multicolumn{2}{c}{\nodata} &    9.5 &    0.7& & \Brg\ max &   6.87 &   0.11 &   4.40 &   0.17 &   4.31 &   0.17 &   11.4 &    1.0\\
H$_{2}$ 1-0S(1)$^{\dag}$ &   0.87 &   0.08 &   1.90 &   0.18 & \multicolumn{2}{c}{\nodata} &    8.9 &    0.7& & H$_{2}$ 1-0S(1)$^{\dag}$ &   6.45 &   0.20 &   9.65 &   0.22 &   5.17 &   0.25 &    7.1 &    0.7\\
$[$FeII] max &   \multicolumn{2}{c}{\nodata} &   \multicolumn{2}{c}{\nodata} &   \multicolumn{2}{c}{\nodata} &    \multicolumn{2}{c}{\nodata}& & [FeII] max &   5.23 &   0.16 &  10.64 &   0.38 &  46.26 &   1.29 &   13.4 &    1.3\\
\noalign{\smallskip}
\cline{1-9} \cline{11-19}
\noalign{\vspace{0.5mm}}
\cline{1-9} \cline{11-19}
\noalign{\smallskip}

\multicolumn{9}{c}{\object{IRASF 17138-1017}} & & \multicolumn{9}{c}{\object{IC 4687}}\\
\cline{1-9} \cline{11-19}
\noalign{\smallskip}
\multicolumn{1}{c}{\multirow{2}{*}{Region}} & \multicolumn{2}{c}{\Brg\ Flux} & \multicolumn{2}{c}{H$_{2}$ 1-0S(1) Flux} & \multicolumn{2}{c}{[Fe II] Flux} & \multicolumn{2}{c}{W$_{\rm CO}$} & &\multicolumn{1}{c}{\multirow{2}{*}{Region}} & \multicolumn{2}{c}{\Brg\ Flux} & \multicolumn{2}{c}{H$_{2}$ 1-0S(1) Flux} & \multicolumn{2}{c}{[Fe II] Flux} & \multicolumn{2}{c}{W$_{\rm CO}$}\\
& \multicolumn{2}{m{2.5cm}}{(erg\,s$^{-1}$cm$^{-2}$)$\times$10$^{-16}$} & \multicolumn{2}{m{2.5cm}}{(erg\,s$^{-1}$cm$^{-2}$)$\times$10$^{-16}$} &\multicolumn{2}{m{2.5cm}}{(erg\,s$^{-1}$cm$^{-2}$)$\times$10$^{-16}$} & \multicolumn{2}{c}{(\AA)} & & & \multicolumn{2}{m{2.5cm}}{(erg\,s$^{-1}$cm$^{-2}$)$\times$10$^{-16}$} & \multicolumn{2}{m{2.5cm}}{(erg\,s$^{-1}$cm$^{-2}$)$\times$10$^{-16}$} &\multicolumn{2}{m{2.5cm}}{(erg\,s$^{-1}$cm$^{-2}$)$\times$10$^{-16}$} & \multicolumn{2}{c}{(\AA)} \\
\noalign{\smallskip}
\cline{1-9} \cline{11-19}
\noalign{\vspace{0.5mm}}
\cline{1-9} \cline{11-19}
\noalign{\smallskip}
Nuclear &   6.35 &   0.11 &   3.06 &   0.09 &   5.05 &   0.14 &   11.1 &    1.0& & Nuclear  &   5.31 &   0.11 &   3.27 &   0.11 &   3.75 &   0.23 &   12.3 &    1.1\\
Integrated &  89.48 &   2.69 &  42.29 &   1.97 &  61.55 &   2.75 & \multicolumn{2}{c}{\nodata}& & Integrated & 146.64 &   3.77 &  59.31 &   2.91 &  78.03 &   7.14 & \multicolumn{2}{c}{\nodata}\\
\Brg\ max &  12.14 &   0.13 &   1.75 &   0.06 &   5.49 &   0.11 &    8.8 &    0.9& & \Brg\ max &   7.28 &   0.10 &   0.97 &   0.03 &   2.64 &   0.11 &    8.3 &    1.0\\
H$_{2}$ 1-0S(1)$^{\dag}$ &   6.35 &   0.11 &   3.06 &   0.09 &   5.05 &   0.14 &   11.1 &    1.0& & H$_{2}$ 1-0S(1)$^{\dag}$ &   4.46 &   0.09 &   2.86 &   0.10 &   3.34 &   0.23 &   11.9 &    1.3\\
$[$FeII] max &   7.74 &   0.10 &   1.90 &   0.04 &   4.13 &   0.08 &    8.7 &    0.6& & [FeII] max &   7.05 &   0.09 &   0.99 &   0.03 &   2.86 &   0.12 &    9.2 &    1.0\\
\noalign{\smallskip}
\cline{1-9} \cline{11-19}
\noalign{\vspace{0.5mm}}
\cline{1-9} \cline{11-19}
\noalign{\smallskip}

\multicolumn{9}{c}{\object{NGC 7130}} & & \multicolumn{9}{c}{\object{IC 5179}}\\
\cline{1-9} \cline{11-19}
\noalign{\smallskip}
\multicolumn{1}{c}{\multirow{2}{*}{Region}} & \multicolumn{2}{c}{\Brg\ Flux} & \multicolumn{2}{c}{H$_{2}$ 1-0S(1) Flux} & \multicolumn{2}{c}{[Fe II] Flux} & \multicolumn{2}{c}{W$_{\rm CO}$} & &\multicolumn{1}{c}{\multirow{2}{*}{Region}} & \multicolumn{2}{c}{\Brg\ Flux} & \multicolumn{2}{c}{H$_{2}$ 1-0S(1) Flux} & \multicolumn{2}{c}{[Fe II] Flux} & \multicolumn{2}{c}{W$_{\rm CO}$}\\
& \multicolumn{2}{m{2.5cm}}{(erg\,s$^{-1}$cm$^{-2}$)$\times$10$^{-16}$} & \multicolumn{2}{m{2.5cm}}{(erg\,s$^{-1}$cm$^{-2}$)$\times$10$^{-16}$} &\multicolumn{2}{m{2.5cm}}{(erg\,s$^{-1}$cm$^{-2}$)$\times$10$^{-16}$} & \multicolumn{2}{c}{(\AA)} & & & \multicolumn{2}{m{2.5cm}}{(erg\,s$^{-1}$cm$^{-2}$)$\times$10$^{-16}$} & \multicolumn{2}{m{2.5cm}}{(erg\,s$^{-1}$cm$^{-2}$)$\times$10$^{-16}$} &\multicolumn{2}{m{2.5cm}}{(erg\,s$^{-1}$cm$^{-2}$)$\times$10$^{-16}$} & \multicolumn{2}{c}{(\AA)} \\
\noalign{\smallskip}
\cline{1-9} \cline{11-19}
\noalign{\vspace{0.5mm}}
\cline{1-9} \cline{11-19}
\noalign{\smallskip}
Nuclear &  17.32 &   0.36 &  25.39 &   0.44 &  49.26 &   1.84 &   11.4 &    0.7& & Nuclear  &  13.70 &   0.19 &   8.69 &   0.32 &   8.48 &   0.54 &   12.2 &    1.2\\
Integrated &  30.62 &   1.43 &  51.12 &   2.33 &  32.95 &   2.40 & \multicolumn{2}{c}{\nodata}& & Integrated &  98.24 &   6.98 &  61.19 &   5.00 &  45.45 &   6.25 & \multicolumn{2}{c}{\nodata}\\
\Brg\ max$^{\dag}$ &  17.32 &   0.36 &  25.39 &   0.44 &  49.26 &   1.84 &   11.4 &    0.7& & \Brg\ max$^{\dag}$ &  13.50 &   0.17 &   8.06 &   0.32 &   8.16 &   0.52 &   12.2 &    1.2\\
H$_{2}$ 1-0S(1)$^{\dag}$ &  16.84 &   0.35 &  25.17 &   0.44 &  49.02 &   1.84 &   11.6 &    0.9& & H$_{2}$ 1-0S(1)$^{\dag}$ &  13.42 &   0.17 &   8.34 &   0.32 &   8.39 &   0.52 &   12.2 &    1.2\\
$[$FeII] max$^{\dag}$ &  17.32 &   0.36 &  25.39 &   0.44 &  49.26 &   1.84 &   11.4 &    0.7& & [FeII] max$^{\dag}$ &  13.12 &   0.16 &   7.60 &   0.31 &   7.98 &   0.50 &   12.2 &    1.1\\
\noalign{\smallskip}
\cline{1-9} \cline{11-19}
\noalign{\vspace{0.5mm}}
\cline{1-9} \cline{11-19}
\noalign{\smallskip}

\end{tabular}
\end{center}
\tablefoot{\tiny Spectra are integrated within a 400$\times$400\,pc aperture, covering the nuclear region defined as the brightest spaxel in the K-band; the integrated emission of the FoV, defined as the integrated flux of those spaxels that contain at least the $\gsim90\%$ of the total continuum flux in each band; and the peaks of the \Brg, H$_2$ 1-0S(1) and [FeII] emission, centred on the brightest spaxel in each of the respective maps. The errors are obtained by a Monte Carlo method of $N=1000$ simulations of each spectra.
$^{\dag}$ Regions that are coincident with the nucleus of the object. The spectra are extracted and the lines are fitted independently.$^{\ddag}$The \Brg\ peak of emission does not coincide with the stellar nucleus of the object but with the region we have adopted as the ``nucleus" (see the main text for a further explanation).}
\label{table:lirgs_fluxes}
\end{table}
\end{landscape}

\addtocounter{table}{-1}
\addtocounter{subtab}{1}

\begin{landscape}
\begin{table}
\caption{\Pa\ and H$_2$ 1-0S(1)  integrated observed fluxes and CO (2-0) equivalent widths of the ULIRGs subsample}
\begin{center}
\tiny
\begin{tabular}[t]{c x{1cm}@{ $\pm$ }z{1cm} x{1cm}@{ $\pm$ }z{1cm}  r@{ $\pm$ }l p{0.01mm} c x{1cm}@{ $\pm$ }z{1cm} x{1cm}@{ $\pm$ }z{1cm} r@{ $\pm$ }l}

\cline{1-7} \cline{9-15}
\noalign{\vspace{0.5mm}}
\cline{1-7} \cline{9-15}
\noalign{\smallskip}
\multicolumn{7}{c}{\object{IRAS 06206-6315}} & & \multicolumn{7}{c}{\object{IRAS 12112+0305}}\\
\cline{1-7} \cline{9-15}
\noalign{\smallskip}
\multicolumn{1}{c}{\multirow{2}{*}{Region}} & \multicolumn{2}{c}{\Pa\ Flux} & \multicolumn{2}{c}{H$_{2}$ 1-0S(1) Flux} & \multicolumn{2}{c}{W$_{\rm CO}$} & &\multicolumn{1}{c}{\multirow{2}{*}{Region}} & \multicolumn{2}{c}{\Pa\ Flux} & \multicolumn{2}{c}{H$_{2}$ 1-0S(1) Flux} & \multicolumn{2}{c}{W$_{\rm CO}$}\\
& \multicolumn{2}{c}{(erg\,s$^{-1}$cm$^{-2}$)$\times$10$^{-16}$} & \multicolumn{2}{c}{(erg\,s$^{-1}$cm$^{-2}$)$\times$10$^{-16}$} & \multicolumn{2}{c}{(\AA)} & & & \multicolumn{2}{c}{(erg\,s$^{-1}$cm$^{-2}$)$\times$10$^{-16}$} & \multicolumn{2}{c}{(erg\,s$^{-1}$cm$^{-2}$)$\times$10$^{-16}$} & \multicolumn{2}{c}{(\AA)} \\
\noalign{\smallskip}
\cline{1-7} \cline{9-15}
\noalign{\vspace{0.5mm}}
\cline{1-7} \cline{9-15}
\noalign{\smallskip}
Nuclear&  32.91 &   0.90 &   6.06 &   0.12 & \multicolumn{2}{c}{\nodata}& & Nuclear &  45.48 &   0.56 &   3.62 &   0.10 & \multicolumn{2}{c}{\nodata}\\
Integrated &  42.47 &  6.54 &  25.13 &   1.42 & \multicolumn{2}{c}{\nodata}& & Integrated & 190.87 &   6.61 &  32.44 &   1.59 & \multicolumn{2}{c}{\nodata}\\
\Pa\  max$^{\dag}$ &  32.34 &   0.90 &   5.85 &   0.12 & \multicolumn{2}{c}{\nodata}& & \Pa\  max$^{\dag}$ &  45.20 &   0.56 &   3.59 &   0.10 & \multicolumn{2}{c}{\nodata}\\
H$_{2}$ 1-0S(1)$^{\dag}$ &  32.91 &   0.90 &   6.06 &   0.12 & \multicolumn{2}{c}{\nodata}& & H$_{2}$ 1-0S(1) &  42.67 &   0.42 &  10.39 &   0.13 & \multicolumn{2}{c}{\nodata}\\
\noalign{\smallskip}
\cline{1-7} \cline{9-15}
\noalign{\vspace{0.5mm}}
\cline{1-7} \cline{9-15}
\noalign{\smallskip}

\multicolumn{7}{c}{\object{IRAS 14348-1447}} & & \multicolumn{7}{c}{\object{IRAS 17208-0014}}\\
\cline{1-7} \cline{9-15}
\noalign{\smallskip}
\multicolumn{1}{c}{\multirow{2}{*}{Region}} & \multicolumn{2}{c}{\Pa\ Flux} & \multicolumn{2}{c}{H$_{2}$ 1-0S(1) Flux} & \multicolumn{2}{c}{W$_{\rm CO}$} & &\multicolumn{1}{c}{\multirow{2}{*}{Region}} & \multicolumn{2}{c}{\Pa\ Flux} & \multicolumn{2}{c}{H$_{2}$ 1-0S(1) Flux} & \multicolumn{2}{c}{W$_{\rm CO}$}\\
& \multicolumn{2}{c}{(erg\,s$^{-1}$cm$^{-2}$)$\times$10$^{-16}$} & \multicolumn{2}{c}{(erg\,s$^{-1}$cm$^{-2}$)$\times$10$^{-16}$} & \multicolumn{2}{c}{(\AA)} & & & \multicolumn{2}{c}{(erg\,s$^{-1}$cm$^{-2}$)$\times$10$^{-16}$} & \multicolumn{2}{c}{(erg\,s$^{-1}$cm$^{-2}$)$\times$10$^{-16}$} & \multicolumn{2}{c}{(\AA)} \\
\noalign{\smallskip}
\cline{1-7} \cline{9-15}
\noalign{\vspace{0.5mm}}
\cline{1-7} \cline{9-15}
\noalign{\smallskip}
Nuclear&  45.51 &   0.36 &  10.21 &   0.19 & \multicolumn{2}{c}{\nodata}& & Nuclear & 437.13 &  14.26 &  60.50 &   1.31 &   11.8 &    1.1\\
Integrated & 144.90 &   2.94 &  39.39 &   0.81 & \multicolumn{2}{c}{\nodata}& & Integrated & 519.96 &  16.00 &  96.47 &   9.99 & \multicolumn{2}{c}{\nodata}\\
\Pa\  max$^{\dag}$ &  45.51 &   0.36 &  10.21 &   0.19 & \multicolumn{2}{c}{\nodata}& & \Pa\  max$^{\dag}$ & 437.13 &  14.26 &  60.50 &   1.31 &   11.8 &    1.1\\
H$_{2}$ 1-0S(1)$^{\dag}$&  45.33 &   0.38 &  10.25 &   0.20 & \multicolumn{2}{c}{\nodata}& & H$_{2}$ 1-0S(1)$^{\dag}$ & 435.91 &  14.68 &  60.70 &   1.35&   11.8 &    1.3\\
\noalign{\smallskip}
\cline{1-7} \cline{9-15}
\noalign{\vspace{0.5mm}}
\cline{1-7} \cline{9-15}
\noalign{\smallskip}

\multicolumn{7}{c}{\object{IRAS 21130-4446}} & & \multicolumn{7}{c}{\object{IRAS 22491-1808}}\\
\cline{1-7} \cline{9-15}
\noalign{\smallskip}
\multicolumn{1}{c}{\multirow{2}{*}{Region}} & \multicolumn{2}{c}{\Pa\ Flux} & \multicolumn{2}{c}{H$_{2}$ 1-0S(1) Flux} & \multicolumn{2}{c}{W$_{\rm CO}$} & &\multicolumn{1}{c}{\multirow{2}{*}{Region}} & \multicolumn{2}{c}{\Pa\ Flux} & \multicolumn{2}{c}{H$_{2}$ 1-0S(1) Flux} & \multicolumn{2}{c}{W$_{\rm CO}$}\\
& \multicolumn{2}{c}{(erg\,s$^{-1}$cm$^{-2}$)$\times$10$^{-16}$} & \multicolumn{2}{c}{(erg\,s$^{-1}$cm$^{-2}$)$\times$10$^{-16}$} & \multicolumn{2}{c}{(\AA)} & & & \multicolumn{2}{c}{(erg\,s$^{-1}$cm$^{-2}$)$\times$10$^{-16}$} & \multicolumn{2}{c}{(erg\,s$^{-1}$cm$^{-2}$)$\times$10$^{-16}$} & \multicolumn{2}{c}{(\AA)} \\
\noalign{\smallskip}
\cline{1-7} \cline{9-15}
\noalign{\vspace{0.5mm}}
\cline{1-7} \cline{9-15}
\noalign{\smallskip}
Nuclear&  26.98 &   0.98 &   3.10 &   0.11 & \multicolumn{2}{c}{\nodata}& & Nuclear &  11.68 &   0.18 &   0.75 &   0.05 & \multicolumn{2}{c}{\nodata}\\
Integrated& 107.65 &   2.62 &   6.00 &   0.25 & \multicolumn{2}{c}{\nodata}& & Integrated &  54.85 &   1.18 &  10.22 &   0.62 & \multicolumn{2}{c}{\nodata}\\
\Pa\  max &  20.76 &   0.19 &   0.57 &   0.04 & \multicolumn{2}{c}{\nodata}& & \Pa\  max &  27.17 &   0.43 &   8.64 &   0.16 & \multicolumn{2}{c}{\nodata}\\
H$_{2}$ 1-0S(1)$^{\dag}$ &  27.75 &   0.98 &   2.96 &   0.10 & \multicolumn{2}{c}{\nodata}& & H$_{2}$ 1-0S(1) &  27.45 &   0.42 &   8.65 &   0.17 & \multicolumn{2}{c}{\nodata}\\
\noalign{\smallskip}
\cline{1-7} \cline{9-15}
\noalign{\vspace{0.5mm}}
\cline{1-7} \cline{9-15}
\noalign{\smallskip}

\multicolumn{7}{c}{\object{IRAS 23128-5919}} & & \multicolumn{7}{c}{ }\\
\cline{1-7}
\noalign{\smallskip}
\multicolumn{1}{c}{\multirow{2}{*}{Region}} & \multicolumn{2}{c}{\Pa\ Flux} & \multicolumn{2}{c}{H$_{2}$ 1-0S(1) Flux} & \multicolumn{2}{c}{W$_{\rm CO}$} & \multicolumn{8}{c}{ }\\
& \multicolumn{2}{c}{(erg\,s$^{-1}$cm$^{-2}$)$\times$10$^{-16}$} & \multicolumn{2}{c}{(erg\,s$^{-1}$cm$^{-2}$)$\times$10$^{-16}$} & \multicolumn{2}{c}{(\AA)} & \multicolumn{8}{c}{ }\\
\noalign{\smallskip}
\cline{1-7}
\noalign{\vspace{0.5mm}}
\cline{1-7}
\noalign{\smallskip}
Nuclear& 336.35 &   6.04 &  11.76 &   0.35 &    8.1 &    0.7&  \multicolumn{8}{c}{ }\\
Integrated& 578.33 &   8.60 &  26.42 &   1.54 &    9.0 &    1.2& \multicolumn{8}{c}{ }\\
\Pa\  max$^{\dag}$ & 335.42 &   6.01 &  11.80 &   0.35 &    8.2 &    0.8& \multicolumn{8}{c}{ }\\
H$_{2}$ 1-0S(1)$^{\dag}$ & 335.42 &   6.01 &  11.80 &   0.35 &    8.2 &    0.8& \multicolumn{8}{c}{ }\\
\noalign{\smallskip}
\cline{1-7}
\noalign{\vspace{0.5mm}}
\cline{1-7}
\noalign{\smallskip}

\end{tabular}
\end{center}
\tablefoot{\tiny Spectra are integrated within a 2$\times$2\,kpc aperture, covering the nuclear region defined as the brightest spaxel in the K-band; the integrated emission of the FoV, defined as the integrated flux of those spaxels that contain at least the $\gsim90\%$ of the total continuum flux in each band; and the peaks of the \Pa\ and H$_2$ 1-0S(1) emission, centred on the brightest spaxel in each of the respective maps. The errors are obtained by a Monte Carlo method of $N=1000$ simulations of each spectra. .
$^{\dag}$ Regions that are coincident with the nucleus of the object. The spectra are extracted and the lines are fitted independently.}
\label{table:ulirgs_fluxes}
\end{table}
\end{landscape}

\renewcommand{\thetable}{\arabic{table}}


\setcounter{fake_table}{\value{table}}
\refstepcounter{fake_table}
\label{table:sigma}
\renewcommand{\thetable}{\arabic{table}\alph{subtab}}
\setcounter{subtab}{1}

\begin{landscape}
\begin{table}
\caption{\Brg, H$_2$ 1-0S(1), and [FeII] velocity dispersion values of the LIRGs subsample}
\begin{center}
\tiny
\begin{tabular}{c x{1cm}@{ $\pm$ }z{1cm} x{1cm}@{ $\pm$ }z{1cm} r@{ $\pm$ }l p{0.01mm} c x{1cm}@{ $\pm$ }z{1cm} x{1cm}@{ $\pm$ }z{1cm} r@{ $\pm$ }l}

\cline{1-7} \cline{9-15}
\noalign{\vspace{0.5mm}}
\cline{1-7} \cline{9-15}
\noalign{\smallskip}
\multicolumn{7}{c}{\object{NGC 2369}} & & \multicolumn{7}{c}{\object{NGC 3110}}\\
\cline{1-7} \cline{9-15}
\noalign{\smallskip}
\multicolumn{1}{c}{\multirow{2}{*}{Region}} & \multicolumn{2}{c}{\Brg\ $\sigma$} & \multicolumn{2}{c}{H$_{2}$ 1-0S(1) $\sigma$} & \multicolumn{2}{c}{[Fe II] $\sigma$} &  &\multicolumn{1}{c}{\multirow{2}{*}{Region}} & \multicolumn{2}{c}{\Brg\ $\sigma$} & \multicolumn{2}{c}{H$_{2}$ 1-0S(1) $\sigma$} & \multicolumn{2}{c}{[Fe II] $\sigma$}\\
& \multicolumn{2}{c}{(km\,s$^{-1}$)} & \multicolumn{2}{c}{(km\,s$^{-1}$)} &\multicolumn{2}{c}{(km\,s$^{-1}$)} &  & & \multicolumn{2}{c}{(km\,s$^{-1}$)} & \multicolumn{2}{c}{(km\,s$^{-1}$)} &\multicolumn{2}{c}{(km\,s$^{-1}$)} \\
\noalign{\smallskip}
\cline{1-7} \cline{9-15}
\noalign{\vspace{0.5mm}}
\cline{1-7} \cline{9-15}
\noalign{\smallskip}
Nuclear &    119 &      5 &    113 &      6 &    113 &      8& & Nuclear &     78 &      2 &     98 &      6 &     69 &     13\\
Integrated &    166 &     25 &    133 &     13 &    174 &     31 & & Integrated &    124 &      4 &    145 &      8 &     91 &   29\\
\Brg\ max$^{\dag}$ &    124 &      5 &    116 &      6 &    120 &      8& & \Brg\ max &     50 &      1 &     48 &      2 &     30 &     11\\
H$_{2}$ 1-0S(1)$^{\dag}$ &    121 &      5 &    113 &      6 &    116 &      8& & H$_{2}$ 1-0S(1)$^{\dag}$ &     78 &      2 &     98 &      6 &     69 &     13\\
$[$FeII] max &    136 &      7 &    111 &      6 &    125 &     13& & [FeII] max$^{\dag}$ &     72 &      2 &     89 &      5 &     62 &     14\\
\noalign{\smallskip}
\cline{1-7} \cline{9-15}
\noalign{\vspace{0.5mm}}
\cline{1-7} \cline{9-15}
\noalign{\smallskip}

\multicolumn{7}{c}{\object{NGC 3256}} & & \multicolumn{7}{c}{\object{ESO 320-G030}}\\
\cline{1-7} \cline{9-15}
\noalign{\smallskip}
\multicolumn{1}{c}{\multirow{2}{*}{Region}} & \multicolumn{2}{c}{\Brg\ $\sigma$} & \multicolumn{2}{c}{H$_{2}$ 1-0S(1) $\sigma$} & \multicolumn{2}{c}{[Fe II] $\sigma$} &  &\multicolumn{1}{c}{\multirow{2}{*}{Region}} & \multicolumn{2}{c}{\Brg\ $\sigma$} & \multicolumn{2}{c}{H$_{2}$ 1-0S(1) $\sigma$} & \multicolumn{2}{c}{[Fe II] $\sigma$}\\
& \multicolumn{2}{c}{(km\,s$^{-1}$)} & \multicolumn{2}{c}{(km\,s$^{-1}$)} &\multicolumn{2}{c}{(km\,s$^{-1}$)} &  & & \multicolumn{2}{c}{(km\,s$^{-1}$)} & \multicolumn{2}{c}{(km\,s$^{-1}$)} &\multicolumn{2}{c}{(km\,s$^{-1}$)} \\
\noalign{\smallskip}
\cline{1-7} \cline{9-15}
\noalign{\vspace{0.5mm}}
\cline{1-7} \cline{9-15}
\noalign{\smallskip}
Nuclear &    110 &      1 &     86 &      1 &     57 &     12& & Nuclear&     91 &     13 &    113 &      4 &     72 &   23\\
Integrated&     82 &      2 &     83 &      1 &     84 &      6& & Integrated &    137 &     10 &    135 &     10 &    119 &     18\\
\Brg\ max$^{\dag}$&    113 &      2 &     87 &      1 &    \multicolumn{2}{c}{\nodata}& & \Brg\ max &     48 &      2 &     51 &      5 &     52 &      9\\
H$_{2}$ 1-0S(1) &    103 &      7 &    130 &      4 & \multicolumn{2}{c}{\nodata}& & H$_{2}$ 1-0S(1)$^{\dag}$ &     87 &     15 &    112 &      4 &     71 &   27\\
$[$FeII] max & \multicolumn{2}{c}{\nodata} & \multicolumn{2}{c}{\nodata} &     56 &     13& & [FeII] max$^{\dag}$ &     89 &     11 &    109 &      4 &     71 &     26\\
\noalign{\smallskip}
\cline{1-7} \cline{9-15}
\noalign{\vspace{0.5mm}}
\cline{1-7} \cline{9-15}
\noalign{\smallskip}

\multicolumn{7}{c}{\object{IRASF 12115-4656}} & & \multicolumn{7}{c}{\object{NGC 5135}}\\
\cline{1-7} \cline{9-15}
\noalign{\smallskip}
\multicolumn{1}{c}{\multirow{2}{*}{Region}} & \multicolumn{2}{c}{\Brg\ $\sigma$} & \multicolumn{2}{c}{H$_{2}$ 1-0S(1) $\sigma$} & \multicolumn{2}{c}{[Fe II] $\sigma$} &  &\multicolumn{1}{c}{\multirow{2}{*}{Region}} & \multicolumn{2}{c}{\Brg\ $\sigma$} & \multicolumn{2}{c}{H$_{2}$ 1-0S(1) $\sigma$} & \multicolumn{2}{c}{[Fe II] $\sigma$}\\
& \multicolumn{2}{c}{(km\,s$^{-1}$)} & \multicolumn{2}{c}{(km\,s$^{-1}$)} &\multicolumn{2}{c}{(km\,s$^{-1}$)} &  & & \multicolumn{2}{c}{(km\,s$^{-1}$)} & \multicolumn{2}{c}{(km\,s$^{-1}$)} &\multicolumn{2}{c}{(km\,s$^{-1}$)} \\
\noalign{\smallskip}
\cline{1-7} \cline{9-15}
\noalign{\vspace{0.5mm}}
\cline{1-7} \cline{9-15}
\noalign{\smallskip}
Nuclear &     82 &     12 &     96 &     11 & \multicolumn{2}{c}{\nodata}& & Nuclear &     84 &      3 &     67 &      2 &     61 &      5\\
Integrated &    155 &     15 &    151 &     13 & \multicolumn{2}{c}{\nodata}& & Integrated &     75 &      6 &     78 &      7 &     58 &     11\\
\Brg\ max &     38 &      4 &     29 &      8 & \multicolumn{2}{c}{\nodata}& & \Brg\ max &     65 &      1 &     66 &      3 &     50 &      4\\
H$_{2}$ 1-0S(1)$^{\dag}$ &     82 &     12 &     97 &     11 & \multicolumn{2}{c}{\nodata}& & H$_{2}$ 1-0S(1)$^{\dag}$ &     84 &      3 &     67 &      2 &     61 &      5\\
$[$FeII] max &\multicolumn{2}{c}{\nodata}&  \multicolumn{2}{c}{\nodata}&  \multicolumn{2}{c}{\nodata}& & [FeII] max &     67 &      3 &     90 &      4 &    205 &      7\\
\noalign{\smallskip}
\cline{1-7} \cline{9-15}
\noalign{\vspace{0.5mm}}
\cline{1-7} \cline{9-15}
\noalign{\smallskip}


\multicolumn{7}{c}{\object{IRASF 17138-1017}} & & \multicolumn{7}{c}{\object{IC 4687}}\\
\cline{1-7} \cline{9-15}
\noalign{\smallskip}
\multicolumn{1}{c}{\multirow{2}{*}{Region}} & \multicolumn{2}{c}{\Brg\ $\sigma$} & \multicolumn{2}{c}{H$_{2}$ 1-0S(1) $\sigma$} & \multicolumn{2}{c}{[Fe II] $\sigma$} &  &\multicolumn{1}{c}{\multirow{2}{*}{Region}} & \multicolumn{2}{c}{\Brg\ $\sigma$} & \multicolumn{2}{c}{H$_{2}$ 1-0S(1) $\sigma$} & \multicolumn{2}{c}{[Fe II] $\sigma$}\\
& \multicolumn{2}{c}{(km\,s$^{-1}$)} & \multicolumn{2}{c}{(km\,s$^{-1}$)} &\multicolumn{2}{c}{(km\,s$^{-1}$)} &  & & \multicolumn{2}{c}{(km\,s$^{-1}$)} & \multicolumn{2}{c}{(km\,s$^{-1}$)} &\multicolumn{2}{c}{(km\,s$^{-1}$)} \\
\noalign{\smallskip}
\cline{1-7} \cline{9-15}
\noalign{\vspace{0.5mm}}
\cline{1-7} \cline{9-15}
\noalign{\smallskip}
Nuclear &     74 &      1 &     67 &      2 &     65 &      3& & Nuclear&     77 &      2 &     75 &      3 &     67 &      6\\
Integrated &    106 &      4 &    100 &      5 &     97 &      5& & Integrated &    117 &      3 &    112 &      5 &     93 &      9\\
\Brg\ max &     73 &      1 &     56 &      3 &     66 &      2& & \Brg\ max &     49 &      1 &     43 &      3 &     45 &      4\\
H$_{2}$ 1-0S(1)$^{\dag}$ &     74 &      1 &     67 &      2 &     65 &      3& & H$_{2}$ 1-0S(1)$^{\dag}$ &     80 &      2 &     70 &      3 &     69 &      6\\
$[$FeII] max &     62 &      1 &     52 &      1 &     59 &      2& & [FeII] max &     50 &      1 &     43 &      3 &     46 &      4\\
\noalign{\smallskip}
\cline{1-7} \cline{9-15}
\noalign{\vspace{0.5mm}}
\cline{1-7} \cline{9-15}
\noalign{\smallskip}


\multicolumn{7}{c}{\object{NGC 7130}} & & \multicolumn{7}{c}{\object{IC 5179}}\\
\cline{1-7} \cline{9-15}
\noalign{\smallskip}
\multicolumn{1}{c}{\multirow{2}{*}{Region}} & \multicolumn{2}{c}{\Brg\ $\sigma$} & \multicolumn{2}{c}{H$_{2}$ 1-0S(1) $\sigma$} & \multicolumn{2}{c}{[Fe II] $\sigma$} &  &\multicolumn{1}{c}{\multirow{2}{*}{Region}} & \multicolumn{2}{c}{\Brg\ $\sigma$} & \multicolumn{2}{c}{H$_{2}$ 1-0S(1) $\sigma$} & \multicolumn{2}{c}{[Fe II] $\sigma$}\\
& \multicolumn{2}{c}{(km\,s$^{-1}$)} & \multicolumn{2}{c}{(km\,s$^{-1}$)} &\multicolumn{2}{c}{(km\,s$^{-1}$)} &  & & \multicolumn{2}{c}{(km\,s$^{-1}$)} & \multicolumn{2}{c}{(km\,s$^{-1}$)} &\multicolumn{2}{c}{(km\,s$^{-1}$)} \\
\noalign{\smallskip}
\cline{1-7} \cline{9-15}
\noalign{\vspace{0.5mm}}
\cline{1-7} \cline{9-15}
\noalign{\smallskip}
Nuclear&     93 &      2 &     96 &      2 &    143 &      8& & Nuclear &     77 &      1 &     96 &      4 &     84 &      7\\
Integrated &     64 &      4 &     78 &      5 &     57 &      9& & Integrated &    130 &      9 &    117 &      9 &    123 &     13\\
\Brg\ max$^{\dag}$ &     93 &      2 &     96 &      2 &    143 &      8& & \Brg\ max$^{\dag}$ &     74 &      1 &     91 &      4 &     80 &      7\\
H$_{2}$ 1-0S(1)$^{\dag}$ &     93 &      2 &     96 &      2 &    143 &      8& & H$_{2}$ 1-0S(1)$^{\dag}$ &     74 &      1 &     93 &      4 &     82 &      7\\
$[$FeII] max$^{\dag}$ &     93 &      2 &     96 &      2 &    143 &      8& & [FeII] max$^{\dag}$ &     72 &      1 &     89 &      4 &     79 &      7\\
\noalign{\smallskip}
\cline{1-7} \cline{9-15}
\noalign{\vspace{0.5mm}}
\cline{1-7} \cline{9-15}
\noalign{\smallskip}

\end{tabular}
\end{center}
\tablefoot{The method of extracting the spectra and the selection criteria for the regions are the same as for Table \ref{table:lirgs_fluxes}.}
\label{table:lirgs_sigma}
\end{table}
\end{landscape}

\addtocounter{table}{-1}
\addtocounter{subtab}{1}

\begin{landscape}
\begin{table}
\caption{\Pa\ and H$_2$ 1-0S(1) velocity dispersion values of the ULIRGs subsample}
\begin{center}
\tiny
\begin{tabular}[t]{c x{1cm}@{ $\pm$ }z{1cm} r@{ $\pm$ }l p{0.01mm} c x{1cm}@{ $\pm$ }z{1cm} r@{ $\pm$ }l}

\cline{1-5} \cline{7-11}
\noalign{\vspace{0.5mm}}
\cline{1-5} \cline{7-11}
\noalign{\smallskip}
\multicolumn{5}{c}{\object{IRAS 06206-6315}} & & \multicolumn{5}{c}{\object{IRAS 12112+0305}}\\
\cline{1-5} \cline{7-11}
\noalign{\smallskip}
\multicolumn{1}{c}{\multirow{2}{*}{Region}} & \multicolumn{2}{c}{\Pa\ $\sigma$} & \multicolumn{2}{c}{H$_{2}$ 1-0S(1) $\sigma$} & &\multicolumn{1}{c}{\multirow{2}{*}{Region}} & \multicolumn{2}{c}{\Pa\ $\sigma$} & \multicolumn{2}{c}{H$_{2}$ 1-0S(1) $\sigma$} \\
& \multicolumn{2}{c}{(km\,s$^{-1}$)} & \multicolumn{2}{c}{(km\,s$^{-1}$)} & & & \multicolumn{2}{c}{(km\,s$^{-1}$)} & \multicolumn{2}{c}{(km\,s$^{-1}$)} \\
\noalign{\smallskip}
\cline{1-5} \cline{7-11}
\noalign{\vspace{0.5mm}}
\cline{1-5} \cline{7-11}
\noalign{\smallskip}
Nuclear&    177 &      4 &    139 &      2& & Nuclear &    134 &      1 &    118 &      3 \\
Integrated&    148 &     93 &    117 &      7& & Integrated &    153 &      5 &    160 &      8\\
\Pa\  max$^{\dag}$&    177 &      4 &    139 &      3& & \Pa\  max$^{\dag}$ &    134 &      1 &    119 &      3\\
H$_{2}$ 1-0S(1)$^{\dag}$&    177 &      4 &    139 &      2& & H$_{2}$ 1-0S(1) &    121 &      1 &    150 &      2\\
\noalign{\smallskip}
\cline{1-5} \cline{7-11}
\noalign{\vspace{0.5mm}}
\cline{1-5} \cline{7-11}
\noalign{\smallskip}

\multicolumn{5}{c}{\object{IRAS 14348-1447}} & & \multicolumn{5}{c}{\object{IRAS 17208-0014}}\\
\cline{1-5} \cline{7-11}
\noalign{\smallskip}
\multicolumn{1}{c}{\multirow{2}{*}{Region}} & \multicolumn{2}{c}{\Pa\ $\sigma$} & \multicolumn{2}{c}{H$_{2}$ 1-0S(1) $\sigma$} & &\multicolumn{1}{c}{\multirow{2}{*}{Region}} & \multicolumn{2}{c}{\Pa\ $\sigma$} & \multicolumn{2}{c}{H$_{2}$ 1-0S(1) $\sigma$} \\
& \multicolumn{2}{c}{(km\,s$^{-1}$)} & \multicolumn{2}{c}{(km\,s$^{-1}$)} & & & \multicolumn{2}{c}{(km\,s$^{-1}$)} & \multicolumn{2}{c}{(km\,s$^{-1}$)} \\
\noalign{\smallskip}
\cline{1-5} \cline{7-11}
\noalign{\vspace{0.5mm}}
\cline{1-5} \cline{7-11}
\noalign{\smallskip}
Nuclear&    109 &      1 &    111 &      2& & Nuclear &    219 &      6 &    187 &      3\\
Integrated&    124 &      3 &    127 &      3& & Integrated &    201 &      5 &    200 &     18\\
\Pa\  max$^{\dag}$&    109 &      1 &    111 &      2& & \Pa\  max$^{\dag}$ &    219 &      6 &    187 &      3\\
H$_{2}$ 1-0S(1)$^{\dag}$ &    108 &      1 &    111 &      2& & H$_{2}$ 1-0S(1)$^{\dag}$ &    222 &      6 &    189 &      3\\
\noalign{\smallskip}
\cline{1-5} \cline{7-11}
\noalign{\vspace{0.5mm}}
\cline{1-5} \cline{7-11}
\noalign{\smallskip}

\multicolumn{5}{c}{\object{IRAS 21130-4446}} & & \multicolumn{5}{c}{\object{IRAS 22491-1808}}\\
\cline{1-5} \cline{7-11}
\noalign{\smallskip}
\multicolumn{1}{c}{\multirow{2}{*}{Region}} & \multicolumn{2}{c}{\Pa\ $\sigma$} & \multicolumn{2}{c}{H$_{2}$ 1-0S(1) $\sigma$} & &\multicolumn{1}{c}{\multirow{2}{*}{Region}} & \multicolumn{2}{c}{\Pa\ $\sigma$} & \multicolumn{2}{c}{H$_{2}$ 1-0S(1) $\sigma$} \\
& \multicolumn{2}{c}{(km\,s$^{-1}$)} & \multicolumn{2}{c}{(km\,s$^{-1}$)} & & & \multicolumn{2}{c}{(km\,s$^{-1}$)} & \multicolumn{2}{c}{(km\,s$^{-1}$)} \\
\noalign{\smallskip}
\cline{1-5} \cline{7-11}
\noalign{\vspace{0.5mm}}
\cline{1-5} \cline{7-11}
\noalign{\smallskip}
Nuclear&    132 &      6 &    131 &      4& & Nuclear &     66 &      1 &     83 &      6 \\
Integrated&    119 &      3 &    113 &      5& & Integrated &     76 &      2 &    102 &      8\\
\Pa\  max&     75 &      0 &     74 &      6& & \Pa\  max &    105 &      2 &    125 &      2\\
H$_{2}$ 1-0S(1)$^{\dag}$ &    137 &      6 &    127 &      4& & H$_{2}$ 1-0S(1) &    105 &      1 &    125 &      2\\
\noalign{\smallskip}
\cline{1-5} \cline{7-11}
\noalign{\vspace{0.5mm}}
\cline{1-5} \cline{7-11}
\noalign{\smallskip}

\multicolumn{5}{c}{\object{IRAS 23128-5919}} & \multicolumn{6}{c}{ }\\
\cline{1-5}
\noalign{\smallskip}
\multicolumn{1}{c}{\multirow{2}{*}{Region}} & \multicolumn{2}{c}{\Pa\ $\sigma$} & \multicolumn{2}{c}{H$_{2}$ 1-0S(1) $\sigma$} & \multicolumn{6}{c}{ }\\
& \multicolumn{2}{c}{(km\,s$^{-1}$)} & \multicolumn{2}{c}{(km\,s$^{-1}$)} & \multicolumn{6}{c}{ }\\
\noalign{\smallskip}
\cline{1-5}
\noalign{\vspace{0.5mm}}
\cline{1-5}
\noalign{\smallskip}
Nuclear&    116 &      2 &     81 &      3&  \multicolumn{6}{c}{ }\\
Integrated&    104 &      2 &     87 &      6& \multicolumn{6}{c}{ }\\
\Pa\  max$^{\dag}$&    116 &      2 &     82 &      3& \multicolumn{6}{c}{ }\\
H$_{2}$ 1-0S(1)$^{\dag}$ &    116 &      2 &     82 &      3& \multicolumn{6}{c}{ }\\
\noalign{\smallskip}
\cline{1-5}
\noalign{\vspace{0.5mm}}
\cline{1-5}
\noalign{\smallskip}

\end{tabular}
\end{center}
\tablefoot{The method for extracting the spectra and the selection criteria for the regions are the same as for Table \ref{table:ulirgs_fluxes}.}
\label{table:ulirgs_sigma}
\end{table}
\end{landscape}

\renewcommand{\thetable}{\arabic{table}}

\clearpage
\setcounter{figure}{0}
\setcounter{fake_fig}{\value{figure}}
\refstepcounter{fake_fig}
\label{figure:LIRG}
\renewcommand{\thefigure}{\arabic{figure}\alph{subfig}}
\setcounter{subfig}{1}

\begin{figure*}[!h]
\begin{center}
\includegraphics[angle=0, width=1\textwidth]{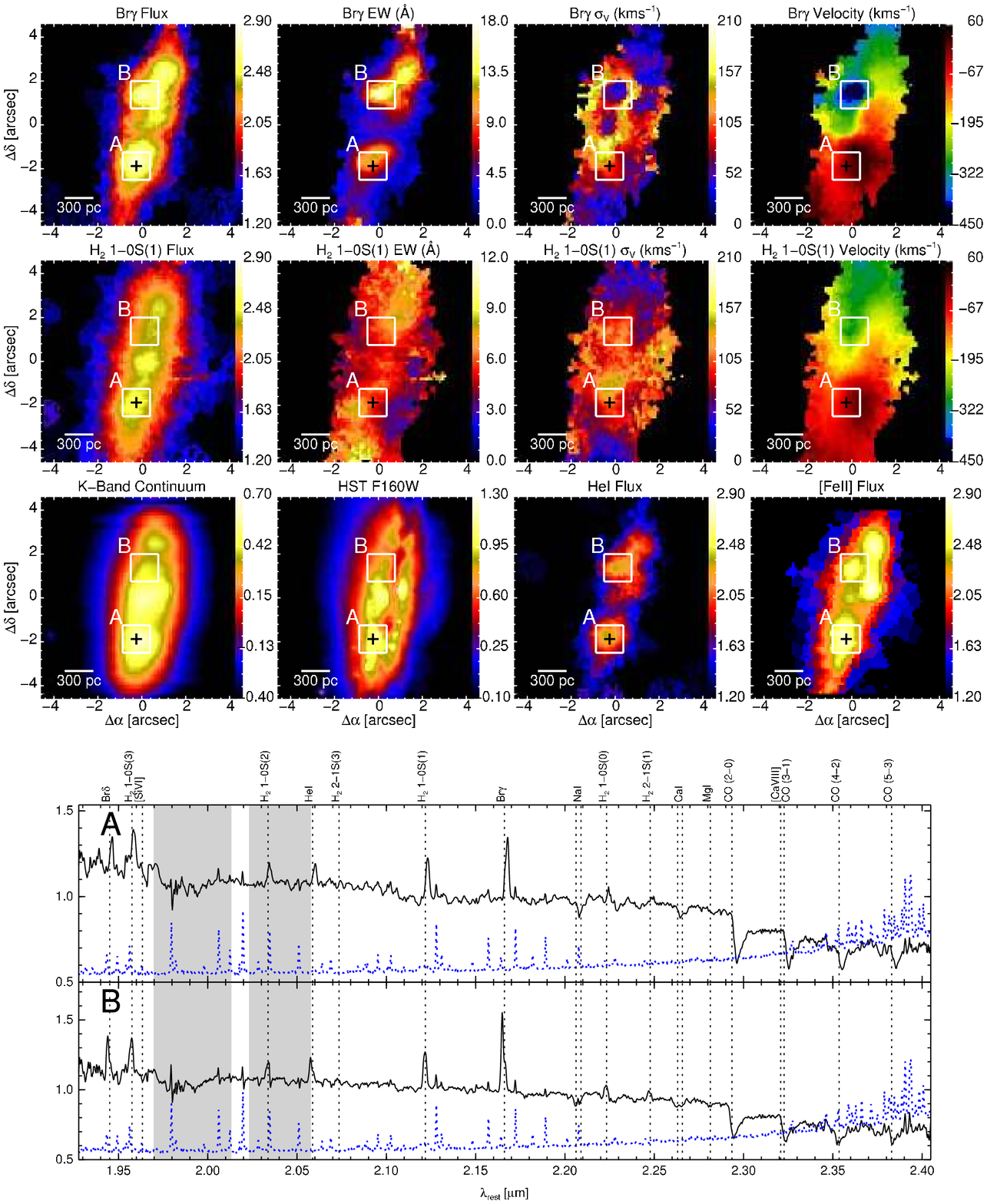}
\end{center}
\caption{\tiny{\object{NGC 2369}. Top and middle panels are SINFONI observed maps (not corrected from extinction) of the lines \Brgl, and \Hcerounol. From left to right: flux, equivalent width, velocity dispersion and velocity. Lower panel shows, from left to right, the K band emission from our SINFONI data, HST/NICMOS F160W continuum image from the archive, \HeI, and \FeII\ emission maps. The brightest spaxel of the SINFONI K band is marked with a cross. The apertures used to extract the spectra at the bottom of the figure are drawn as white squares and labelled accordingly. At the bottom, the two rest-frame spectra extracted from apertures ``A" and ``B" are in black. The most relevant spectral features are labelled at the top and marked with a dotted line. The sky spectrum is overplotted as a dashed blue line, and the wavelength ranges of the water vapour atmospheric absorptions are marked in light grey.}}
\label{figure:NGC2369}
\end{figure*}

\addtocounter{figure}{-1}
\addtocounter{subfig}{1}
\begin{figure*}
\begin{center}
\includegraphics[angle=0, width=1\textwidth]{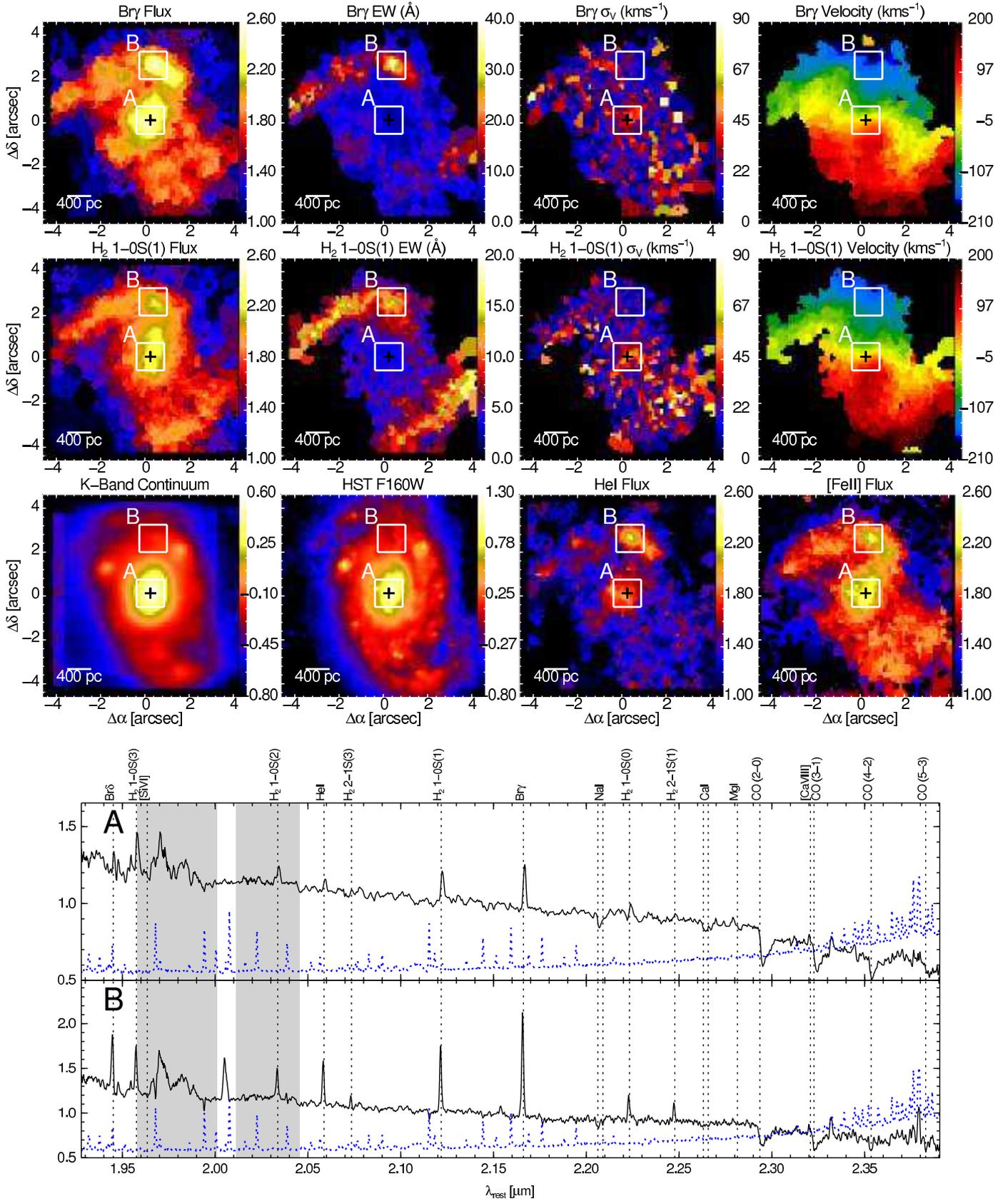}
\end{center}
\caption{As Fig. \ref{figure:NGC2369} but for \object{NGC 3110}.}
\label{figure:NGC3110}
\end{figure*}

\addtocounter{figure}{-1}
\addtocounter{subfig}{1}
\begin{figure*}
\begin{center}
\includegraphics[angle=0, width=1\textwidth]{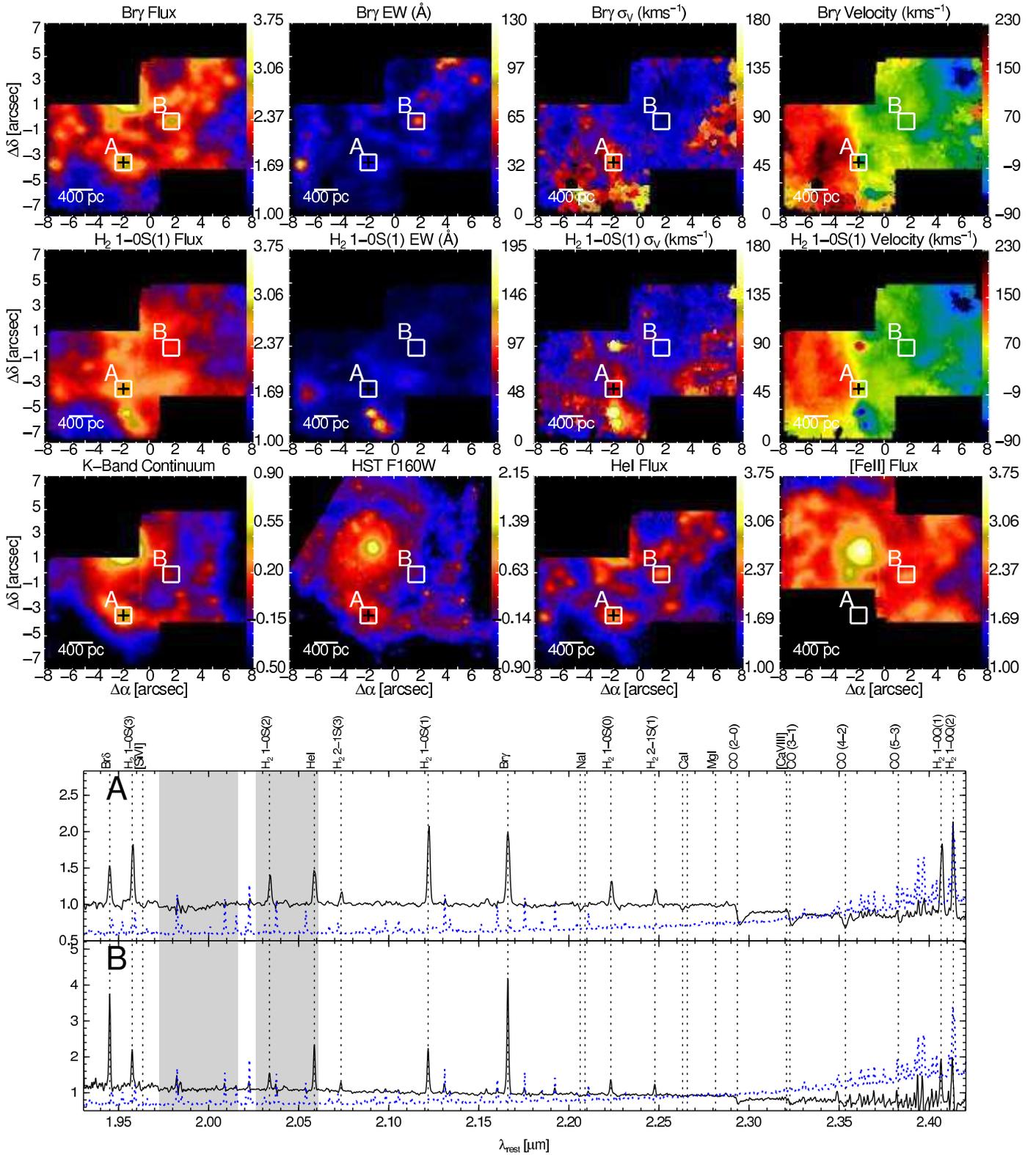}
\end{center}
\caption{As Fig. \ref{figure:NGC2369} but for \object{NGC 3256}.}
\label{figure:NGC3256}
\end{figure*}

\addtocounter{figure}{-1}
\addtocounter{subfig}{1}
\begin{figure*}
\begin{center}
\includegraphics[angle=0, width=1\textwidth]{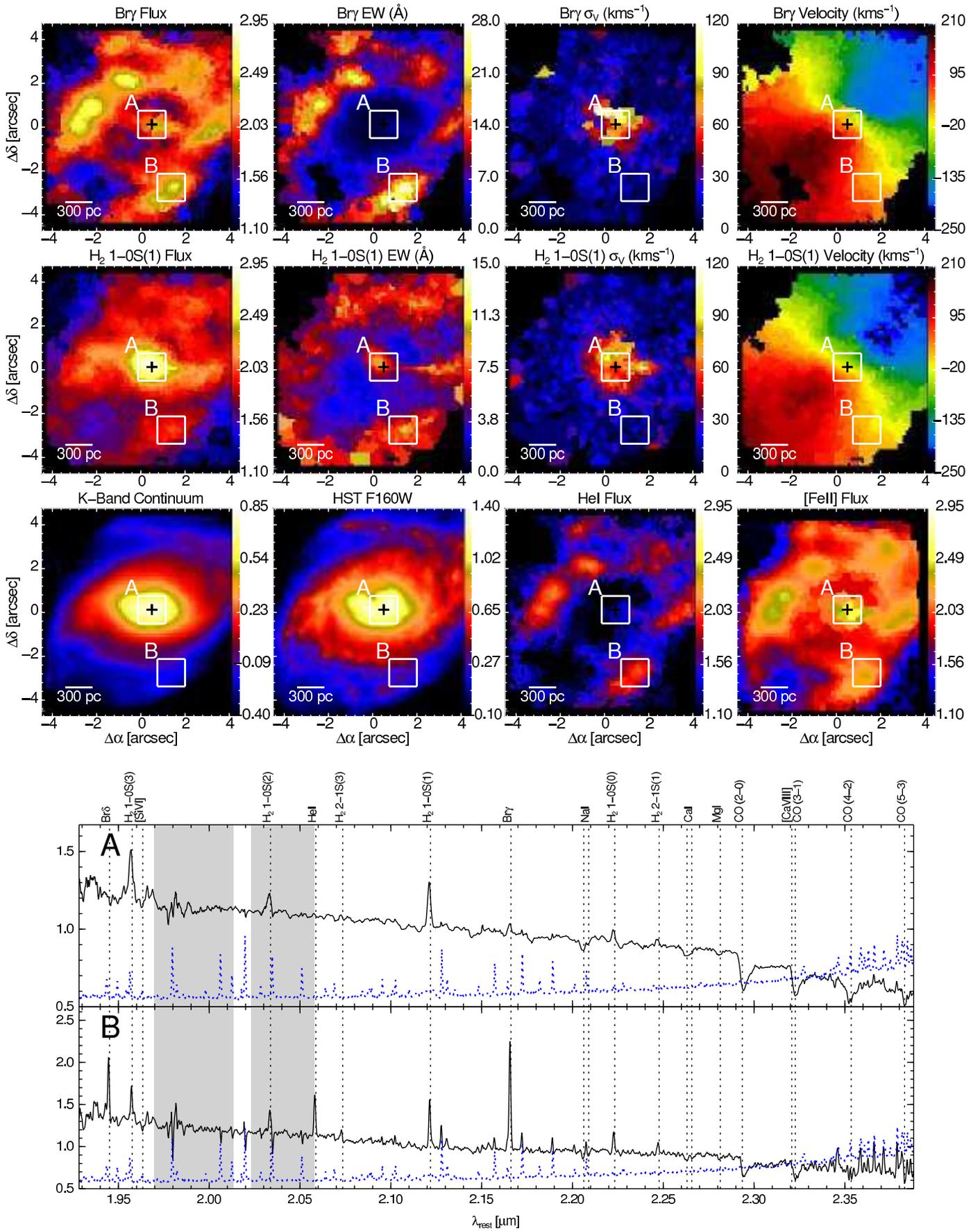}
\end{center}
\caption{As Fig. \ref{figure:NGC2369} but for \object{ESO 320-G030}.}
\label{figure:ESO320}
\end{figure*}

\addtocounter{figure}{-1}
\addtocounter{subfig}{1}
\begin{figure*}
\begin{center}
\includegraphics[angle=0, width=1\textwidth]{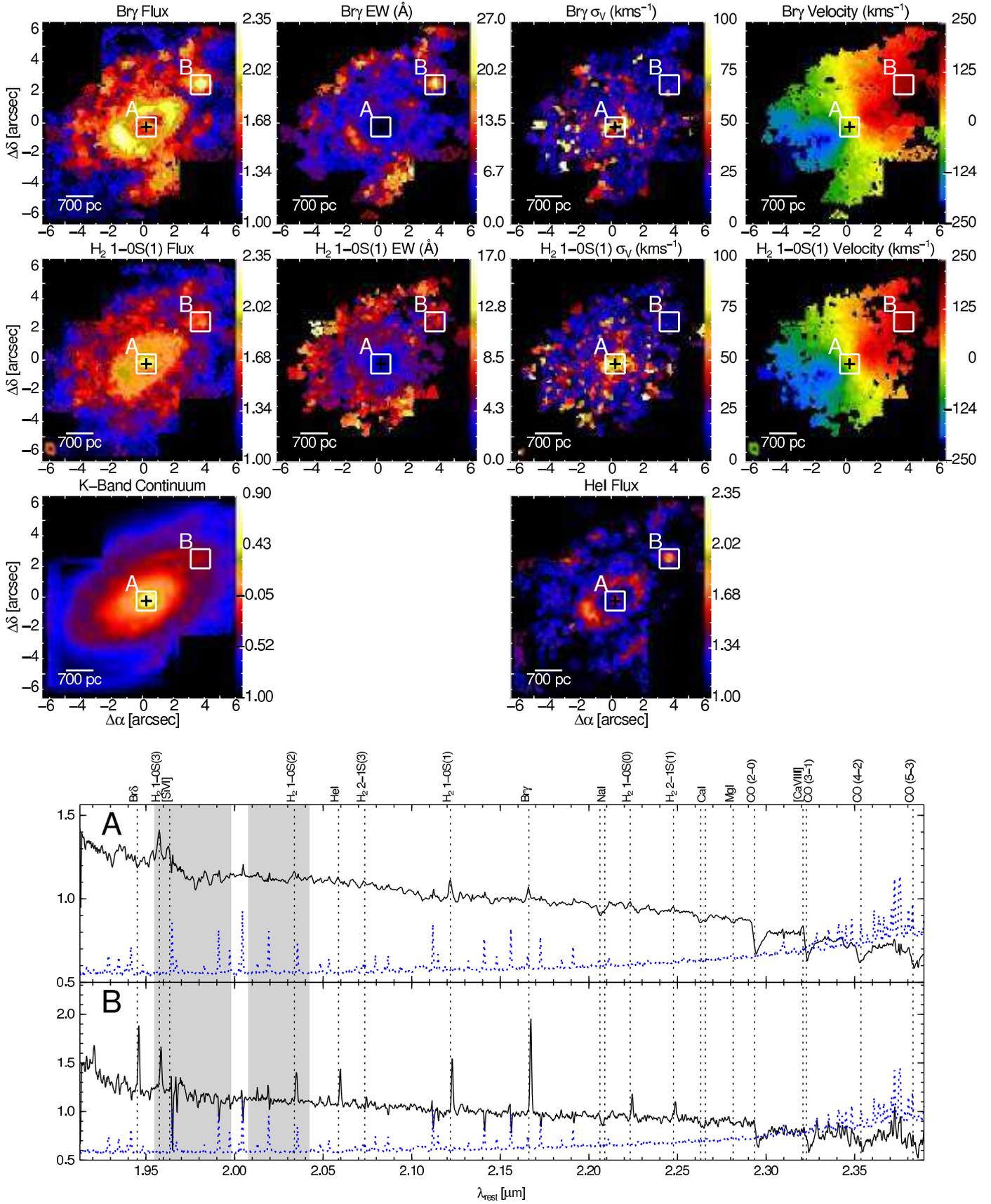}
\end{center}
\caption{As Fig. \ref{figure:NGC2369} but for \object{IRASF 12115-4656}.}
\label{figure:IRASF12115}
\end{figure*}

\addtocounter{figure}{-1}
\addtocounter{subfig}{1}
\begin{figure*}
\begin{center}
\includegraphics[angle=0, width=1\textwidth]{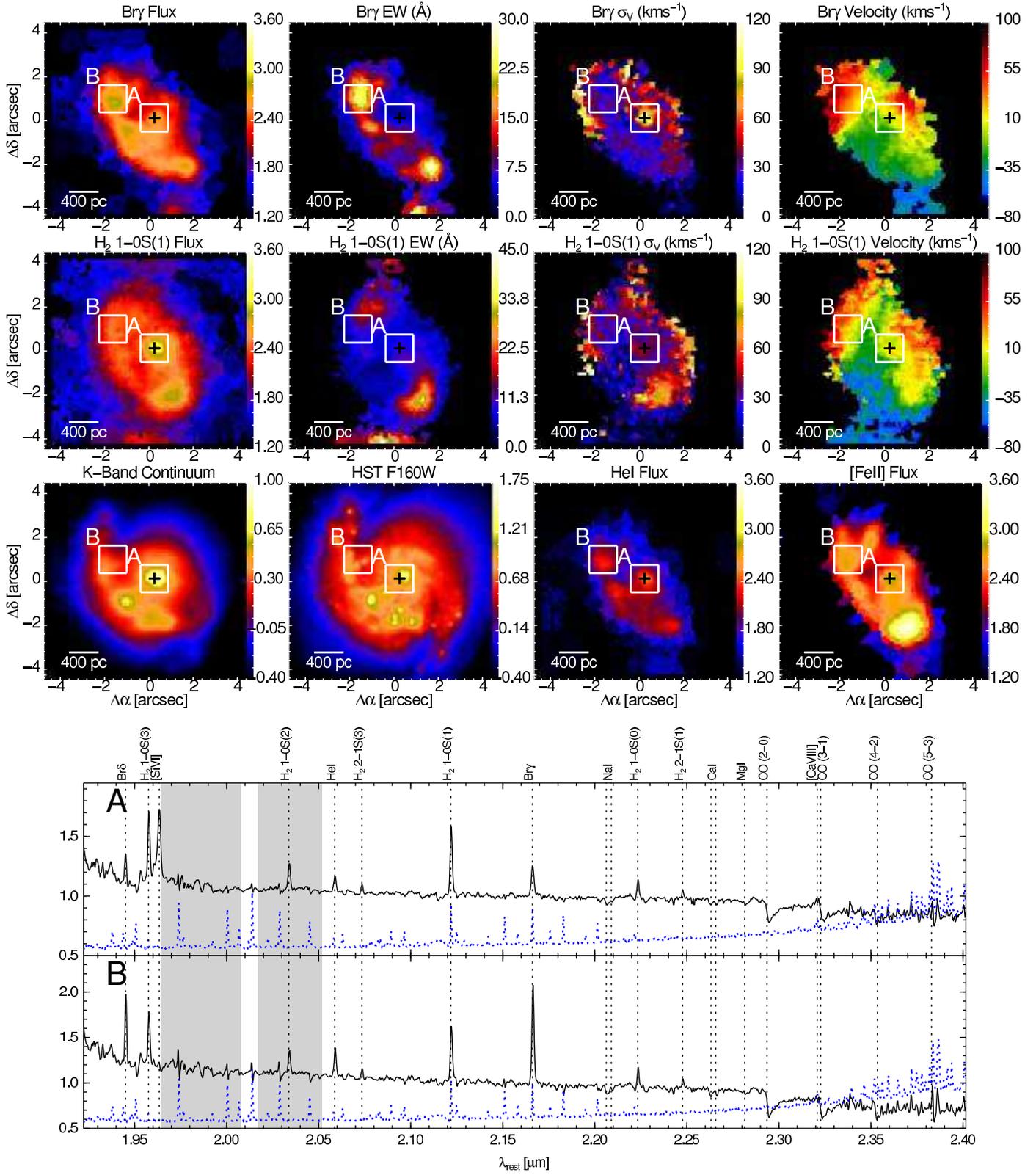}
\end{center}
\caption{As Fig. \ref{figure:NGC2369} but for \object{NGC 5135}.}
\label{figure:NGC5135}
\end{figure*}

\addtocounter{figure}{-1}
\addtocounter{subfig}{1}
\begin{figure*}
\begin{center}
\includegraphics[angle=0, width=1\textwidth]{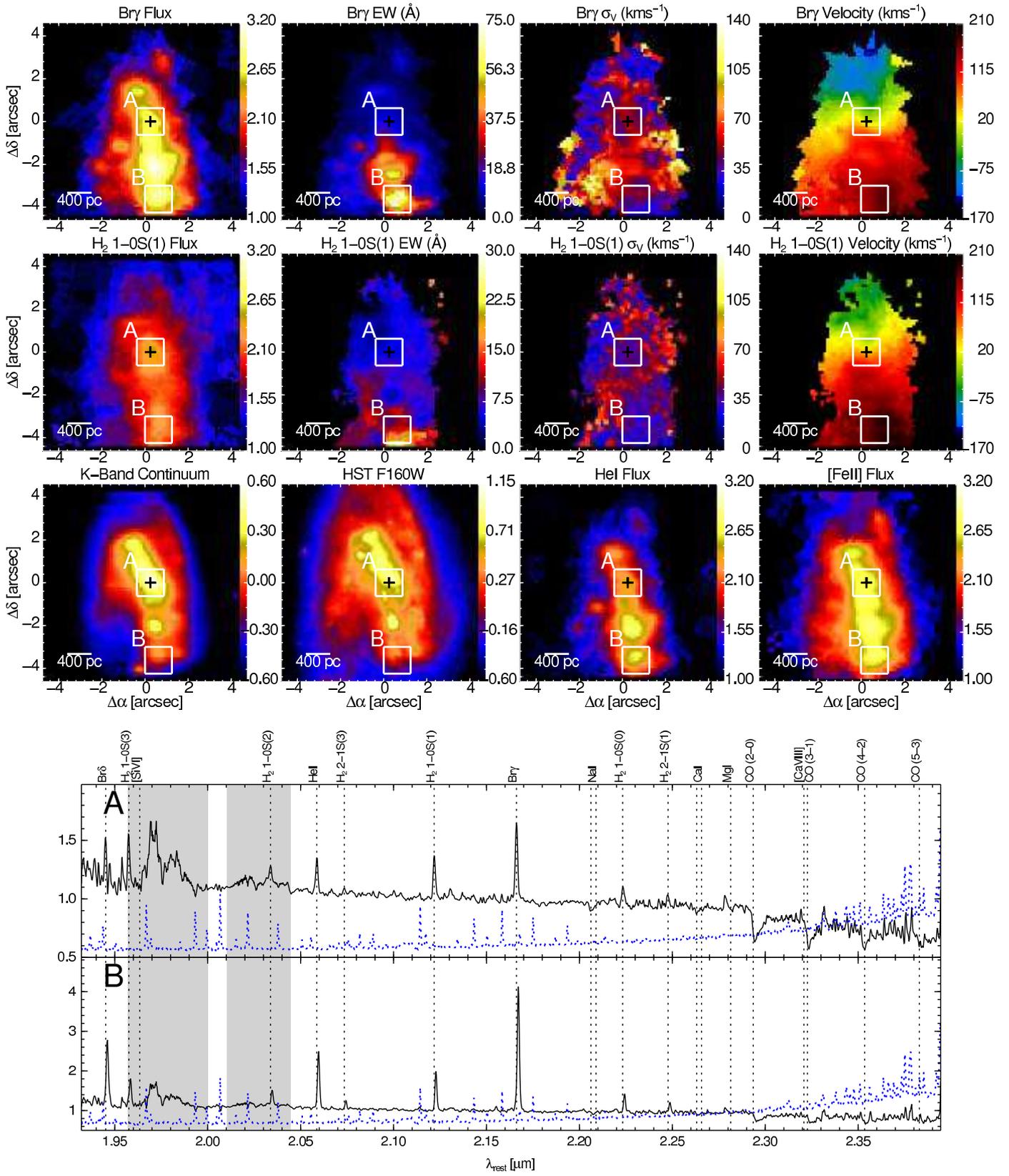}
\end{center}
\caption{As Fig. \ref{figure:NGC2369} but for \object{IRASF 17138-1017}.}
\label{figure:IRASF17138}
\end{figure*}

\addtocounter{figure}{-1}
\addtocounter{subfig}{1}
\begin{figure*}
\begin{center}
\includegraphics[angle=0, width=1\textwidth]{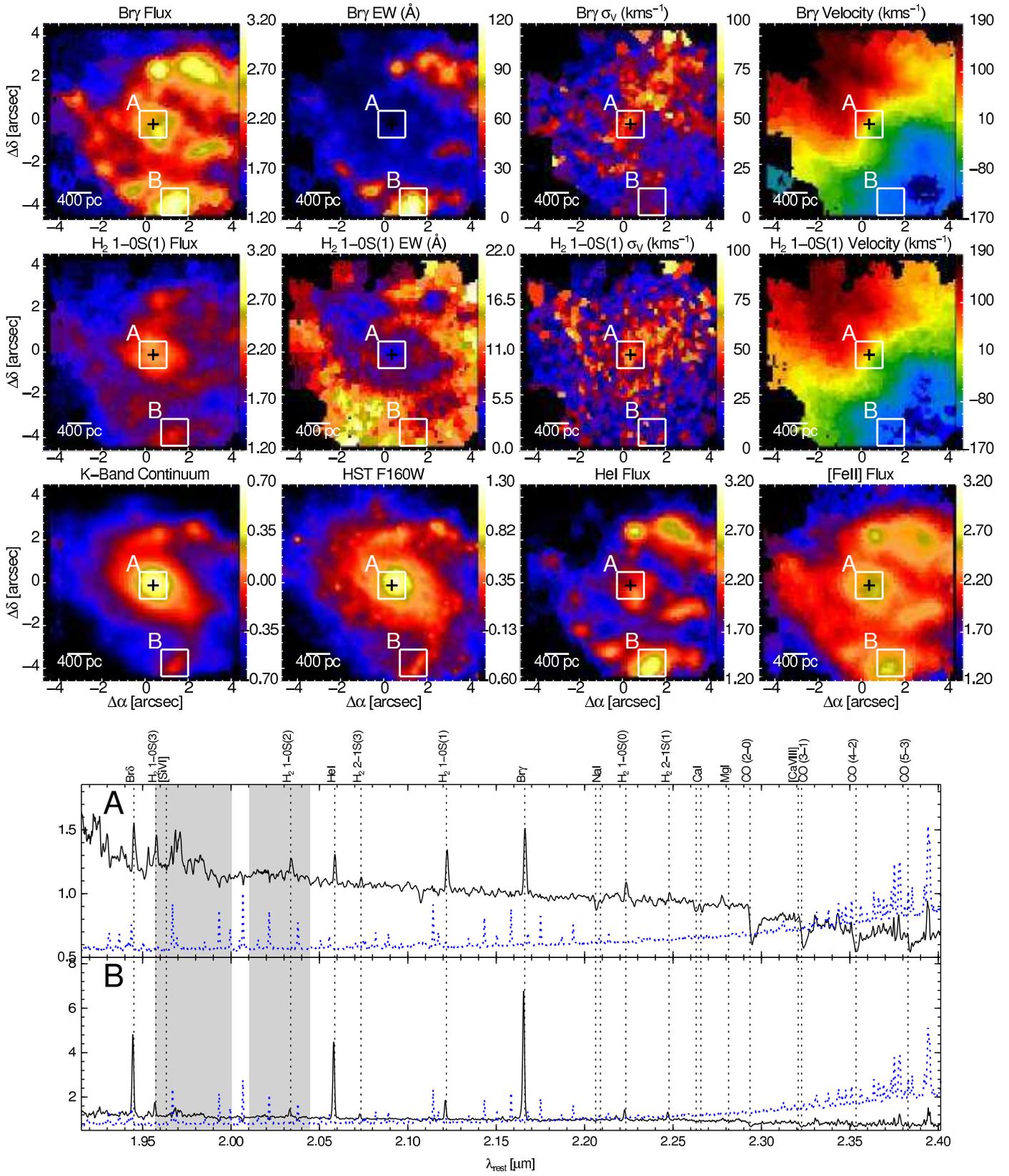}
\end{center}
\caption{As Fig. \ref{figure:NGC2369} but for \object{IC 4687}.}
\label{figure:IC4687}
\end{figure*}

\addtocounter{figure}{-1}
\addtocounter{subfig}{1}
\begin{figure*}
\begin{center}
\includegraphics[angle=0, width=0.75\textwidth]{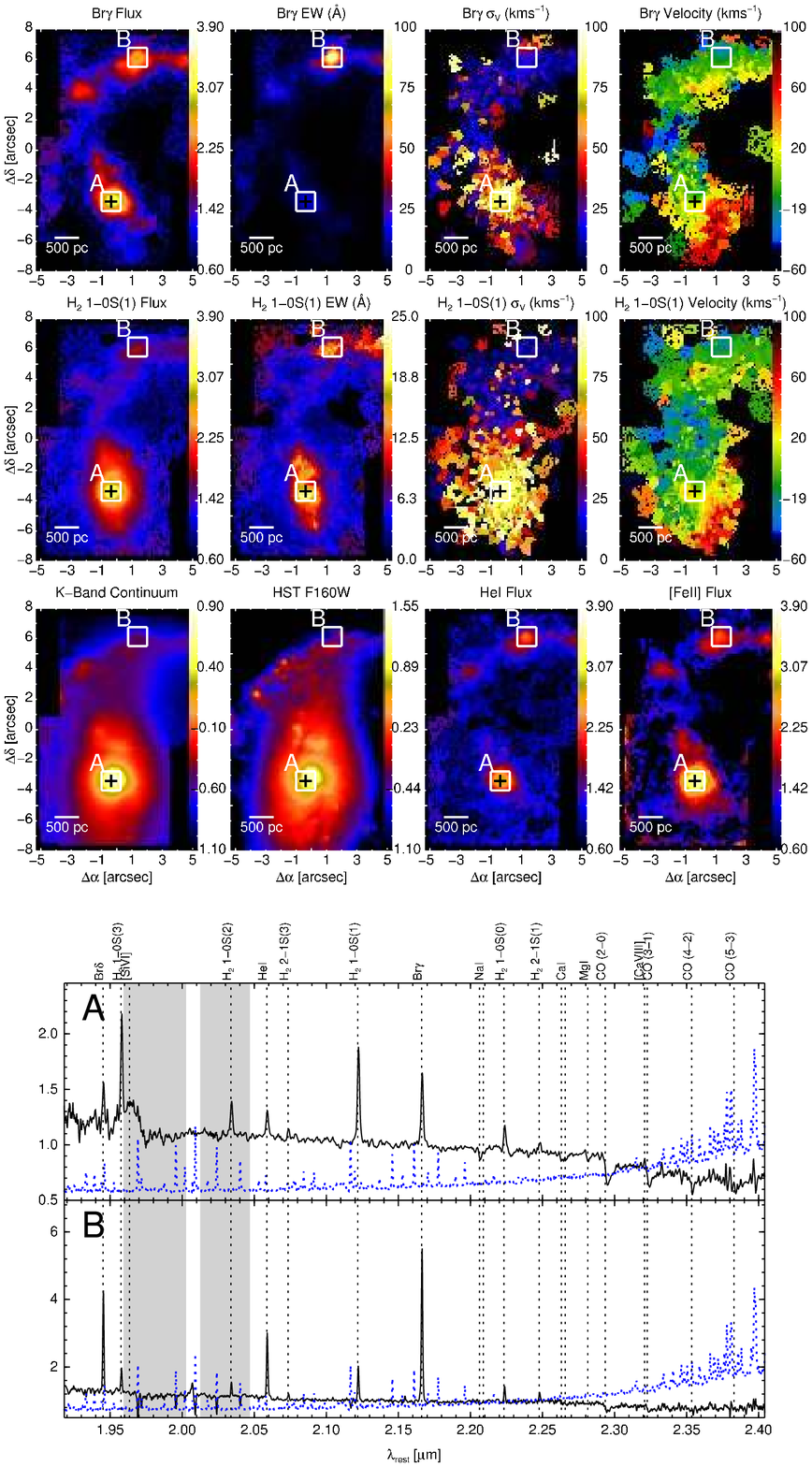}
\end{center}
\caption{As Fig. \ref{figure:NGC2369} but for \object{NGC 7130}.}
\label{figure:NGC7130}
\end{figure*}

\addtocounter{figure}{-1}
\addtocounter{subfig}{1}
\begin{figure*}
\begin{center}
\includegraphics[angle=0, width=1\textwidth]{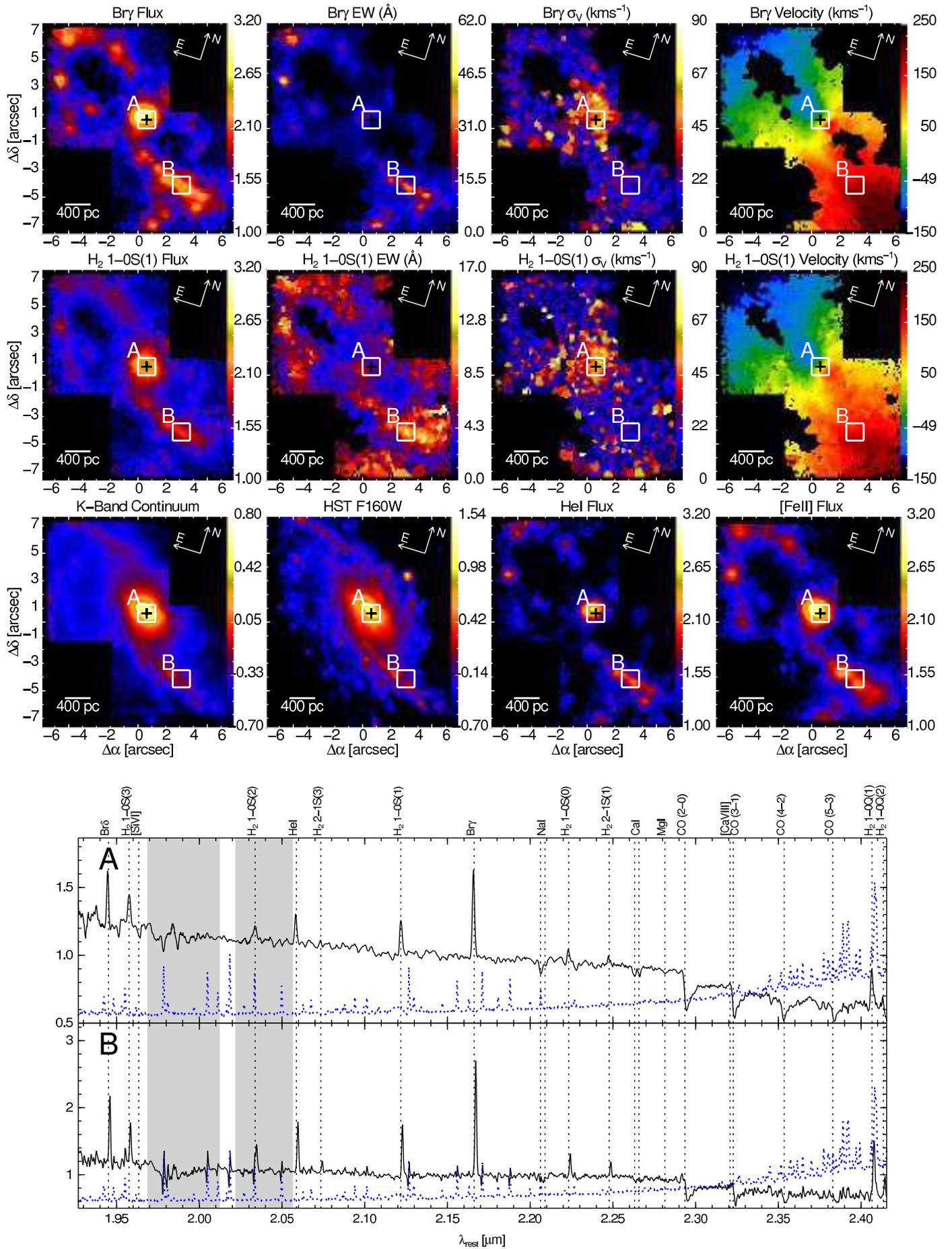}
\end{center}
\caption{As Fig. \ref{figure:NGC2369} but for \object{IC 5179}.}
\label{figure:IC5179}
\end{figure*}


\refstepcounter{fake_fig}
\label{figure:ULIRG}

\addtocounter{figure}{0}
\setcounter{subfig}{1}
\begin{figure*}
\begin{center}
\includegraphics[angle=0, width=1\textwidth]{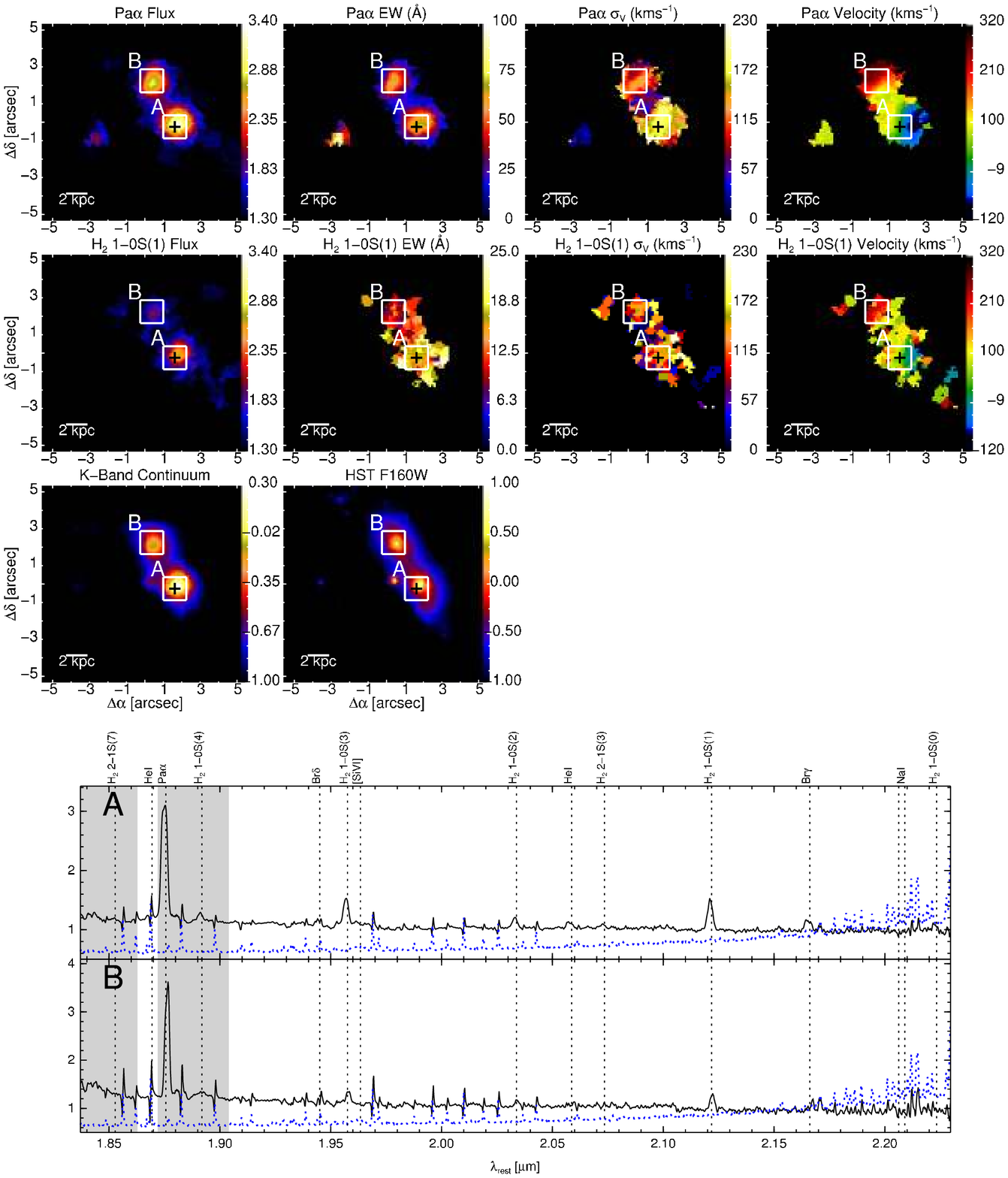}
\end{center}
\caption{\tiny{\object{IRAS 06206-6315}. Top and middle panels are SINFONI observed maps (not corrected from extinction) of the lines \Pal, and \Hcerounol. From left to right: flux, equivalent width, velocity dispersion, and velocity. Lower panel shows, from left to right, the K band emission from our SINFONI data, HST continuum image from the archive, and \HeI\ (when available). The brightest spaxel of the SINFONI K band is marked with a cross. The apertures used to extract the spectra at the bottom of the figure are drawn as white squares and labelled accordingly. At the bottom, the two rest-frame spectra extracted from apertures ``A" and ``B" in black. The most relevant spectral features are labelled at the top and marked with a dotted line. The sky spectrum is overplotted as a dashed blue line, and the wavelength ranges of the water vapour atmospheric absorptions are marked in light grey.}}
\label{figure:IRAS06206}
\end{figure*}

\addtocounter{figure}{-1}
\addtocounter{subfig}{1}
\begin{figure*}
\begin{center}
\includegraphics[angle=0, width=1\textwidth]{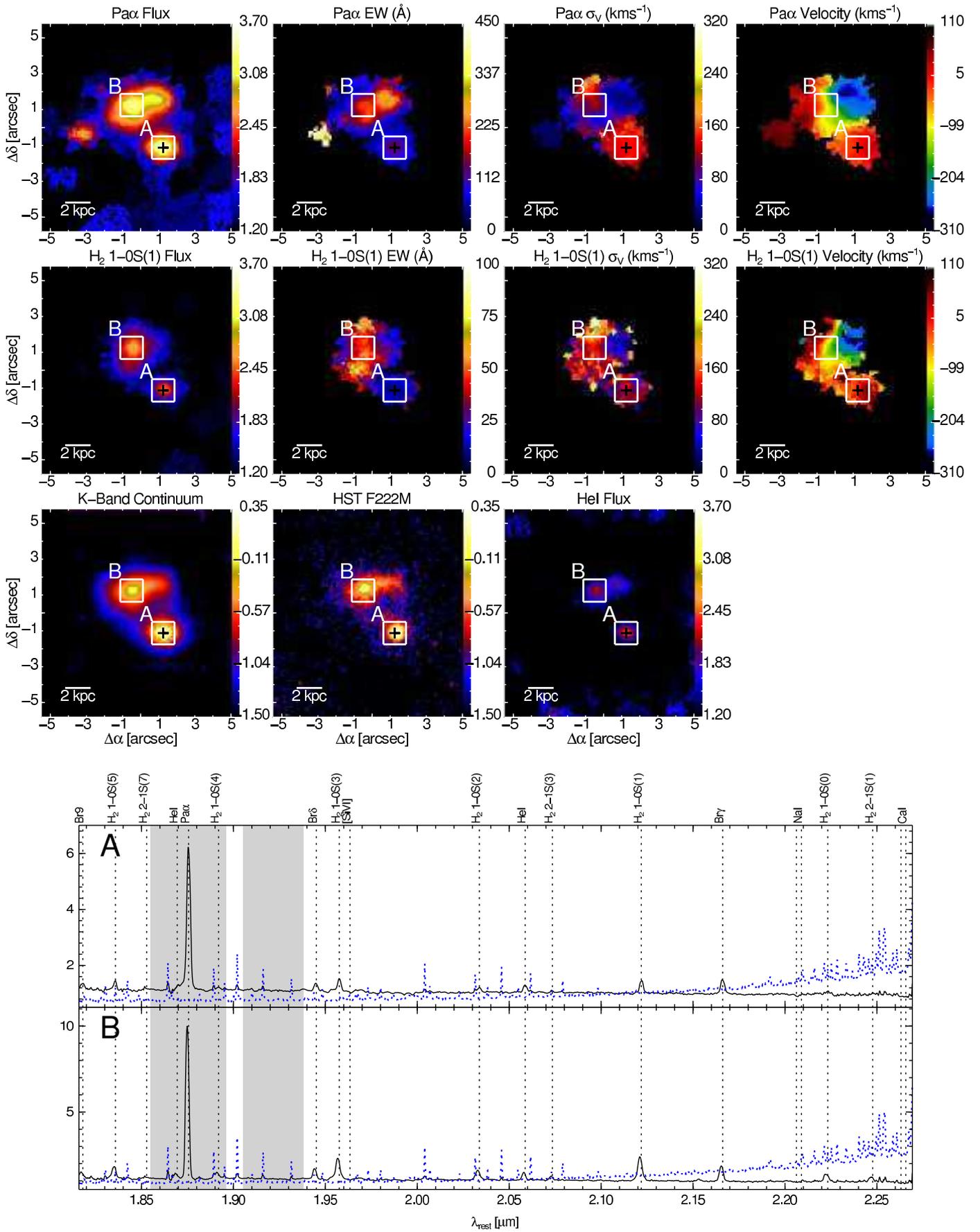}
\end{center}
\caption{As Fig. \ref{figure:IRAS06206} but for \object{IRAS 12112+0305}.}
\label{figure:IRAS12112}
\end{figure*}

\addtocounter{figure}{-1}
\addtocounter{subfig}{1}
\begin{figure*}
\begin{center}
\includegraphics[angle=0, width=1\textwidth]{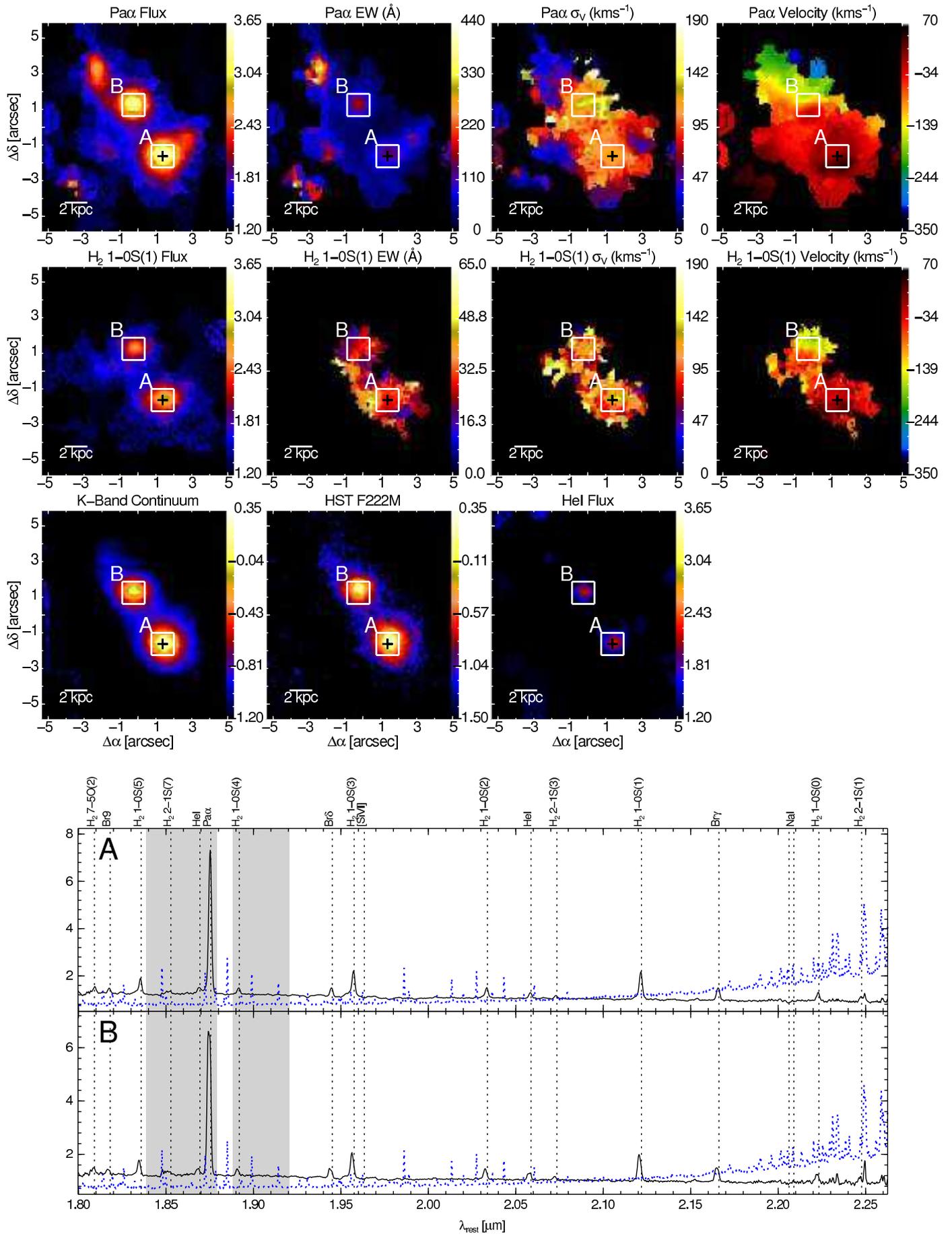}
\end{center}
\caption{As Fig. \ref{figure:IRAS06206} but for \object{IRAS 14348-1447}.}
\label{figure:IRAS14348}
\end{figure*}

\addtocounter{figure}{-1}
\addtocounter{subfig}{1}
\begin{figure*}
\begin{center}
\includegraphics[angle=0, width=1\textwidth]{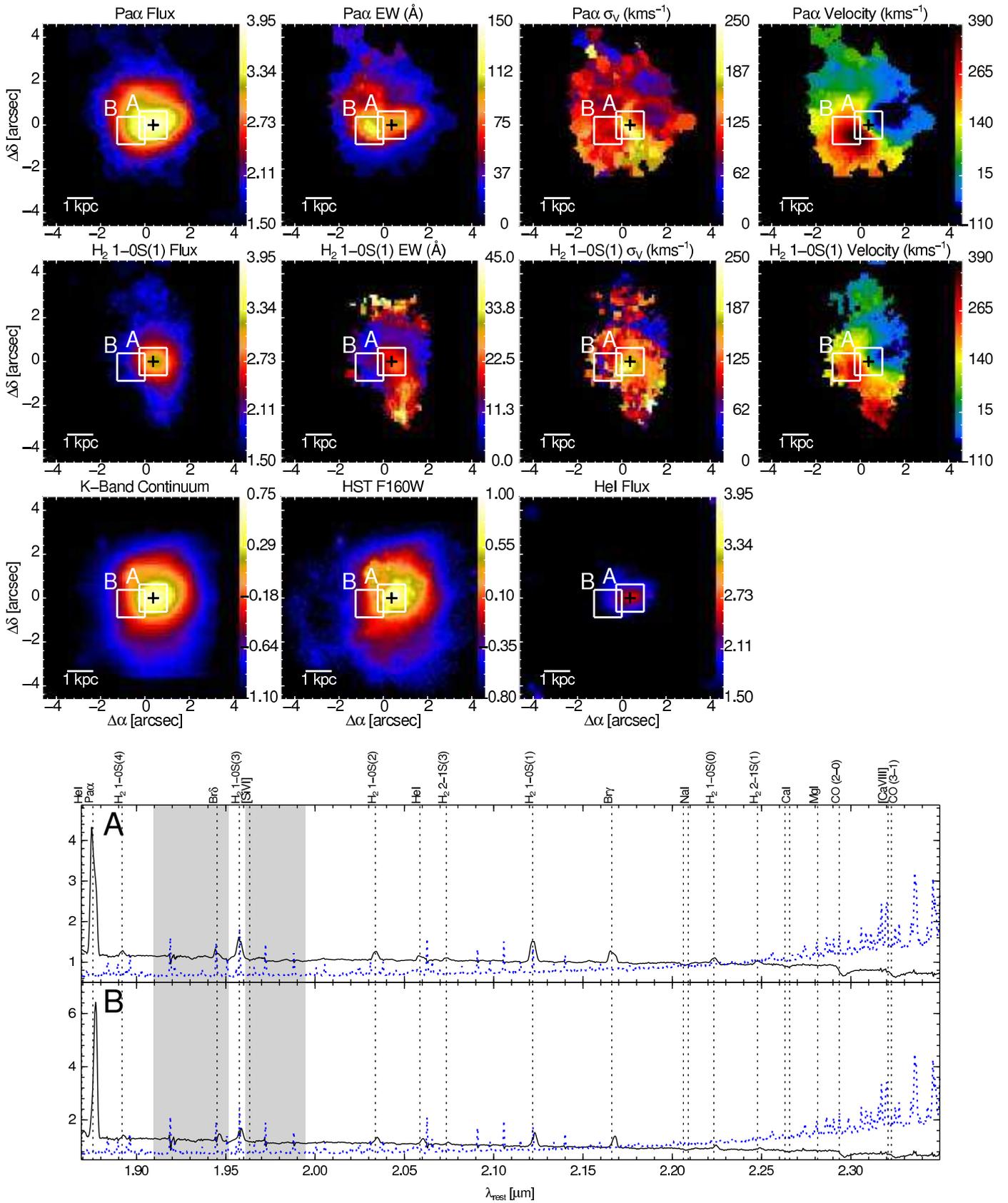}
\end{center}
\caption{As Fig. \ref{figure:IRAS06206} but for \object{IRAS 17208-0014}.}
\label{figure:IRAS17208}
\end{figure*}

\addtocounter{figure}{-1}
\addtocounter{subfig}{1}
\begin{figure*}
\begin{center}
\includegraphics[angle=0, width=1\textwidth]{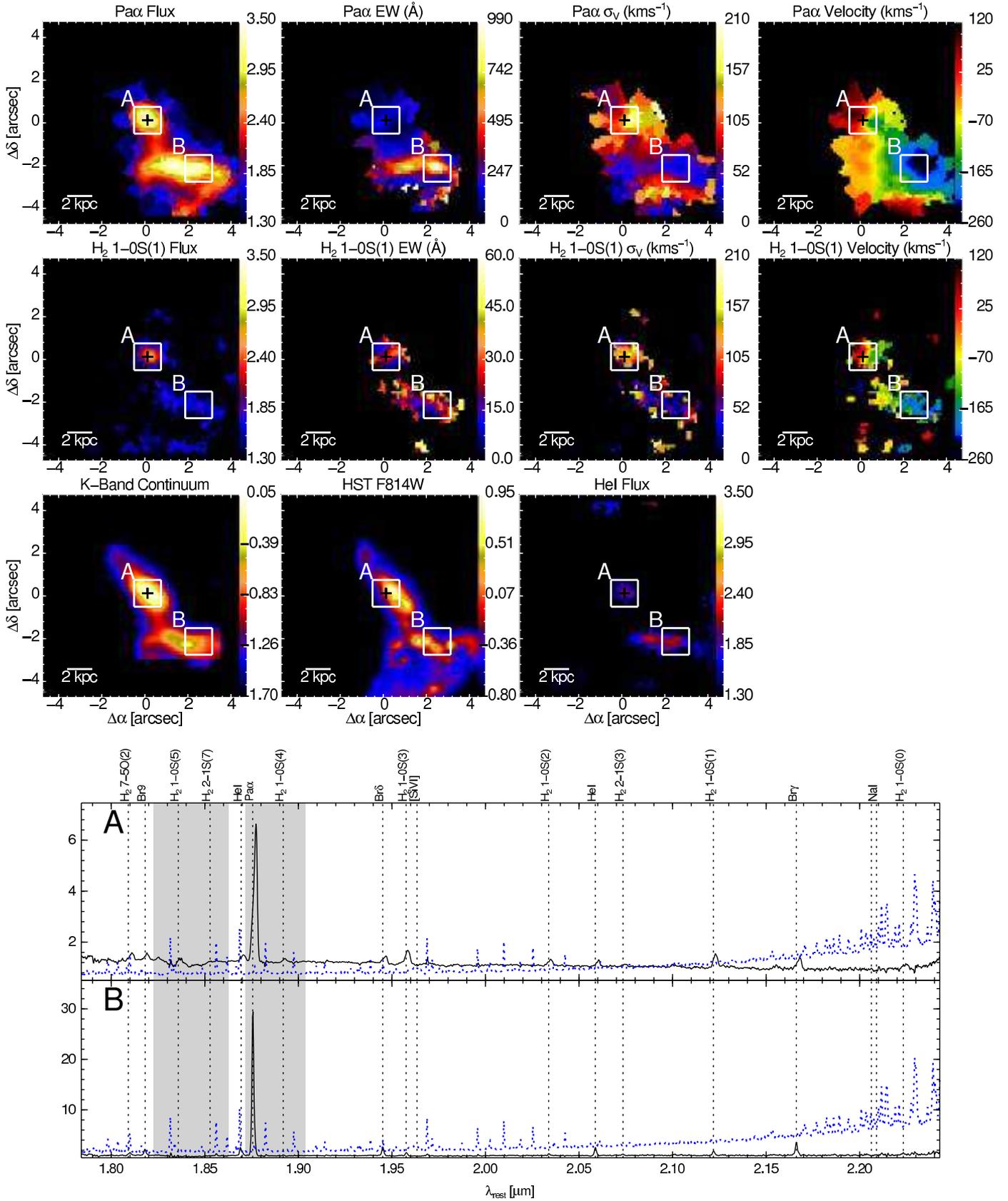}
\end{center}
\caption{As Fig. \ref{figure:IRAS06206} but for \object{IRAS 21130-4446}.}
\label{figure:IRAS21130}
\end{figure*}

\addtocounter{figure}{-1}
\addtocounter{subfig}{1}
\begin{figure*}
\begin{center}
\includegraphics[angle=0, width=1\textwidth]{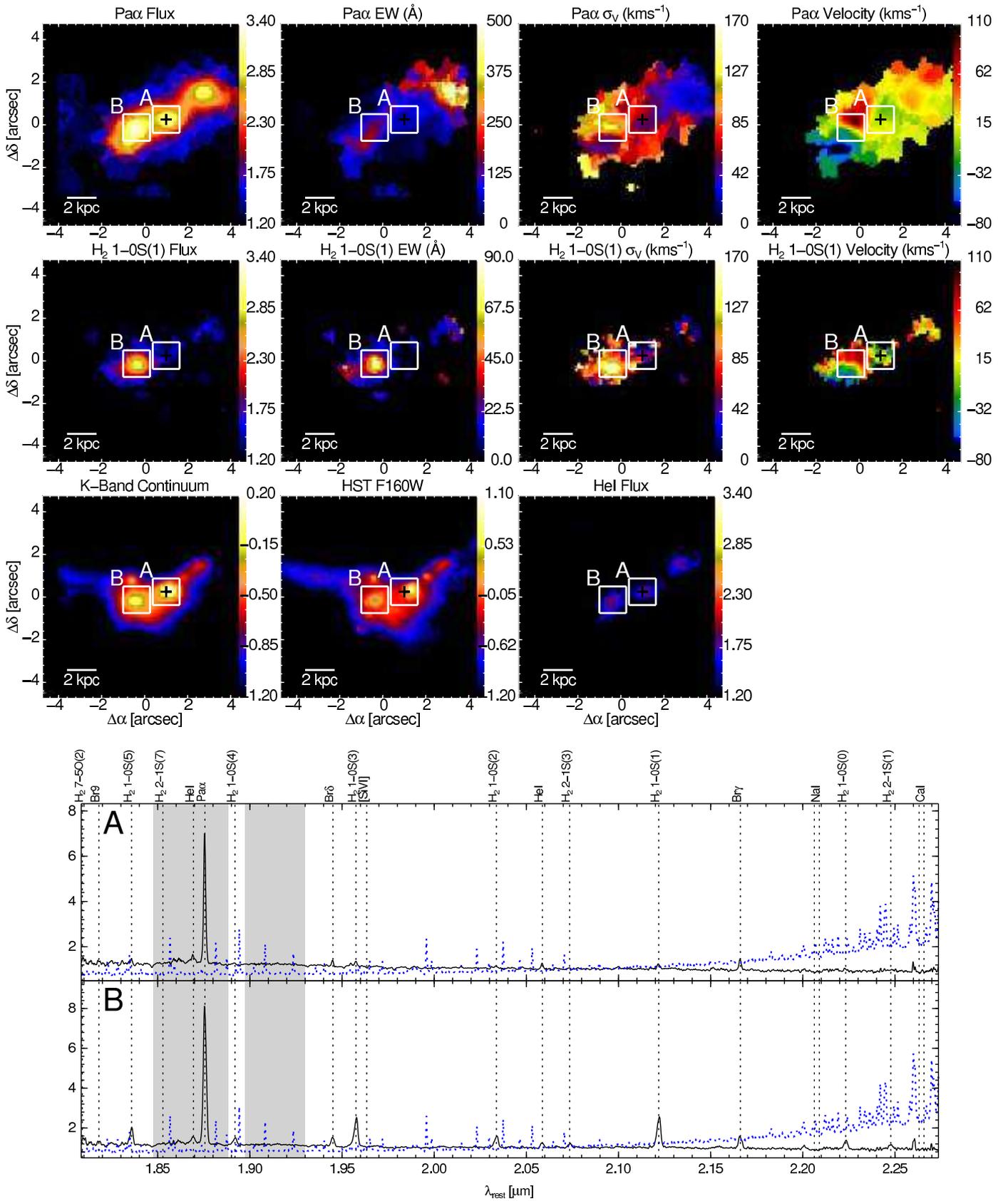}
\end{center}
\caption{As Fig. \ref{figure:IRAS06206} but for \object{IRAS 22491-1808}.}
\label{figure:IRAS22491}
\end{figure*}

\addtocounter{figure}{-1}
\addtocounter{subfig}{1}
\begin{figure*}
\begin{center}
\includegraphics[angle=0, width=0.8\textwidth]{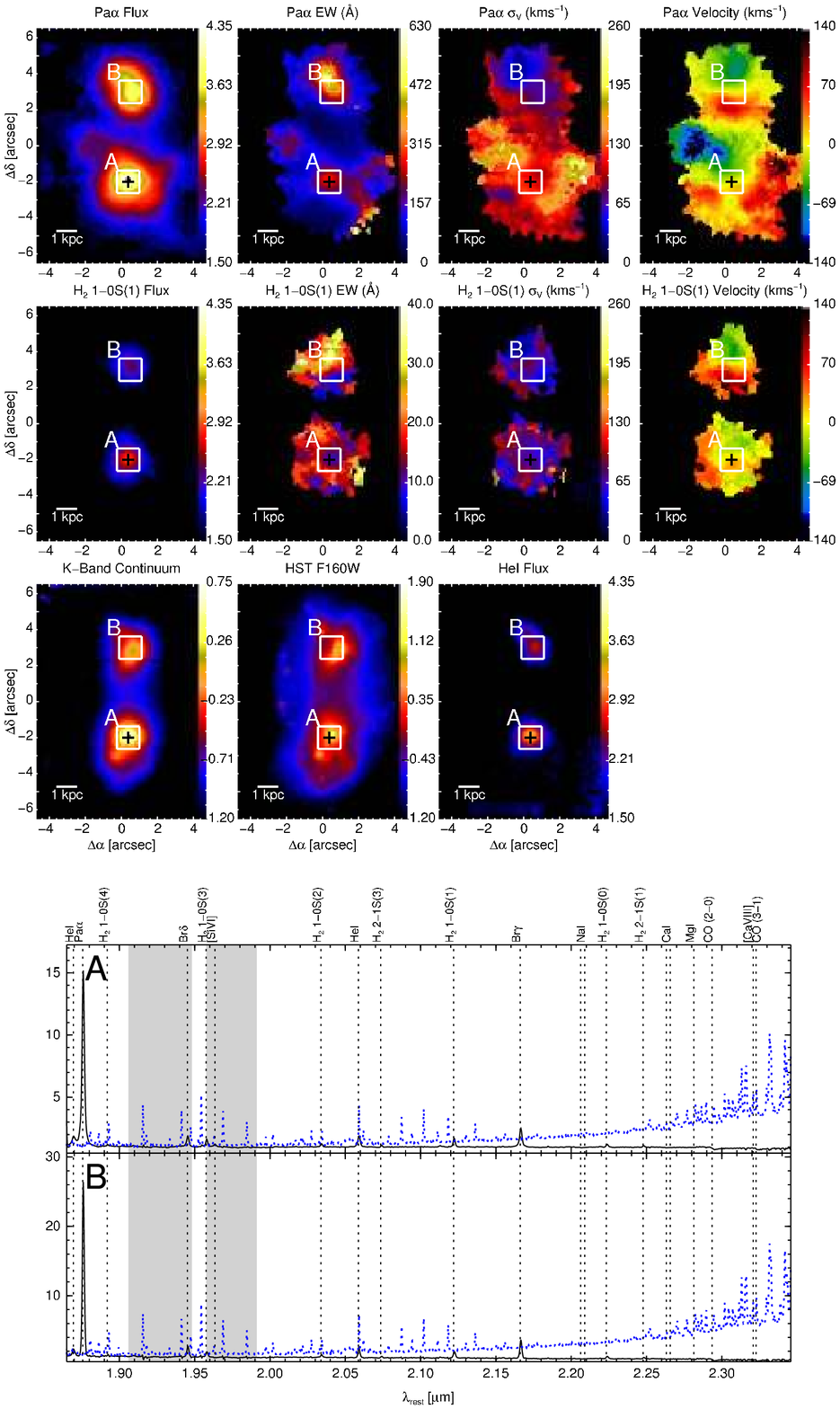}
\end{center}
\caption{As Fig. \ref{figure:IRAS06206} but for \object{IRAS 23128-5919}.}
\label{figure:IRAS23128}
\end{figure*}

\renewcommand{\thefigure}{\arabic{figure}}

\appendix
\section{Notes on individual sources}
\label{notes}
\begin{itemize}

\item[$\circ$]{\bf \object{IRAS 06206-6315}}: This object is a ULIRG classified as a Seyfert 2 galaxy according to its optical spectrum \citep{Duc:1997p7880}. The NICMOS F160W image \citep{Bushouse:2002p8176} shows a double nuclei structure with a tidal tail starting at the north and bending towards the south-east, which is not completely covered by our SINFONI data (Fig. \ref{figure:IRAS06206}). The projected separation between both nuclei is $\sim4.5$\,kpc, and the FoV sampled by SINFONI is $\sim17\times17$\,kpc. The southern nucleus (labelled as ``A" in Fig. \ref{figure:IRAS06206}) is the brightest source in the continuum, \Pa\ and H$_2$ 1-0S(1) lines, and it contributes to the $\sim55$\% of the total \Pa\ emission within the FoV. A local peak of \Pa\ emission is visible at the end of the tidal tail. The northern nucleus (labelled as ``B" in the figure), although similar in \Pa\ brightness to the southern one, is $\sim75$\% less bright in the H$_2$ 1-0S(1) emission.

The kinematics of the ionised gas show the distinct velocity gradients of both progenitors of the interacting system.

\item[$\circ$]{\bf \object{NGC 2369}} (\object{IRAS07160-6215}):  The SINFONI field of view covers the central $\sim2\times2$\,kpc of this almost edge-on spiral LIRG. The NICMOS F160W image \citep{AlonsoHerrero:2006p4703} shows a very complex morphology in the inner regions of the galaxy, with multiple clumps that are not completely resolved in our SINFONI data. One of the brightest sources of \Brg\ emission is covered by aperture ``B" in Fig. \ref{figure:NGC2369}. It is also a bright source in [FeII], but no counterpart is detected in either the continuum image or in the H$_2$ 1-0S(1) map.

The ionised gas kinematics show a strong velocity gradient between regions ``A" and ``B", which may indicate the presence of a warp rotating disk. 

\item[$\circ$]{\bf \object{NGC 3110}} (\object{IRAS10015-0614}): The NICMOS F160W image \citep{AlonsoHerrero:2006p4703} of this almost face-on spiral LIRG shows two well-defined arms that extend for $\sim30$\,kpc, of which $\sim3\times3$\,kpc are covered by our SINFONI data. The arms are outlined by diffuse gas emission that concentrates at the nucleus and in a bright complex to the north of the FoV. The nucleus dominates the emission of both the ionised and molecular gas ($\lesssim20$\% of the total flux), although the star-forming region labelled ``B" in Fig. \ref{figure:NGC3110} has a comparable brightness in \Brg, and even higher in HeI. The [FeII] emission shows the same structure as the ionised and molecular hydrogen.

The kinematics of the different phases of the gas are very similar and show the typical pattern of a thin rotating disk.

\item[$\circ$]{\bf \object{NGC 3256}} (\object{IRAS10257-4338}): Our SINFONI data cover the central $\sim2.5\times3.2$\,kpc of this extreme starburst LIRG. The \Brg\ map shows a very clumpy morphology, with multiple knots of strong emission spread along the southern spiral arm of the object (see \citealt{Lipari:2000p120}, \citealt{Lipari:2004p354}, \citealt{AlonsoHerrero:2006p4703} and references therein). The H$_2$ emission also shows a clumpy distribution with a more diffuse component and two bright spots of strong compact emission in the south of the secondary nucleus (aperture ``A", Fig. \ref{figure:NGC3256}). These knots are also clearly visible in the velocity dispersion map, with values up to $\sim$180\,km\,s$^{-1}$, together with a bright spot east of the southern nucleus (aperture ``B").

\item[$\circ$]{\bf \object{ESO 320-G030}} (\object{IRAS11506-3851}): The NICMOS F160W image \citep{AlonsoHerrero:2006p4703} of this SBA LIRG \citep{Erwin:2004fd} reveals an overall spiral structure that extends over $\sim$32\,kpc. Our SINFONI data cover the inner $\sim2\times2$\,kpc. The \Brg\ map shows a ring-like structure of star formation not observed in the continuum images, with well-defined bright regions (Fig. \ref{figure:ESO320}), as observed in \Pa\ by \cite{AlonsoHerrero:2006p4703}. The maximum of \Brg\ emission is reached by the easternmost region and accounts for $\sim$40\% of the integrated flux. The nucleus is a faint source in \Brg\ but dominates the emission of the molecular hydrogen, with up to $\sim$45\% of the total emission of the H$_2$ 1-0S(1) in the inner $\sim2\times2$\,kpc. The molecular hydrogen emission reveals the structure of the nuclear bar, which connects both sides of the star-forming ring. The [FeII] emission has a similar morphology as the \Brg.

The orientation of the galaxy is almost face-on, and the gas kinematics show a well-defined rotation pattern.

\item[$\circ$]{\bf \object{IRAS 12112+0305}}: This system is a close interacting pair classified as ULIRG and separated $\sim$4.2\,kpc with a bright tidal tail extending 18\,kpc to the north \citep{Surace:2000p8164}. Our SINFONI data covers the central $\sim14\times14$\,kpc of the system (Fig. \ref{figure:IRAS12112}). The southern nucleus is the brightest source in the K-band and in \Pa\ ($\sim35$\% of the integrated flux), although the large \Pa\ EW ($\sim300$\,\AA) measured along the tidal tail suggests a recent burst of star formation in the northern component. Also remarkable is the bright spot of \Pa\ emission in the east of the nuclei, not detected in the continuum, that traces a young massive HII complex that could represent a tidally induced giant extranuclear star-forming region \citep{Colina:2000ApJ533}. The H$_2$ emission is highly concentrated in the northern nucleus, accounting for $\sim52$\% of the total flux of the FoV. The southern nucleus is $\sim42$\% less bright than the northern component, and no significative emission is measured along the tail.

The kinematics of the gas reveal two different velocity gradients for each progenitor, with a steep gradient along the tidal tail and the northern nucleus of more than $\Delta v\sim400$\,km\,s$^{-1}$. 

\item[$\circ$]{\bf \object{IRASF 12115-4656}}:  This spiral-like LIRG might be interacting with IRAS12112-4659, located at $\sim100$\,kpc south-west \citep{Arribas:2008p4403}. Our data cover the central $\sim5\times5$\,kpc and show that a star-forming ring dominates the \Brg\ emission, clearly outlined in the ionised gas map, whereas the nucleus seems to be highly obscured (see Piqueras L\'opez et al. 2012a). The emission from the arms is diffuse, with the exception of a very compact source, labelled ``B" in Fig. \ref{figure:IRASF12115}, where the \Brg\ emission reaches its maximum. The ring is not so clearly visible on the H$_2$ 1-0S(1) map, where the emission is more diffuse, with a peak at the nucleus of the galaxy. As shown in Figs. \ref{figure:SiVI} and \ref{figure:CaVIII}, the [SiVI] line at 1.963\,$\mu$m and the [CaVIII] line at 2.321\,$\mu$m are detected in the nucleus of the galaxy and suggest there is an AGN. The gas kinematics show a smooth velocity gradient along the whole FoV, as is typical of a rotating disk.

\item[$\circ$]{\bf \object{NGC 5135}} (\object{IRAS13229-2934}): This LIRG is an SBab starburst galaxy classified as Seyfert 2 \citep{Bedregal:2009p2426}. Our SINFONI data sample the inner $\sim3\times3$\,kpc of the galaxy, and reveal a high excitation ionisation cone centred on the AGN, and extending up to $\sim$600\,pc radius \citep{Bedregal:2009p2426}. The brightest source of \Brg\ emission (aperture ``B" in Fig. \ref{figure:NGC5135}) do not coincide with the H$_2$ and [FeII] maxima, which are located at the nucleus and at the bright region $\sim2$\,arcsec south-west, respectively. For a detailed study of the ionisation of the different phases of the gas and the star formation activity of NGC~5135 with these SINFONI data, see \cite{Bedregal:2009p2426}.

\item[$\circ$]{\bf \object{IRAS 14348-1447}}: Our SINFONI data sample the central $\sim15\times15$\,kpc of this ULIRG (see Fig. \ref{figure:IRAS14348}). The K-band image shows two bright sources and a diffuse component extending towards the north-east of the FoV, where the \Pa\ map reveals a very bright knot of emission. This region shows high values of \Pa\ EW  up to $\sim$400\AA\ that suggest a young starburst in an inner tidal tail \citep{Colina:2005p3767}. The southern nucleus represents the $\sim52$\% of the integrated \Pa\ and H$_2$ emission, whereas the northern component accounts for the $\sim33$\% and the $\sim26$\% respectively. The molecular emission is highly concentrated at both of the nuclei.

The kinematics of the diffuse ionised gas shows an overall rotation pattern along the system.
The high velocity dispersion measured for the northern component of the system is due to the steep velocity gradient of the gas.

\item[$\circ$]{\bf \object{IRASF 17138-1017}}:  The K-band image of this galaxy reveals that its central regions are rather complex. Our SINFONI data cover the central $\sim3\times3$\,kpc of this almost edge-on spiral LIRG that extends beyond $\sim9$\,kpc. We have identified the nucleus of the galaxy with the brightest spaxel of the K-band image (labelled ``A" in Fig. \ref{figure:IRASF17138}). The \Brg\ emission show a clumpy morphology, with its maximum ($\sim25$\% of the total \Brg\ flux in our FoV) at a bright knot $\sim700$\,pc south of the nucleus, that also reveals a strong emission in HeI. The H$_2$ emission is more diffuse, reaching its maximum at the nucleus of the galaxy. The [FeII] emission has a similar morphology to the \Brg\ and shows different bright sources along the central region of the galaxy, with an underlying diffuse emission. The brightest source is identified with aperture ``B" in Fig. \ref{figure:IRASF17138}.

The kinematics of the gas are, to first order, compatible with a thin rotating disk.

\item[$\circ$]{\bf \object{IRAS 17208-0014}}: The \Pa\ map of this starburst ULIRG \citep{Arribas:2003p4397}, which samples the $\sim7\times7$\,kpc of the object, shows that the ionised gas emission is highly concentrated at the nucleus, and it reveals a young burst of star formation to the south-east (aperture ``B" in Fig. \ref{figure:IRAS17208}), with values of \Pa\ EW $\sim130$\,\AA. The H$_2$ emission is more compact than the \Pa, appears highly concentrated in the nucleus, and extended perpendicularly to the projected plane of the disk.

The kinematics of the gas show a very steep rotation pattern, typical of a disk, and high velocity dispersion values of $\sim250$\,km\,s$^{-1}$ in the nuclear regions.

\item[$\circ$]{\bf \object{IC 4687}} (\object{IRAS18093-5744}): This LIRG is part of a system that involves a group of three galaxies in close interaction \citep{West:1976p7919}. The nuclear separations between the northern (IC~4687) and the central galaxies (IC~4686) and between the central and the southern galaxies (IC~4689) are $\sim$10\,kpc and $\sim$20\,kpc, respectively. The FoV of our SINFONI observations covers the central $\sim3\times3$\,kpc of IC~4687. This object shows a spiral-like morphology with several knots of enhanced \Brg\ emission along its arms and nucleus, in close agreement with the \Pa\ emission from \cite{AlonsoHerrero:2006p4703}. The brightest \Brg\ region is located south of the FoV (labelled ``B" in Fig. \ref{figure:IC4687}), and accounts for the $\sim$25\% of the total \Brg\ emission within the inner $\sim3\times3$\,kpc. Although this region and the nucleus are equally bright, the EW of the \Brg\ line is much higher, up to 120\,\AA, and the intense HeI emission suggests that it is a young star-forming complex. The morphology of the H$_2$ emission is rather different from the observed in the ionised gas, and the peak of emission is located at the nucleus. The [FeII] emission has a similar morphology to the \Brg\ and shows different circumnuclear sources. The peak of emission coincides with the brightest \Brg\ region, located south of the FoV.

The kinematics of the ionised gas on large scales show a smooth velocity gradient along the FoV, with evident signs of deviations from a rotation pattern. 

\item[$\circ$]{\bf \object{IRAS 21130-4446}}: This ULIRG is identified as a double nucleus system with a nuclear separation $\sim5$ kpc, however, the complex morphology of the galaxy makes it hard to locate the exact position of the two nuclei \citep{Cui:2001p8170}. The WFPC2 F814W image of this source shows different condensations and tails, some of them unresolved in our SINFONI data, which cover $\sim14\times14$\,kpc centred in the northern nucleus of the galaxy.  The \Pa\ emission is concentrated at the northern nucleus, marked as ``A" in Fig. \ref{figure:IRAS21130}, and extended along the southern part of the system. It reaches its maximum in these concentrations that extends towards aperture ``B", where the \Pa\ EW is up to $\sim900$\,\AA. The emission from warm molecular gas comes, on the other hand, mostly from the northern nucleus, and only a weak diffuse emission from the southern region is detected. The properties of this southern area suggest an extremely young burst of star formation spreading over the region.

\item[$\circ$]{\bf \object{NGC 7130}} (\object{IRAS~21453-3511}): Our $\sim3\times5$\,kpc SINFONI FoV covers the nucleus and part of the northern spiral arm / tidal tail of NGC~7130. The gas emission is highly concentrated at the nucleus and in a bright star-forming region in the arm (aperture ``B" in Fig. \ref{figure:NGC7130}), three times less bright than the nucleus (see also \citealt{AlonsoHerrero:2006p4703} and \citealt{DiazSantos:2010p4685}). We have tentatively detected [SiVI] emission at 1.963\,$\mu$m and [CaVIII] at 2.321\,$\mu$m, as shown in Figs. \ref{figure:SiVI} and \ref{figure:CaVIII}, in agreement with the LINER and Seyfert-like features observed  in the nuclear, optical spectrum of this LIRG \citep{Veilleux:1995p98}.

The kinematics of the gas reveal a rotation pattern in the nucleus of the galaxy and a steady velocity gradient along the arm / tidal tail.

\item[$\circ$]{\bf \object{IC 5179}} (\object{IRAS~22132-3705}): Our SINFONI data sample the central $\sim4\times2$\,kpc of this spiral LIRG. The gas emission maps show a very clumpy distribution of compact knots of star formation spread along the spiral arms and a very compact nucleus that dominates the emission (see Fig. \ref{figure:IC5179}). It accounts for the $\sim40$\% of the total \Brg\ emission, up to the $\sim30$\% of the H$_2$ 1-0S(1) and $\sim45$\% of the [FeII] emission within our FoV.

The general kinematics of the gas reveal a smooth velocity field, compatible with the rotation pattern of a thin disk.

\item[$\circ$]{\bf \object{IRAS 22491-1808}}: \cite{Cui:2001p8170} propose a multiple merger origin for this ULIRG based on its morphology. The K-band image, which samples the central $\sim12\times12$\,kpc, reveals two nuclei, separated by a projected distance of $\sim2.3$\,kpc, with several knots and condensations in the central regions and along the tidal tails, which extend beyond $\sim5$\,kpc to the east and north-west of the system (see Fig. \ref{figure:IRAS22491}). The \Pa\ map shows three clearly distinguished concentrations extending towards the north-west of the FoV. Two of them coincide with the two nuclei observed in the continuum, whereas the third, located to the north-west, could be associated with the knots of emission that extend along the tail, with values of \Pa\ EW $\sim500$\,\AA. On the other hand, the H$_2$ emission is very concentrated in the eastern nucleus, accounting for $\sim85$\% of the total H$_2$ 1-0S(1) flux. Although the mid-infrared spectrum of this galaxy is consistent with a starburst and an AGN \citep{Farrah:2003iy}, we have not detected [SiVI] emission.

The gas kinematics of the eastern nucleus shows a well differentiated steep gradient of $\Delta v\sim200$\,km\,s$^{-1}$ in $\sim2$\,kpc. Due to beam smearing effects, this gradient enhances the measured values of the velocity dispersion up to $\sim170$\,km\,s$^{-1}$.

\item[$\circ$]{\bf \object{IRAS 23128-5919}}: This ULIRG is a strong interacting system, classed as a mixture of starburst, LINER, and Sy2 (\citealt{Kewley:2001ApJS132}, \citealt{Bushouse:2002p8176}). The projected distance between its double nucleus is $\sim4$\,kpc, and it presents prominent tidal tails, containing many bright knots of emission, extending to the north and south-east over $\sim50$\,kpc \citep{Bushouse:2002p8176}. Our SINFONI data cover the central $\sim7\times11$\,kpc of the system. The double nucleus structure is clearly visible in both the K-band image and the ionised gas map (Fig. \ref{figure:IRAS23128}). The southern nucleus, which coincides with the AGN, is the brightest \Pa\ source, and it accounts for $\sim58$\% of the integrated \Pa\ flux. In contrast, the \Pa\ emission from the northern nucleus seems to be dominated by star-forming activity, with \Pa\ EW up to $\sim600$\,\AA. The presence of an AGN in the southern nucleus is supported by the strong compact [SiVI] emission detected (Fig. \ref{figure:SiVI}). The H$_2$ morphology shows a very peaked and concentrated emission in the southern nucleus ($\sim45$ of the total H$_2$ 1-0S(1) flux), whereas the northern nucleus accounts for half the flux of the southern nucleus.

The gas kinematics of the northern nucleus show a velocity gradient of $\Delta v\sim140$\,km\,s$^{-1}$ in the north-south direction, whereas the southern nucleus kinematics seem to be dominated by the AGN. The gas kinematics show extremely high velocities ($\sim1000$\,,km\,s$^{-1}$) in the ionised gas at radial distances of $\sim2$\,kpc from the southern nucleus, and the presence of molecular gas outflows in the same regions. The high velocity dispersion and the blue/red wings in the \Pa\ and H$_2$ 1-0S(1) line profiles suggest a cone-like structure, centred on the AGN, and extending $\sim$3--4\,kpc to the north-east and south-west (Fig. \ref{figure:IRAS23128}).

\end{itemize}

\bibliographystyle{aa}
\bibliography{bib_file}

\begin{thebibliography}{97}
\expandafter\ifx\csname natexlab\endcsname\relax\def\natexlab#1{#1}\fi

\bibitem[{Alonso-Herrero {et~al.}(2009)Alonso-Herrero, Garc{\'\i}a-Mar{\'\i}n,
  Monreal-Ibero, Colina, Arribas, Alfonso-Garz{\'o}n, \&
  Labiano}]{AlonsoHerrero:2009p3373}
Alonso-Herrero, A., Garc{\'\i}a-Mar{\'\i}n, M., Monreal-Ibero, A., {et~al.}
  2009, A\&A, 506, 1541

\bibitem[{Alonso-Herrero {et~al.}(2012)Alonso-Herrero, Pereira-Santaella,
  Rieke, \& Rigopoulou}]{Alonso-Herrero:2012p744}
Alonso-Herrero, A., Pereira-Santaella, M., Rieke, G.~H., \& Rigopoulou, D.
  2012, ApJ, 744, 2

\bibitem[{Alonso-Herrero {et~al.}(2006)Alonso-Herrero, Rieke, Rieke, Colina,
  P{\'e}rez-Gonz{\'a}lez, \& Ryder}]{AlonsoHerrero:2006p4703}
Alonso-Herrero, A., Rieke, G.~H., Rieke, M.~J., {et~al.} 2006, ApJ, 650, 835

\bibitem[{Alonso-Herrero {et~al.}(2003)Alonso-Herrero, Rieke, Rieke, \&
  Kelly}]{AlonsoHerrero:2003p3431}
Alonso-Herrero, A., Rieke, G.~H., Rieke, M.~J., \& Kelly, D.~M. 2003, AJ, 125,
  1210

\bibitem[{Alonso-Herrero {et~al.}(1997)Alonso-Herrero, Rieke, Rieke, \&
  Ruiz}]{AlonsoHerrero:1997p5041}
Alonso-Herrero, A., Rieke, M.~J., Rieke, G.~H., \& Ruiz, M. 1997, ApJ, 482, 747

\bibitem[{Arribas {et~al.}(2004)Arribas, Bushouse, Lucas, Colina, \&
  Borne}]{Arribas:2004p127}
Arribas, S., Bushouse, H., Lucas, R.~A., Colina, L., \& Borne, K.~D. 2004, AJ,
  127, 2522

\bibitem[{Arribas {et~al.}(1998)Arribas, Carter, Cavaller, del Burgo, Edwards,
  Fuentes, Garcia, Herreros, {et~al.}}]{Arribas:1998p8476}
Arribas, S., Carter, D., Cavaller, L., {et~al.} 1998, SPIE, 3355, 821

\bibitem[{Arribas \& Colina(2003)}]{Arribas:2003p4397}
Arribas, S. \& Colina, L. 2003, ApJ, 591, 791

\bibitem[{Arribas {et~al.}(2012)Arribas, Colina, Alonso-Herrero,
  Rosales-Ortega, Monreal-Ibero, Garc{\'\i}a-Mar{\'\i}n, Garcia-Burillo, \&
  Rodriguez-Zaur{\'\i}n}]{Arribas:2012p1203}
Arribas, S., Colina, L., Alonso-Herrero, A., {et~al.} 2012, A\&A, 541, A20

\bibitem[{Arribas {et~al.}(2008)Arribas, Colina, Monreal-Ibero, Alfonso,
  Garc{\'\i}a-Mar{\'\i}n, \& Alonso-Herrero}]{Arribas:2008p4403}
Arribas, S., Colina, L., Monreal-Ibero, A., {et~al.} 2008, A\&A, 479, 687

\bibitem[{Bedregal {et~al.}(2009)Bedregal, Colina, Alonso-Herrero, \&
  Arribas}]{Bedregal:2009p2426}
Bedregal, A.~G., Colina, L., Alonso-Herrero, A., \& Arribas, S. 2009, ApJ, 698,
  1852

\bibitem[{Blietz {et~al.}(1994)Blietz, Cameron, Drapatz, Genzel, Krabbe, Van
  Der~Werf, Sternberg, \& Ward}]{Blietz:1994ApJ421}
Blietz, M., Cameron, M., Drapatz, S., {et~al.} 1994, ApJ, 421, 92

\bibitem[{Borne {et~al.}(2000)Borne, Bushouse, Lucas, \&
  Colina}]{Borne:2000ApJ529}
Borne, K.~D., Bushouse, H., Lucas, R.~A., \& Colina, L. 2000, ApJ, 529, L77

\bibitem[{Bushouse {et~al.}(2002)Bushouse, Borne, Colina, Lucas,
  Rowan-Robinson, Baker, Clements, Lawrence, {et~al.}}]{Bushouse:2002p8176}
Bushouse, H.~A., Borne, K.~D., Colina, L., {et~al.} 2002, ApJS, 138, 1

\bibitem[{Cappellari \& Copin(2003)}]{Cappellari:2003p4908}
Cappellari, M. \& Copin, Y. 2003, MNRAS, 342, 345

\bibitem[{Cappellari \& Emsellem(2004)}]{Cappellari:2004p4916}
Cappellari, M. \& Emsellem, E. 2004, PASP, 116, 138

\bibitem[{Cohen {et~al.}(2003)Cohen, Wheaton, \& Megeath}]{Cohen:2003p3372}
Cohen, M., Wheaton, W.~A., \& Megeath, S.~T. 2003, AJ, 126, 1090

\bibitem[{Colina(1993)}]{Colina:1993p6004}
Colina, L. 1993, ApJ, 411, 565

\bibitem[{Colina {et~al.}(2000)Colina, Arribas, Borne, \&
  Monreal}]{Colina:2000ApJ533}
Colina, L., Arribas, S., Borne, K.~D., \& Monreal, A. 2000, ApJ, 533, L9

\bibitem[{Colina {et~al.}(2005)Colina, Arribas, \&
  Monreal-Ibero}]{Colina:2005p3767}
Colina, L., Arribas, S., \& Monreal-Ibero, A. 2005, ApJ, 621, 725

\bibitem[{Cui {et~al.}(2001)Cui, Xia, Deng, Mao, \& Zou}]{Cui:2001p8170}
Cui, J., Xia, X.-Y., Deng, Z.-G., Mao, S., \& Zou, Z.-L. 2001, AJ, 122, 63

\bibitem[{Dale {et~al.}(2004)Dale, Roussel, Contursi, Helou, Dinerstein,
  Hunter, Hollenbach, Egami, {et~al.}}]{Dale:2004ApJ601}
Dale, D.~A., Roussel, H., Contursi, A., {et~al.} 2004, ApJ, 601, 813

\bibitem[{Dasyra {et~al.}(2006)Dasyra, Tacconi, Davies, Naab, Genzel, Lutz,
  Sturm, Baker, {et~al.}}]{Dasyra:2006ApJ651}
Dasyra, K.~M., Tacconi, L.~J., Davies, R.~I., {et~al.} 2006, ApJ, 651, 835

\bibitem[{Davies(2007)}]{Davies:2007p2525}
Davies, R.~I. 2007, MNRAS, 375, 1099

\bibitem[{Davies {et~al.}(2003)Davies, Sternberg, Lehnert, \&
  Tacconi-Garman}]{Davies:2003p3644}
Davies, R.~I., Sternberg, A., Lehnert, M., \& Tacconi-Garman, L.~E. 2003, ApJ,
  597, 907

\bibitem[{Davies {et~al.}(2005)Davies, Sternberg, Lehnert, \&
  Tacconi-Garman}]{Davies:2005p3646}
Davies, R.~I., Sternberg, A., Lehnert, M.~D., \& Tacconi-Garman, L.~E. 2005,
  ApJ, 633, 105

\bibitem[{D{\'\i}az-Santos {et~al.}(2010)D{\'\i}az-Santos, Alonso-Herrero,
  Colina, Packham, Levenson, Pereira-Santaella, Roche, \&
  Telesco}]{DiazSantos:2010p4685}
D{\'\i}az-Santos, T., Alonso-Herrero, A., Colina, L., {et~al.} 2010, ApJ, 711,
  328

\bibitem[{D{\'\i}az-Santos {et~al.}(2008)D{\'\i}az-Santos, Alonso-Herrero,
  Colina, Packham, Radomski, \& Telesco}]{DiazSantos:2008p685}
D{\'\i}az-Santos, T., Alonso-Herrero, A., Colina, L., {et~al.} 2008, ApJ, 685,
  211

\bibitem[{Doherty {et~al.}(1995)Doherty, Puxley, Lumsden, \&
  Doyon}]{Doherty:1995MNRAS277}
Doherty, R.~M., Puxley, P.~J., Lumsden, S.~L., \& Doyon, R. 1995, MNRAS, 277,
  577

\bibitem[{Duc {et~al.}(1997)Duc, Mirabel, \& Maza}]{Duc:1997p7880}
Duc, P.-A., Mirabel, I.~F., \& Maza, J. 1997, A\&AS, 124, 533

\bibitem[{Eisenhauer {et~al.}(2003)Eisenhauer, Abuter, Bickert,
  Biancat-Marchet, Bonnet, Brynnel, Conzelmann, Delabre,
  {et~al.}}]{Eisenhauer:2003p8484}
Eisenhauer, F., Abuter, R., Bickert, K., {et~al.} 2003, SPIE, 4841, 1548

\bibitem[{Epinat {et~al.}(2012)Epinat, Tasca, Amram, Contini, Le~Fevre,
  Queyrel, Vergani, Garilli, {et~al.}}]{Epinat:2012ga}
Epinat, B., Tasca, L., Amram, P., {et~al.} 2012, A\&A, 539, A92

\bibitem[{Erwin(2004)}]{Erwin:2004fd}
Erwin, P. 2004, A\&A, 415, 941

\bibitem[{Farrah {et~al.}(2003)Farrah, Afonso, Efstathiou, Rowan-Robinson, Fox,
  \& Clements}]{Farrah:2003iy}
Farrah, D., Afonso, J., Efstathiou, A., {et~al.} 2003, MNRAS, 343, 585

\bibitem[{Ferland {et~al.}(2008)Ferland, Fabian, Hatch, Johnstone, Porter, van
  Hoof, \& Williams}]{Ferland:2008p8001}
Ferland, G.~J., Fabian, A.~C., Hatch, N.~A., {et~al.} 2008, MNRAS, 386, L72

\bibitem[{F{\"o}rster~Schreiber(2000)}]{ForsterSchreiber:2000AJ120}
F{\"o}rster~Schreiber, N.~M. 2000, AJ, 120, 2089

\bibitem[{F{\"o}rster~Schreiber {et~al.}(2009)F{\"o}rster~Schreiber, Genzel,
  Bouch{\'e}, Cresci, Davies, Buschkamp, Shapiro, Tacconi,
  {et~al.}}]{ForsterSchreiber:2009p706}
F{\"o}rster~Schreiber, N.~M., Genzel, R., Bouch{\'e}, N., {et~al.} 2009, ApJ,
  706, 1364

\bibitem[{F{\"o}rster~Schreiber {et~al.}(2006)F{\"o}rster~Schreiber, Genzel,
  Lehnert, Bouch{\'e}, Verma, Erb, Shapley, Steidel,
  {et~al.}}]{ForsterSchreiber:2006ApJ645}
F{\"o}rster~Schreiber, N.~M., Genzel, R., Lehnert, M.~D., {et~al.} 2006, ApJ,
  645, 1062

\bibitem[{F{\"o}rster~Schreiber {et~al.}(2011)F{\"o}rster~Schreiber, Shapley,
  Erb, Genzel, Steidel, Bouch{\'e}, Cresci, \&
  Davies}]{ForsterSchreiber:2011p731}
F{\"o}rster~Schreiber, N.~M., Shapley, A.~E., Erb, D.~K., {et~al.} 2011, ApJ,
  731, 65

\bibitem[{Garc{\'\i}a-Mar{\'\i}n
  {et~al.}(2009{\natexlab{a}})Garc{\'\i}a-Mar{\'\i}n, Colina, \&
  Arribas}]{GarciaMarin:2009p8459}
Garc{\'\i}a-Mar{\'\i}n, M., Colina, L., \& Arribas, S. 2009{\natexlab{a}},
  A\&A, 505, 1017

\bibitem[{Garc{\'\i}a-Mar{\'\i}n
  {et~al.}(2009{\natexlab{b}})Garc{\'\i}a-Mar{\'\i}n, Colina, Arribas, \&
  Monreal-Ibero}]{GarciaMarin:2009p4348}
Garc{\'\i}a-Mar{\'\i}n, M., Colina, L., Arribas, S., \& Monreal-Ibero, A.
  2009{\natexlab{b}}, A\&A, 505, 1319

\bibitem[{Haan {et~al.}(2011)Haan, Surace, Armus, Evans, Howell, Mazzarella,
  Kim, Vavilkin, {et~al.}}]{Haan:2011p141}
Haan, S., Surace, J.~A., Armus, L., {et~al.} 2011, AJ, 141, 100

\bibitem[{Kennicutt(1998)}]{Kennicutt:1998p36}
Kennicutt, R. C.~J. 1998, ARA\&A, 36, 189

\bibitem[{Kewley {et~al.}(2001)Kewley, Heisler, Dopita, \&
  Lumsden}]{Kewley:2001ApJS132}
Kewley, L.~J., Heisler, C.~A., Dopita, M.~A., \& Lumsden, S. 2001, ApJS, 132,
  37

\bibitem[{Kotilainen {et~al.}(1996)Kotilainen, Moorwood, Ward, \&
  Forbes}]{Kotilainen:1996p305}
Kotilainen, J.~K., Moorwood, A. F.~M., Ward, M.~J., \& Forbes, D.~A. 1996,
  A\&A, 305, 107

\bibitem[{Labrie \& Pritchet(2006)}]{Labrie:2006p166}
Labrie, K. \& Pritchet, C.~J. 2006, ApJS, 166, 188

\bibitem[{Law {et~al.}(2009)Law, Steidel, Erb, Larkin, Pettini, Shapley, \&
  Wright}]{Law:2009p697}
Law, D.~R., Steidel, C.~C., Erb, D.~K., {et~al.} 2009, ApJ, 697, 2057

\bibitem[{LeF{\`e}vre {et~al.}(2003)LeF{\`e}vre, Saisse, Mancini, Brau-Nogue,
  Caputi, Castinel, D'Odorico, Garilli, {et~al.}}]{LeFevre:2003p8480}
LeF{\`e}vre, O., Saisse, M., Mancini, D., {et~al.} 2003, SPIE, 4841, 1670

\bibitem[{Leitherer {et~al.}(1999)Leitherer, Schaerer, Goldader,
  Gonz{\'a}lez~Delgado, Robert, Kune, de~Mello, Devost,
  {et~al.}}]{Leitherer:1999p6938}
Leitherer, C., Schaerer, D., Goldader, J.~D., {et~al.} 1999, ApJS, 123, 3

\bibitem[{L{\'\i}pari {et~al.}(2000)L{\'\i}pari, D{\'\i}az, Taniguchi,
  Terlevich, Dottori, \& Carranza}]{Lipari:2000p120}
L{\'\i}pari, S., D{\'\i}az, R., Taniguchi, Y., {et~al.} 2000, AJ, 120, 645

\bibitem[{L{\'\i}pari {et~al.}(2004)L{\'\i}pari, D{\'\i}az, Forte, Terlevich,
  Taniguchi, Aguero, Alonso-Herrero, Mediavilla, {et~al.}}]{Lipari:2004p354}
L{\'\i}pari, S.~L., D{\'\i}az, R.~J., Forte, J.~C., {et~al.} 2004, MNRAS, 354,
  L1

\bibitem[{Lonsdale {et~al.}(2006)Lonsdale, Farrah, \&
  Smith}]{Lonsdale:2006p4228}
Lonsdale, C.~J., Farrah, D., \& Smith, H.~E. 2006, Astrophysics Update 2, 285

\bibitem[{Lumsden {et~al.}(2001)Lumsden, Puxley, \&
  Hoare}]{Lumsden:2001MNRAS320}
Lumsden, S.~L., Puxley, P.~J., \& Hoare, M.~G. 2001, MNRAS, 320, 83

\bibitem[{Lumsden {et~al.}(2003)Lumsden, Puxley, Hoare, Moore, \&
  Ridge}]{Lumsden:2003MNRAS340}
Lumsden, S.~L., Puxley, P.~J., Hoare, M.~G., Moore, T. J.~T., \& Ridge, N.~A.
  2003, MNRAS, 340, 799

\bibitem[{Lutz {et~al.}(1999)Lutz, Veilleux, \& Genzel}]{Lutz:1999ApJ517L.13L}
Lutz, D., Veilleux, S., \& Genzel, R. 1999, ApJ, 517, L13

\bibitem[{Maraston(1998)}]{Maraston:1998p2886}
Maraston, C. 1998, MNRAS, 300, 872

\bibitem[{Maraston(2005)}]{Maraston:2005p2885}
Maraston, C. 2005, MNRAS, 362, 799

\bibitem[{Markwardt(2009)}]{Markwardt:2009p7399}
Markwardt, C.~B. 2009, Astronomical Data Analysis Software and Systems XVIII
  ASP Conference Series, 411, 251

\bibitem[{Monreal-Ibero {et~al.}(2010)Monreal-Ibero, Arribas, Colina,
  Rodriguez-Zaur{\'\i}n, Alonso-Herrero, \&
  Garc{\'\i}a-Mar{\'\i}n}]{Monreal-Ibero:2010p517}
Monreal-Ibero, A., Arribas, S., Colina, L., {et~al.} 2010, A\&A, 517, 28

\bibitem[{Mouri {et~al.}(2000)Mouri, Kawara, \& Taniguchi}]{Mouri:2000p6654}
Mouri, H., Kawara, K., \& Taniguchi, Y. 2000, ApJ, 528, 186

\bibitem[{Murphy {et~al.}(1996)Murphy, Armus, Matthews, Soifer, Mazzarella,
  Shupe, Strauss, \& Neugebauer}]{Murphy:1996p8249}
Murphy, T.~W., Armus, L., Matthews, K., {et~al.} 1996, AJ, 111, 1025

\bibitem[{Nardini {et~al.}(2010)Nardini, Risaliti, Watabe, Salvati, \&
  Sani}]{Nardini:2010p405}
Nardini, E., Risaliti, G., Watabe, Y., Salvati, M., \& Sani, E. 2010, MNRAS,
  405, 2505

\bibitem[{Pereira-Santaella {et~al.}(2010)Pereira-Santaella, Alonso-Herrero,
  Rieke, Colina, D{\'\i}az-Santos, Smith, P{\'e}rez-Gonz{\'a}lez, \&
  Engelbracht}]{PereiraSantaella:2010p4662}
Pereira-Santaella, M., Alonso-Herrero, A., Rieke, G.~H., {et~al.} 2010, ApJS,
  188, 447

\bibitem[{Pereira-Santaella {et~al.}(2011)Pereira-Santaella, Alonso-Herrero,
  Santos-Lleo, Colina, Jim{\'e}nez-Bail{\'o}n, Longinotti, Rieke, Ward,
  {et~al.}}]{PereiraSantaella:2011p535}
Pereira-Santaella, M., Alonso-Herrero, A., Santos-Lleo, M., {et~al.} 2011,
  A\&A, 535, 93

\bibitem[{P{\'e}rez-Gonz{\'a}lez {et~al.}(2005)P{\'e}rez-Gonz{\'a}lez, Rieke,
  Egami, Alonso-Herrero, Dole, Papovich, Blaylock, Jones,
  {et~al.}}]{PerezGonzalez:2005p4031}
P{\'e}rez-Gonz{\'a}lez, P.~G., Rieke, G.~H., Egami, E., {et~al.} 2005, ApJ,
  630, 82

\bibitem[{Prieto {et~al.}(2005)Prieto, Marco, \& Gallimore}]{Prieto:2005p364}
Prieto, M.~A., Marco, O., \& Gallimore, J. 2005, MNRAS, 364, L28

\bibitem[{Riffel {et~al.}(2010)Riffel, Storchi-Bergmann, \&
  Nagar}]{Riffel:2010MNRAS404}
Riffel, R.~A., Storchi-Bergmann, T., \& Nagar, N.~M. 2010, MNRAS, 404, 166

\bibitem[{Rodr{\'\i}guez-Ardila {et~al.}(2004)Rodr{\'\i}guez-Ardila, Pastoriza,
  Viegas, Sigut, \& Pradhan}]{Rodriguez-Ardila:2004p425}
Rodr{\'\i}guez-Ardila, A., Pastoriza, M.~G., Viegas, S., Sigut, T. A.~A., \&
  Pradhan, A.~K. 2004, A\&A, 425, 457

\bibitem[{Rodr{\'\i}guez-Ardila {et~al.}(2011)Rodr{\'\i}guez-Ardila, Prieto,
  Portilla, \& Tejeiro}]{RodriguezArdila2011ApJ743}
Rodr{\'\i}guez-Ardila, A., Prieto, M.~A., Portilla, J.~G., \& Tejeiro, J.~M.
  2011, ApJ, 743, 100

\bibitem[{Rodr{\'\i}guez-Ardila {et~al.}(2006)Rodr{\'\i}guez-Ardila, Prieto,
  Viegas, \& Gruenwald}]{Rodriguez-Ardila:2006ApJ653}
Rodr{\'\i}guez-Ardila, A., Prieto, M.~A., Viegas, S., \& Gruenwald, R. 2006,
  ApJ, 653, 1098

\bibitem[{Rodr{\'\i}guez-Ardila {et~al.}(2005)Rodr{\'\i}guez-Ardila, Riffel, \&
  Pastoriza}]{Rodriguez-Ardila:2005p364}
Rodr{\'\i}guez-Ardila, A., Riffel, R., \& Pastoriza, M.~G. 2005, MNRAS, 364,
  1041

\bibitem[{Rodriguez-Zaur{\'\i}n {et~al.}(2011)Rodriguez-Zaur{\'\i}n, Arribas,
  Monreal-Ibero, Colina, Alonso-Herrero, \&
  Alfonso-Garz{\'o}n}]{2011A&A...527A..60R}
Rodriguez-Zaur{\'\i}n, J., Arribas, S., Monreal-Ibero, A., {et~al.} 2011, A\&A,
  527, 60

\bibitem[{Rosales-Ortega {et~al.}(2012)Rosales-Ortega, Arribas, \&
  Colina}]{RosalesOrtega:2012A&A539}
Rosales-Ortega, F.~F., Arribas, S., \& Colina, L. 2012, A\&A, 539, 73

\bibitem[{Rosenberg {et~al.}(2012)Rosenberg, van~der Werf, \&
  Israel}]{Rosenberg:2012A&A540}
Rosenberg, M. J.~F., van~der Werf, P.~P., \& Israel, F.~P. 2012, A\&A, 540, 116

\bibitem[{Roth {et~al.}(2005)Roth, Kelz, Fechner, Hahn, Bauer, Becker,
  B{\"o}hm, Christensen, {et~al.}}]{Roth:2005p4504}
Roth, M.~M., Kelz, A., Fechner, T., {et~al.} 2005, PASP, 117, 620

\bibitem[{Sanders {et~al.}(1995)Sanders, Egami, L{\'\i}pari, Mirabel, \&
  Soifer}]{Sanders:1995AJ110}
Sanders, D.~B., Egami, E., L{\'\i}pari, S., Mirabel, I.~F., \& Soifer, B.~T.
  1995, AJ, 110, 1993

\bibitem[{Sanders {et~al.}(2003)Sanders, Mazzarella, Kim, Surace, \&
  Soifer}]{Sanders:2003p1433}
Sanders, D.~B., Mazzarella, J.~M., Kim, D.-C., Surace, J.~A., \& Soifer, B.~T.
  2003, AJ, 126, 1607

\bibitem[{Sanders \& Mirabel(1996)}]{Sanders:1996p845}
Sanders, D.~B. \& Mirabel, I.~F. 1996, ARA\&A, 34, 749

\bibitem[{Sargent {et~al.}(2012)Sargent, B{\'e}thermin, Daddi, \&
  Elbaz}]{Sargent:2012ApJ747}
Sargent, M.~T., B{\'e}thermin, M., Daddi, E., \& Elbaz, D. 2012, ApJ, 747, L31

\bibitem[{Shields(1993)}]{Shields:1993ApJ419}
Shields, J.~C. 1993, ApJ, 419, 181

\bibitem[{Skrutskie {et~al.}(2006)Skrutskie, Cutri, Stiening, Weinberg,
  Schneider, Carpenter, Beichman, Capps, {et~al.}}]{Skrutskie:2006p5780}
Skrutskie, M.~F., Cutri, R.~M., Stiening, R., {et~al.} 2006, AJ, 131, 1163

\bibitem[{Soifer {et~al.}(1989)Soifer, Boehmer, Neugebauer, \&
  Sanders}]{Soifer:1989AJ98}
Soifer, B.~T., Boehmer, L., Neugebauer, G., \& Sanders, D.~B. 1989, AJ, 98, 766

\bibitem[{Soifer {et~al.}(1984)Soifer, Neugebauer, Helou, Lonsdale, Hacking,
  Rice, Houck, Low, {et~al.}}]{Soifer:1984p8204}
Soifer, B.~T., Neugebauer, G., Helou, G., {et~al.} 1984, ApJ, 283, L1

\bibitem[{Surace {et~al.}(2000)Surace, Sanders, \& Evans}]{Surace:2000p8164}
Surace, J.~A., Sanders, D.~B., \& Evans, A.~S. 2000, ApJ, 529, 170

\bibitem[{Valencia-S {et~al.}(2012)Valencia-S, Zuther, Eckart,
  Garc{\'\i}a-Mar{\'\i}n, Iserlohe, \& Wright}]{ValenciaS:2012up}
Valencia-S, M., Zuther, J., Eckart, A., {et~al.} 2012, arXiv, astro-ph.CO

\bibitem[{van~der Werf(2000)}]{vanderWerf:2000vb}
van~der Werf, P.~P. 2000, arXiv, astro-ph

\bibitem[{Veilleux {et~al.}(2002)Veilleux, Kim, \& Sanders}]{Veilleux:2002p760}
Veilleux, S., Kim, D.-C., \& Sanders, D.~B. 2002, ApJS, 143, 315

\bibitem[{Veilleux {et~al.}(1995)Veilleux, Kim, Sanders, Mazzarella, \&
  Soifer}]{Veilleux:1995p98}
Veilleux, S., Kim, D.-C., Sanders, D.~B., Mazzarella, J.~M., \& Soifer, B.~T.
  1995, ApJS, 98, 171

\bibitem[{Vergani {et~al.}(2012)Vergani, Epinat, Contini, Tasca, Tresse, Amram,
  Garilli, Kissler-Patig, {et~al.}}]{2012arXiv1202.3107V}
Vergani, D., Epinat, B., Contini, T., {et~al.} 2012, arXiv, 1202, 3107

\bibitem[{V{\'e}ron-Cetty \& V{\'e}ron(2006)}]{VeronCetty:2006p4241}
V{\'e}ron-Cetty, M.-P. \& V{\'e}ron, P. 2006, A\&A, 455, 773

\bibitem[{Wamsteker {et~al.}(1985)Wamsteker, Prieto, Vitores, Schuster, Danksa,
  Gonzalez, \& Rodriguez}]{Wamsteker:1985p7881}
Wamsteker, W., Prieto, A., Vitores, A., {et~al.} 1985, A\&AS, 62, 255

\bibitem[{West(1976)}]{West:1976p7919}
West, R.~M. 1976, A\&A, 46, 327

\bibitem[{Winge {et~al.}(2009)Winge, Riffel, \&
  Storchi-Bergmann}]{Winge:2009p4732}
Winge, C., Riffel, R.~A., \& Storchi-Bergmann, T. 2009, ApJS, 185, 186

\bibitem[{Wisnioski {et~al.}(2011)Wisnioski, Glazebrook, Blake, Wyder, Martin,
  Poole, Sharp, Couch, {et~al.}}]{Wisnioski:2011p417}
Wisnioski, E., Glazebrook, K., Blake, C., {et~al.} 2011, MNRAS, 417, 2601

\bibitem[{Wright {et~al.}(2003)Wright, Egan, Kraemer, \&
  Price}]{Wright:2003p4322}
Wright, C.~O., Egan, M.~P., Kraemer, K.~E., \& Price, S.~D. 2003, AJ, 125, 359

\bibitem[{Wright(2006)}]{Wright:2006p4236}
Wright, E.~L. 2006, PASP, 118, 1711

\bibitem[{Wright {et~al.}(2009)Wright, Larkin, Law, Steidel, Shapley, \&
  Erb}]{Wright:2009p699}
Wright, S.~A., Larkin, J.~E., Law, D.~R., {et~al.} 2009, ApJ, 699, 421

\end{thebibliography}
\end{document}